\documentclass[aps,prx,onecolumn,10pt,superscriptaddress,longbibliography,nofootinbib]{revtex4-2}
\usepackage{amsmath}
\usepackage{amsfonts}
\usepackage{amsbsy} 
\usepackage{amsthm}
\usepackage{algorithm}
\usepackage{amssymb}
\usepackage{graphicx}
\usepackage{color}
\usepackage{changes}
\usepackage{parskip}
\usepackage{qcircuit}
\usepackage{braket}
\usepackage{bbm}
\usepackage{bm}
\usepackage{comment}

\usepackage{blindtext}
\usepackage{algorithmic}
\usepackage{array}
\usepackage[makeroom]{cancel}
\usepackage{comment}
\usepackage{qcircuit}
\usepackage{braket}
\usepackage{bbm}
\usepackage{bm}
\usepackage{enumitem}   
\usepackage{multirow}
\usepackage{setspace}

\usepackage[version=4]{mhchem} % Chemistry Formula

\usepackage{blkarray}

\usepackage{lipsum}

\usepackage{indentfirst}
\usepackage{subfigure} % For subfiguress
\usepackage{makecell}

\usepackage[colorlinks=true,allcolors=blue]{hyperref}

\DeclareMathOperator*{\argmin}{arg\,min}
\DeclareMathOperator{\E}{\mathrm{e}}
\DeclareMathOperator{\I}{\mathrm{i}}

\newtheorem{theorem}{Theorem}
\newtheorem{corollaryinner}{Corollary}[theorem] % just the internal version

\newtheorem{lemma}{Lemma}
\newtheorem{remark}{Remark}[theorem]
\ExplSyntaxOn
\NewDocumentEnvironment{corollary}{o}
 {
  \IfValueT { #1 }
   {
    \giobrach_corollary_setup:n { #1 }
   }
  \corollaryinner
 }
 { \endcorollaryinner }
\prop_new:N \g_giobrach_corollary_prop
\cs_new_protected:Nn \giobrach_corollary_setup:n
 {
  \prop_gput:Nnx \g_giobrach_corollary_prop { #1 }
   {
    \int_eval:n { \prop_item:Nn \g_giobrach_corollary_prop { #1 } + 1 }
   }
  \cs_set:Npn \thecorollaryinner {\ref {#1}.\prop_item:Nn \g_giobrach_corollary_prop { #1 }}
 }
\ExplSyntaxOff

\begin{document}

\title{Filtered Quantum Phase Estimation}

\author{Gwonhak Lee}
\affiliation{SKKU Advanced Institute of Nanotechnology (SAINT), Sungkyunkwan University, Suwon 16419, Republic of Korea}
\author{Minhyeok Kang}
\affiliation{SKKU Advanced Institute of Nanotechnology (SAINT), Sungkyunkwan University, Suwon 16419, Republic of Korea}
\author{Jungsoo Hong}
\affiliation{SKKU Advanced Institute of Nanotechnology (SAINT), Sungkyunkwan University, Suwon 16419, Republic of Korea}
\author{Stepan Fomichev}
\affiliation{Xanadu, Toronto, ON, M5G2C8, Canada}
\author{Joonsuk Huh}
\email{joonsukhuh@gmail.com}
\affiliation{Department of Chemistry, Yonsei University, Seoul 03722, Republic of Korea}
\affiliation{Department of Quantum Information, Yonsei University, Incheon 21983, Republic of Korea}
\date{May 15, 2026}

\begin{abstract}
    % [REVISION-v2] Reframed the abstract to emphasize the cost-aware framework, the modified Krylov construction, and the regime-dependent scope of the numerical claims.
    Accurate state preparation is a critical bottleneck in many quantum algorithms, particularly those for ground-state energy estimation.
    Even in fault-tolerant quantum computing, preparing a quantum state with sufficient overlap with the desired eigenstate remains a major challenge.
    To address this, we develop a unified cost-aware framework for filtered-state preparation that enhances the overlap of a given input state through spectral filtering.
    The framework covers polynomial and trigonometric realizations of filters and makes explicit the trade-off among overlap amplification, preparation success probability, and filter-implementation cost.
    As representative examples, we analyze Gaussian filters and introduce a modified Krylov-subspace-based filter that improves the success-probability/overlap trade-off relevant to filtered state preparation.
    Within this framework, we study a filtered variant of quantum phase estimation (FQPE) that mitigates the unfavorable dependence on the initial overlap present in standard QPE.
    Numerical experiments on Fermi–Hubbard models show that FQPE reduces the total runtime by more than two orders of magnitude in the high-precision regime, with overlap amplification exceeding a factor of one hundred.
\end{abstract}

\maketitle

\section{Introduction}
One of the most promising applications of quantum computing is to deepen our understanding of many-body quantum phenomena, with significant impact expected in chemistry, material science, and condensed matter physics.
A central task in these domains is to solve the eigenvalue problem of many-body Hamiltonians, which determine essential properties, such as the ground-state energy, excited spectra, and correlation functions.
In this context, quantum phase estimation (QPE)~\cite{Seth1996, Nielsen_Chuang_2010} remains the most accurate and asymptotically optimal method for eigenvalue estimation, provided that a quantum state exhibiting significant overlap with the desired eigenstate can be prepared.
Specifically, the success probability of QPE scales with the initial squared overlap $|\gamma_0|^2=|\braket{E_0|\phi_0}|^2$, where $\ket{\phi_0}$ is the prepared input state and $\ket{E_0}$ is the target eigenstate.

However, preparing such a state is often the main bottleneck in quantum simulation.
In many practical settings, especially in large molecular systems or strongly correlated electron models, the overlap is either unknown or extremely small.
When $|\gamma_0|^2\ll1$, QPE requires many repetitions or postselections to succeed, which results in a prohibitive cost even on fault-tolerant quantum computing devices.

This challenge is not only practical but also fundamental.
For generic many-body systems, the initial overlap decays exponentially with system size if the input state is drawn randomly.
Even when the input is physically motivated, such as the Hartree–Fock ground state, the overlap remains small in a large system because the ground state is highly sensitive to perturbations.
This perturbation is mathematically analogous to the correlation part of the Hamiltonian, since both represent modifications of the underlying Hamiltonian.
This phenomenon, known as the orthogonality catastrophe~\cite{Gebert2014}, implies that even weak changes can drastically reduce the fidelity between the approximate and exact ground states.
As a result, the sampling complexity and measurement overhead in QPE and related algorithms can scale exponentially unless the state preparation is improved.

Consequently, much attention has been devoted to improving state preparation by taking advantage of symmetry adaptation~\cite{SUGISAKI2019100002, Gard2020, PhysRevLett.125.230502}, low-entanglement ansatz using tensor network~\cite{Melnikov_2023, PhysRevLett.132.040404}, adiabatic continuation~\cite{PhysRevA.105.032403, doi:10.1126/science.1113479, doi:10.1126/science.1057726, farhi2000numericalstudyperformancequantum}, classical approximate diagonalization~\cite{toloui2013quantumalgorithmsquantumchemistry, tubman2018postponingorthogonalitycatastropheefficient} and variational quantum methods~\cite{TILLY20221, Cerezo2021}.
However, each of these approaches still faces critical challenges.
Symmetry-adapted strategies do not guarantee high overlap with the target eigenstate and primarily serve to reduce the effective Hilbert space by restricting it within a block-diagonal sector of the Hamiltonian.
Tensor network-based state preparation often requires classical variational optimization, which becomes prohibitively costly in high-dimensional or critical systems, and their circuit compilation can be nontrivial.
Adiabatic continuation suffers from the well-known spectral gap bottleneck, where the required evolution time scales inversely with the square of the minimum gap, making the protocol fragile to gap closings and decoherence.
Classical diagonalization methods, such as truncated configuration interaction, can fail to capture essential many-body correlations, and compiling such multi-determinant states into quantum circuits causes high gate overhead.
Finally, variational methods such as variational quantum eigensolver (VQE) are limited by barren plateaus, optimization instability, and no assurance of convergence in the strongly correlated regime.

An alternative approach is to post-process a given input state via a quantum transformation that filters out unwanted eigencomponents.
This concept originates from classical signal processing, where filter functions selectively pass certain frequencies while attenuating others.
In the quantum context, such filtering can be implemented through polynomial or trigonometric transformations of the Hamiltonian, using techniques such as quantum singular value transformation (QSVT)~\cite{camps2023explicitquantumcircuitsblock, gilyen2019quantum, tang2023, sunderhauf2023GQSVT} or quantum signal processing (QSP)~\cite{lin2022QETU, motlagh2024GQSP}.
These techniques enable us to implement Hamiltonian functions $f(\hat{H})$ and their applications to the initial state, where the function $f$ is chosen to amplify eigencomponents near the desired energy while suppressing others.
Recently, quantum algorithms for state preparation in this paradigm have been developed~\cite{lin2020optimal, lin2022QETU, He2022_GaussianFilter, Wang2022statepreparation, Lin2020nearoptimalground, x2v8-jx1h}, especially for early fault-tolerant machines~\cite{PRXQuantum.5.020101}.
Also, a recent work~\cite{PRXQuantum.6.020327} demonstrated efficient initial-state preparation for quantum phase estimation using either matrix product states or Kaiser-window filtering, and quantified the associated cost.

However, the systematic design and analysis of filter functions remains insufficiently explored despite these advances.
%[Revision v-2] start
Prior works have already established that polynomial-based~\cite{lin2020optimal,Lin2020nearoptimalground,gilyen2019quantum}, Fourier- or time-evolution-based~\cite{Somma_2019,parrish2019quantumfilterdiagonalization,lin2022QETU,He2022_GaussianFilter}, and signal-processing-based transformations~\cite{gilyen2019quantum,motlagh2024GQSP,sunderhauf2023GQSVT} can be used for state preparation and spectral filtering.
Our contribution is therefore not the generic idea of ``filtering before QPE,'' but rather a unified and cost-aware treatment of how filtering helps in QPE and how the relevant trade-offs should be quantified.
In particular, many existing works rely on fixed-shape filters, such as Gaussians, and focus on asymptotic behavior without explicitly analyzing the combined effect of circuit cost, postselection success probability, and overlap amplification on the downstream algorithmic cost.
%[Revision v-2] end
It also remains unclear how to construct filters effectively when only partial spectral information is available, and how robust such filters are in the presence of Hamiltonian simulation errors or imperfect block-encodings.

To address these questions, we develop a unified framework for filtered-state preparation based on signal-processing-inspired filter functions implemented through polynomial or trigonometric transformations of the Hamiltonian.
Within this framework, we analyze how filter bandwidth, rejection ratio, and circuit depth affect the success probability and overlap gain, and we derive bounds that make the trade-offs among overlap amplification, postselection, and filter cost explicit.
We also establish error-propagation guarantees showing how imperfections in the filter implementation affect the fidelity of the prepared state.

%[Revision v-2] start
As a second ingredient, we revisit Krylov subspace diagonalization (KSD) from the viewpoint of state preparation.
Existing KSD-based approaches primarily focus on eigenvalue extraction and do not directly yield a high-fidelity approximation of the ground state~\cite{saad_ksd, epperly2022theory}.
We therefore propose an application of the KSD algorithm tailored explicitly for state preparation, which is seamlessly integrated into the filtering-based framework.
In contrast to fixed-shape filter designs that merely adjust peak location and width, the Krylov-based filter flexibly adapts to the spectral structure of the Hamiltonian, suppressing excited-state components with greater selectivity, thereby enabling rapid convergence toward the ground state.
However, the success probability of a Krylov-filtered state typically decays exponentially as the filter sharpens, making direct application impractical for large systems.
To overcome this limitation, we develop a modified KSD protocol that establishes a tunable and gentle trade-off between convergence sharpness and success probability, allowing efficient and reliable ground-state preparation within realistic circuit depth and repetitions.

The main contributions of this work are as follows:
\begin{enumerate}
    \item We formulate a unified cost-aware framework for filtered-state preparation within QPE, covering polynomial and trigonometric realizations on the same footing.
    \item We analyze Gaussian FQPE in the high-precision regime, both with coarse prior spectral information and in a two-stage setting where such priors are obtained within the procedure, thereby establishing an end-to-end cost advantage over standard QPE.
    \item We introduce a modified Krylov-based filter tailored to state preparation, which improves the success-probability/overlap trade-off relative to the standard Krylov construction.
\end{enumerate}
The expected advantage of FQPE is governed by a cost balance rather than by overlap amplification alone.
FQPE is most favorable when the repetition overhead of standard QPE, caused by a small initial overlap, dominates the total cost.
In this condition, the additional depth and postselection overhead of filtered-state preparation remain smaller than the cost saved by the improved overlap.
This typically occurs in a spectrally resolved, high-precision regime, where the target accuracy is smaller than the relevant gap, the filter can be constructed from sufficiently accurate prior information, and the postselection probability remains non-negligible.
%[Revision v-2] end

The remainder of this paper is organized as follows.
Section~\ref{sec:method} develops the filtered-state framework.
Section~\ref{subsec:filtering_a_state_mod} defines filtered states, and the cost--overlap trade-off; Sec.~\ref{subsec:fqpe} formulates filtered quantum phase estimation and derives the generic FQPE cost theorem; Sec.~\ref{subsec:gaussian_filter_and_gaussian_FQPE} analyzes Gaussian FQPE with coarse spectral estimates; Sec.~\ref{subsec:gaussian_FQPE_without_prior_estimates} extends this result to a two-stage Gaussian FQPE procedure without prior estimates; and Sec.~\ref{subsec:krylov_filters} introduces modified Krylov filters for cost-aware state preparation.
Section~\ref{sec:results} numerically compares Gaussian, Krylov, and modified Krylov filters.
Section~\ref{sec:discussion} discusses the scope, limitations, and possible extensions of the proposed framework, and Sec.~\ref{sec:conclusion} concludes the paper.
Formal theorem statements, proofs, classically inspired filter constructions, robustness analyses, and additional numerical experiments are provided in the Supplementary Notes.

Throughout this work, all matrix and operator norms are assumed to be spectral norms unless otherwise noted.
% [Revision v-2 start]
Furthermore, throughout this work, we measure circuit depth or computational cost by the number of calls, i.e., the query complexity, to either the unit-time propagator $\E^{-\I\pi\hat{H}}$ or the qubitization operator for $\hat{H}$.
We use $D$ for the depth or query cost of a single circuit block, and $C$ for the cumulative cost including all repetitions or postselection attempts.
Explicit T-count or wall-clock estimates would require fixing a concrete block encoding or Hamiltonian-simulation oracle, together with hardware-dependent compilation assumptions.
Thus, we focus on query complexity, from which lower-level resource estimates can be obtained once the cost of each oracle call is specified.
% [Revision v-2 end]

\section{Method}\label{sec:method}

\subsection{Filtered states and cost--overlap trade-off}\label{subsec:filtering_a_state_mod}

We first introduce the analytical setting used throughout the paper.
Let
\begin{equation}
    \hat{H}=\sum_{i=0}^{d-1}E_i\ket{E_i}\bra{E_i}
\end{equation}
be a $d$-dimensional Hamiltonian with rescaled eigenvalues
$-1\le E_0\le\cdots\le E_{d-1}\le 1$, and $\ket{E_i}$ is the corresponding eigenstate.
Unless otherwise stated, the target state is the ground state $\ket{E_0}$, and we denote the first spectral gap by
$\Delta E_0:=E_1-E_0$.
The input state is written as
\begin{equation}
    \ket{\phi_0}=\sum_{i=0}^{d-1}\gamma_i\ket{E_i},
    \qquad
    \gamma_i=\braket{E_i|\phi_0}.
\end{equation}
The squared initial overlap with the target eigenstate is therefore $|\gamma_0|^2$.

A filtered state is obtained by applying a bounded function of the Hamiltonian to $\ket{\phi_0}$.
Throughout the main text, we assume that a filter function $f:[-1,1]\to\mathbb{C}$ satisfies $|f(x)|\le 1$ and that $f(\hat{H})$ can be implemented as a block-encoded non-unitary transformation.
Typical realizations use polynomial or trigonometric expansions,
\begin{equation}\label{eq:ham_func_series}
    f(\hat{H})=\sum_{k=0}^{N}c_k b_k(\hat{H}),
\end{equation}
where $b_k(x)$ may be, for example, Chebyshev polynomials $T_k(x)$ or Fourier modes $\E^{\I(k-N/2)\pi x}$.
The corresponding implementation cost is denoted by $D_{\mathrm{sp},f}$.
The technical construction of such Hamiltonian functions using QSVT, QETU, and GQSP is reviewed in Supplementary Note~\ref{suppl:hamiltonian_funciton_implementation}.

\begin{figure*}[t]
    \centering
    \includegraphics[width=0.96\linewidth]{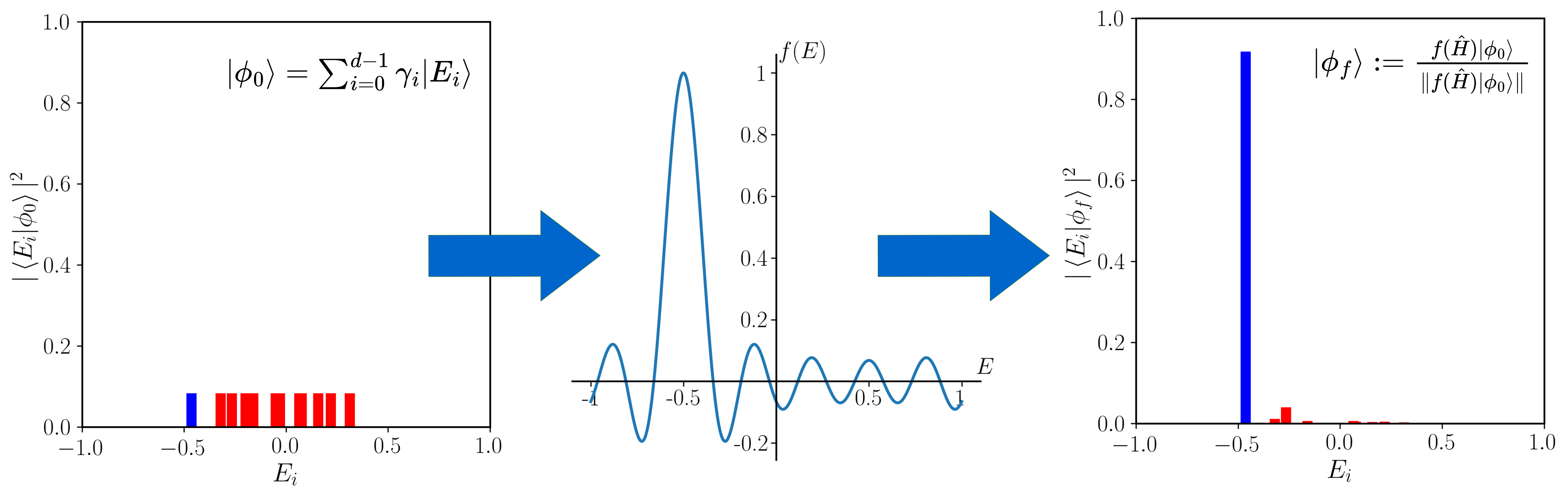}
    \caption{
    Schematic illustration of filtered-state preparation.
    A reference state $\ket{\phi_0}$, initially supported on several eigenstates of $\hat{H}$, is transformed by a filter function $f(\hat{H})$.
    A properly designed filter suppresses undesired spectral components while preserving the target component, thereby increasing the target-state overlap of the normalized filtered state $\ket{\phi_f}$.
    }
    \label{fig:simple_filtered_state}
\end{figure*}

Figure~\ref{fig:simple_filtered_state} summarizes the basic idea of filtered-state preparation.
A filter function $f(\hat H)$ reshapes the spectral amplitudes of the input state by suppressing undesired eigencomponents while retaining the target component.
The resulting improvement in the target-state overlap must be balanced against the postselection success probability and the implementation cost of the filter.

Because $f(\hat{H})$ is generally non-unitary, applying it through a block-encoded circuit succeeds only probabilistically.
The success probability of filtered-state preparation is
\begin{equation}\label{eq:succ_prob}
    p_f
    =
    \braket{\phi_0|f(\hat{H})^{\dagger}f(\hat{H})|\phi_0}
    =
    \sum_{i=0}^{d-1}|\gamma_i f(E_i)|^2 .
\end{equation}
Conditioned on success, the normalized filtered state is
\begin{equation}\label{eq:filtered_state}
    \ket{\phi_f}
    =
    \frac{f(\hat{H})\ket{\phi_0}}
    {\|f(\hat{H})\ket{\phi_0}\|}
    =
    p_f^{-1/2}
    \sum_{i=0}^{d-1}\gamma_i f(E_i)\ket{E_i}.
\end{equation}
The resulting target-state overlap is therefore
\begin{equation}\label{eq:filtered_overlap}
    |\gamma_{f0}|^2
    :=
    |\braket{E_0|\phi_f}|^2
    =
    \frac{|\gamma_0 f(E_0)|^2}{p_f}.
\end{equation}
Thus, a filter improves the input state when
\begin{equation}\label{eq:overlap_amplification}
    |\gamma_{f0}|^2>|\gamma_0|^2,
\end{equation}
which occurs when the filter preserves the target component while suppressing the remaining spectral weight.

The quantities $p_f$ and $|\gamma_{f0}|^2$ cannot be optimized independently.
Combining Eqs.~\eqref{eq:succ_prob} and \eqref{eq:filtered_overlap} gives
\begin{equation}\label{eq:prob_over_tradeoff}
    p_f
    =
    \frac{|\gamma_0|^2}{|\gamma_{f0}|^2}
    |f(E_0)|^2
    \le
    \frac{|\gamma_0|^2}{|\gamma_{f0}|^2},
\end{equation}
where the inequality follows from the boundedness condition $|f(E_0)|\le 1$.
Equation~\eqref{eq:prob_over_tradeoff} is the basic cost--overlap trade-off of filtered-state preparation: a filter function that significantly amplifies the overlap necessarily exhibits a reduced success probability, which is bounded by the inverse of the amplification factor.
The central question is therefore not merely whether filtering increases $|\gamma_0|^2$, but whether the resulting reduction in QPE repetitions outweighs the additional cost of filtering, namely the circuit depth $D_{\mathrm{sp},f}$ per preparation attempt and the postselection overhead $p_f^{-1}$.

\subsection{Filtered quantum phase estimation}\label{subsec:fqpe}
\begin{figure*}
    \centering
    % First subfigure
    \subfigure[~Standard QPE]{
        \includegraphics[width=0.38\textwidth]{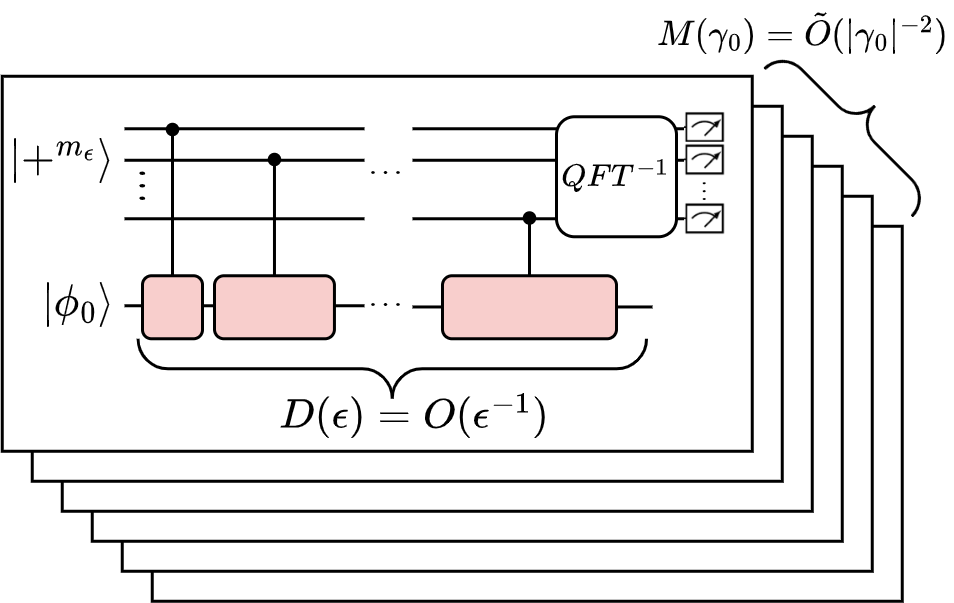}
        \label{fig:qpe_circuit}
    }
    \hfill
    % Second subfigure
    \subfigure[~Filtered QPE]{
        \includegraphics[width=0.58\textwidth]{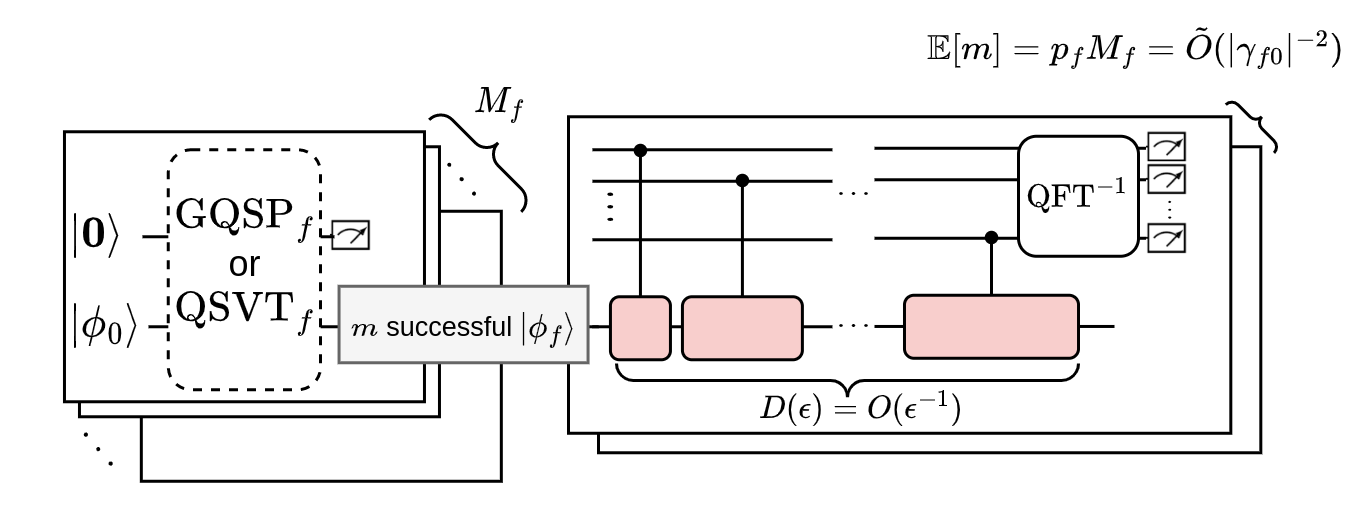}
        \label{fig:fqpe_circuit}
    }
    \caption{
    Circuit diagrams of standard QPE and filtered QPE, both achieving an algorithm accuracy of $\epsilon$ with the identical confidence level.
    (a) In standard QPE, $M(\gamma_0)$ repetitions of QPE circuit, each with a depth of $D(\epsilon)$, are executed.
    (b) Filtered QPE involves filter operators that probabilistically generate the filtered state $\ket{\phi_f}$.
    The diagram illustrates a scenario where $m$ filtered states are successfully generated out of $M_f$ trials.
    }
    \label{fig:qpe_circuit_comparison}
\end{figure*}

We now specialize the cost--overlap trade-off in Eq.~\eqref{eq:prob_over_tradeoff} to quantum phase estimation.
For an input state $\ket{\phi_0}$, a single QPE experiment samples the eigenvalue $E_i$ with probability $|\gamma_i|^2$, up to the finite precision and failure probability of the phase-estimation circuit itself.
Therefore, when the target is $E_0$, the repetition cost of standard QPE is governed by the inverse target-state overlap $|\gamma_0|^{-2}$.
Filtered QPE instead first prepares the filtered state $\ket{\phi_f}$ in Eq.~\eqref{eq:filtered_state}, and then applies QPE to $\ket{\phi_f}$.
As illustrated in Fig.~\ref{fig:qpe_circuit_comparison}, this replaces the reference-state preparation in standard QPE by a probabilistic filtered-state preparation step.
The relevant overlap is changed from $|\gamma_0|^2$ to $|\gamma_{f0}|^2$, but this comes with the postselection overhead $p_f^{-1}$ and the filtering depth $D_{\mathrm{sp},f}$.

Let $D_{\phi_0}$ denote the circuit depth required to prepare the reference state $\ket{\phi_0}$, and let $D_{\mathrm{QPE}}(\epsilon)$ denote the depth of a single QPE experiment that estimates an eigenvalue to precision $\epsilon$, excluding the preparation of the input state.
We state the generic cost comparison in terms of the expected total circuit depth.

\begin{theorem}[Cost of filtered quantum phase estimation]\label{thm:generic_fqpe}
Consider the Hamiltonian, input state, and filter function defined in Sec.~\ref{subsec:filtering_a_state_mod}.
Assume that a single QPE experiment of depth $D_{\mathrm{QPE}}(\epsilon)$ estimates the eigenvalue of an input eigenstate to precision $\epsilon$ with constant success probability.
Then standard QPE, initialized with $\ket{\phi_0}$, estimates $E_0$ with failure probability at most $\delta$ using expected total depth
\begin{equation}\label{eq:standard_qpe_cost}
    C_{\mathrm{QPE}}
    =
    \mathcal{O}\!\left(
    |\gamma_0|^{-2}
    \log(\delta^{-1})
    \left[
        D_{\phi_0}
        +
        D_{\mathrm{QPE}}(\epsilon)
    \right]
    \right).
\end{equation}
By contrast, filtered QPE estimates $E_0$ with failure probability at most $\delta$ using expected total depth
\begin{equation}\label{eq:fqpe_cost}
    \bar{C}_{\mathrm{FQPE}}
    =
    \mathcal{O}\!\left(
    |\gamma_{f0}|^{-2}
    \log(\delta^{-1})
    \left[
        p_f^{-1}
        \left(
            D_{\phi_0}
            +
            D_{\mathrm{sp},f}
        \right)
        +
        D_{\mathrm{QPE}}(\epsilon)
    \right]
    \right).
\end{equation}
Equivalently, filtering is advantageous at the level of expected circuit depth when
\begin{equation}\label{eq:fqpe_advantage_condition}
    \frac{|\gamma_{f0}|^2}{|\gamma_0|^2}
    \gtrsim
    \frac{
        p_f^{-1}
        \left(
            D_{\phi_0}
            +
            D_{\mathrm{sp},f}
        \right)
        +
        D_{\mathrm{QPE}}(\epsilon)
    }{
        D_{\phi_0}
        +
        D_{\mathrm{QPE}}(\epsilon)
    } .
\end{equation}
\end{theorem}

Theorem~\ref{thm:generic_fqpe} makes explicit the two competing effects of filtered-state preparation.
The number of QPE repetitions is reduced from $\mathcal{O}(|\gamma_0|^{-2})$ to $\mathcal{O}(|\gamma_{f0}|^{-2})$, while each QPE repetition now requires the successful preparation of $\ket{\phi_f}$.
Consequently, an overlap increase alone is not sufficient for a practical advantage; the gain in QPE repetitions must compensate for the postselection overhead $p_f^{-1}$ and the depth $D_{\mathrm{sp},f}$ required to implement the filter.
The proof of Theorem~\ref{thm:generic_fqpe}, including the standard concentration bound for a fixed number of postselection attempts, is given in Supplementary Note~\ref{suppl:generic_FQPE}.

\subsection{Gaussian FQPE with coarse spectral estimates}
\label{subsec:gaussian_filter_and_gaussian_FQPE}

We next instantiate the generic FQPE cost theorem with a Gaussian filter, which is a natural choice for band-pass filtering~\cite{He2022_GaussianFilter,low2017hamiltoniansimulationuniformspectral}.
Its role in FQPE is to preserve the ground-state component while suppressing excited-state components separated by the spectral gap. Because we need to determine the position and the width of the Gaussian filter, we assume access to coarse estimates of the ground- and first-excited-state energies.
Such coarse spectral estimates can be obtained from low-resolution QPE~\cite{PRXQuantum.5.020101}, alternative quantum filtering or subspace methods~\cite{lin2020optimal,Wang2022statepreparation,Lin2020nearoptimalground,parrish2019quantumfilterdiagonalization}, or classical Krylov/iterative diagonalization methods~\cite{liesen2012krylov,saad2003iterative}.
Specifically, let $\tilde{E}_0$ and $\tilde{E}_1$ satisfy
\begin{equation}
\label{eq:gaussian_prior_assumption}
    |\tilde{E}_0-E_0|
    \le
    \epsilon' \Delta E_0,
    \qquad
    |\tilde{E}_1-E_1|
    \le
    \epsilon' \Delta E_0,
\end{equation}
where $\Delta E_0=E_1-E_0$ is the first spectral gap.
For simplicity, we impose the conservative condition
\begin{equation}
\label{eq:gaussian_epsilon_prime_condition}
    0<\epsilon'<\frac{1}{5}.
\end{equation}
This condition ensures that the filter width determined from $\tilde E_1-\tilde E_0$ remains separated from the first excited state by a constant fraction of the true spectral gap.
The particular constant is not essential; it is used to obtain a simple closed-form bound in the proof.

Given these estimates, we use the Gaussian filter
\begin{equation}
\label{eq:gaussian_filter_function_approx}
    g(x)
    =
    \exp
    \left[
        -\frac{
            4(x-\mu)^2
            \log \epsilon_g^{-1}
        }{
            \Delta^2
        }
    \right],
\end{equation}
where the parameters are chosen as
\begin{equation}
\label{eq:gaussian_filter_params}
    \mu
    =
    \tilde{E}_0,
    \qquad
    \frac{\Delta}{2}
    =
    (1-\epsilon')
    (\tilde{E}_1-\tilde{E}_0),
    \qquad
    \epsilon_g
    =
    \sqrt{
        \frac{
            c\epsilon
        }{
            \Delta E_0
        }
    } .
\end{equation}
Here, $c\simeq 0.110$ is a numerical constant.
The factor $(1-\epsilon')$ narrows the filter window so that the first excited state is rejected even in the presence of errors in $\tilde{E}_0$ and $\tilde{E}_1$.
The choice of $\epsilon_g$ balances two competing effects: a larger $\epsilon_g$ increases spectral leakage into excited states, whereas a smaller $\epsilon_g$ requires a sharper and therefore more costly filter.

\begin{figure}[t]
    \centering
    \includegraphics[width=0.70\linewidth]{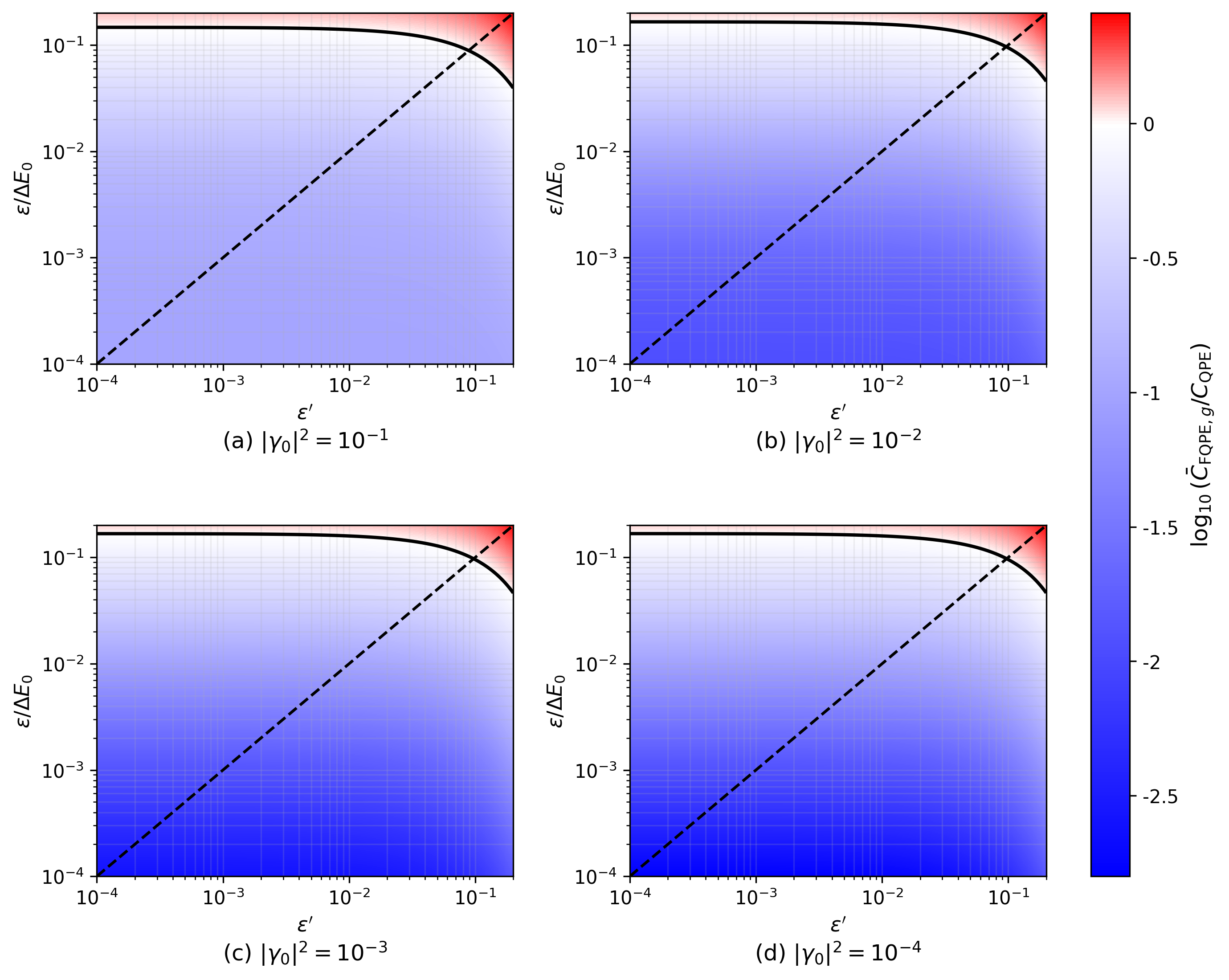}
    \caption{
    Sufficient advantage regime of Gaussian FQPE over standard QPE.
    The color indicates the logarithmic cost ratio $\log_{10}(\bar C_{\mathrm{FQPE},g}/C_{\mathrm{QPE}})$, where $\bar C_{\mathrm{FQPE},g}$ is the expected Gaussian FQPE cost and $C_{\mathrm{QPE}}$ is the standard QPE cost.
    Blue regions correspond to $\bar C_{\mathrm{FQPE},g}<C_{\mathrm{QPE}}$, white indicates equal cost, and red regions correspond to $\bar C_{\mathrm{FQPE},g}>C_{\mathrm{QPE}}$.
    Both axes are shown on logarithmic scales.
    The black solid curve denotes the equal-cost boundary, and the black dashed line denotes the refinement condition $\epsilon/\Delta E_0=\epsilon'$.
    Panels (a)--(d) correspond to initial overlaps $|\gamma_0|^2=10^{-1},10^{-2},10^{-3},10^{-4}$, respectively.
    The plot shows that the advantage becomes more pronounced as the initial overlap decreases and as the target precision becomes smaller relative to the spectral gap.
    }
    \label{fig:gaussian_fqpe_advantage_map}
\end{figure}

The Gaussian filter can be implemented either by a trigonometric expansion or by a polynomial approximation.
For the trigonometric construction, reviewed in Supplementary Note~\ref{suppl:gaussian_fourier}, the required number of basis functions scales as
\begin{equation}
\label{eq:gaussian_filter_depth}
    N
    =
    O
    \left(
        \Delta^{-1}
        \log \epsilon_g^{-1}
    \right).
\end{equation}
Equivalently, the filtering depth scales as
$D_{\mathrm{sp},g}=
\tilde{O}(\Delta E_0^{-1})$
for the parameter choice in Eq.~\eqref{eq:gaussian_filter_params}.
The polynomial construction gives the same asymptotic cost scaling,
as shown in Supplementary Note~\ref{suppl:gaussian_FQPE}.

\begin{theorem}[Gaussian FQPE with coarse spectral estimates]
\label{thm:gaussian_fqpe}
    Suppose that the coarse estimates $\tilde{E}_0$ and $\tilde{E}_1$ satisfy Eq.~\eqref{eq:gaussian_prior_assumption} with $0<\epsilon'<1/5$.
    For a target precision in the high-precision regime $\epsilon\ll \Delta E_0$, Gaussian FQPE with the parameters in Eq.~\eqref{eq:gaussian_filter_params} estimates the ground-state energy with failure probability at most $\delta$ using expected cumulative circuit depth over all repetitions
    \begin{equation}
    \label{eq:gaussian_fqpe_cost}
        \bar{C}_{\mathrm{FQPE},g}
        (\epsilon,\delta;\epsilon')
        =
        \tilde{O}
        \left(
            \epsilon^{-1}
            +
            |\gamma_0|^{-2}
            \epsilon^{-\epsilon'}
            \Delta E_0^{\epsilon'-1}
        \right).
    \end{equation}
\end{theorem}

The first term in Eq.~\eqref{eq:gaussian_fqpe_cost} is the cost of the final high-overlap QPE stage after successful filtering.
The second term is the cost of preparing the Gaussian-filtered state.
Compared with standard QPE, whose cost scales as $\tilde{O}(|\gamma_0|^{-2}\epsilon^{-1})$, Gaussian FQPE reduces the dependence on the target precision from $\epsilon^{-1}$ to $\epsilon^{-\epsilon'}$ in the filtering-dominated part.
Thus, when the prior estimates are sufficiently accurate ($\epsilon'\ll 1$), the filtering overhead approaches $\tilde{O}(|\gamma_0|^{-2}\Delta E_0^{-1})$, so the dominant dependence shifts from the final precision $\epsilon$ to the spectral gap $\Delta E_0$.

The comparison with standard QPE also yields a sufficient advantage regime.
For fixed $\epsilon'$ and $|\gamma_0|^2$, this regime can be written as
\begin{equation}
    \frac{\epsilon}{\Delta E_0}
    <
    r_\ast(\epsilon',|\gamma_0|^2),    
\end{equation}
where $r_\ast$ is obtained by equating the Gaussian FQPE cost bound with the standard QPE cost. The explicit expression for this condition is derived in Supplementary Note~\ref{suppl:gaussian_FQPE}.

Figure~\ref{fig:gaussian_fqpe_advantage_map} visualizes this sufficient advantage regime by plotting the logarithmic cost ratio $\log_{10}(\bar C_{\mathrm{FQPE},g}/C_{\mathrm{QPE}})$ as a function of the coarse-estimate error $\epsilon'$ and the precision ratio $\epsilon/\Delta E_0$.
The blue region indicates $\bar C_{\mathrm{FQPE},g}<C_{\mathrm{QPE}}$, while the red region indicates the opposite.
The black solid curve marks the equal-cost boundary, and the dashed line denotes the refinement condition $\epsilon/\Delta E_0=\epsilon'$.
The figure illustrates that the advantage region expands as $|\gamma_0|^2$ decreases and as the target precision becomes smaller relative to the spectral gap.

A detailed proof of Theorem~\ref{thm:gaussian_fqpe}, including the trigonometric and polynomial constructions, is given in Supplementary Note~\ref{suppl:gaussian_FQPE}.

\subsection{Two-stage Gaussian FQPE without prior estimates}
\label{subsec:gaussian_FQPE_without_prior_estimates}

The Gaussian FQPE theorem above assumes coarse estimates of both $E_0$ and $E_1$.
This assumption is useful for isolating the effect of filtering, but it should not be hidden in an end-to-end resource comparison.
We therefore consider a two-stage procedure in which the required coarse spectral estimates are first obtained by standard QPE and are then used to construct the Gaussian filter.

The first stage runs standard QPE with precision $\epsilon'\Delta E_0$ and confidence $1-\delta_1$ until estimates $\tilde E_0$ and $\tilde E_1$ satisfying Eq.~\eqref{eq:gaussian_prior_assumption} are obtained.
The second stage applies Gaussian FQPE with these estimates, target precision $\epsilon$, and conditional confidence $1-\delta_2$.
The failure probabilities are chosen so that $(1-\delta_1)(1-\delta_2)\ge 1-\delta$.

\begin{corollaryinner}[Two-stage Gaussian FQPE without prior estimates]
\label{cor:two_stage_gaussian_fqpe}
    Assume that the initial state has nonzero overlap with both the ground state and the first excited state, and define
    \begin{equation}
        |\gamma_{\min}|^2
        :=
        \min\{|\gamma_0|^2,|\gamma_1|^2\}.
    \end{equation}
    In the high-precision regime $\epsilon/\Delta E_0<1/5$, there exists a two-stage Gaussian FQPE procedure that estimates $E_0$ within accuracy $\epsilon$ with failure probability at most $\delta$ and total expected cost
    \begin{equation}
    \label{eq:two_stage_gaussian_fqpe_cost_explicit}
        C_{\mathrm{2stage}}
        =
        \tilde{O}
        \left(
            \epsilon^{-1}
            +
            |\gamma_{\min}|^{-2}\Delta E_0^{-1}
        \right).
    \end{equation}
    In particular, when $|\gamma_1|=\Omega(|\gamma_0|)$, this simplifies to
    \begin{equation}
    \label{eq:two_stage_gaussian_fqpe_cost}
        C_{\mathrm{2stage}}
        =
        \tilde{O}
        \left(
            \epsilon^{-1}
            +
            |\gamma_0|^{-2}\Delta E_0^{-1}
        \right).
    \end{equation}
\end{corollaryinner}

The first term in Eq.~\eqref{eq:two_stage_gaussian_fqpe_cost_explicit} is the final high-overlap QPE cost after filtering.
The second term includes both the cost of obtaining coarse spectral estimates and the cost of preparing the Gaussian-filtered state.
A balanced choice of the coarse-estimation accuracy is
\begin{equation}
    \epsilon'
    =
    \Theta
    \left[
        \left(
            \log \frac{\Delta E_0}{\epsilon}
        \right)^{-1}
    \right],
\end{equation}
up to logarithmic corrections and constants specified in Supplementary Note~\ref{suppl:FQPE_cost_gaussian_without_prior}.
With this choice, the coarse-estimation stage does not reintroduce the full $|\gamma_0|^{-2}\epsilon^{-1}$ cost of standard QPE.

Corollary~\ref{cor:two_stage_gaussian_fqpe} shows that the prior information required by Gaussian FQPE can be generated within the same asymptotic cost bound.
Thus, compared with standard QPE, whose cost scales as $\tilde{O}(|\gamma_0|^{-2}\epsilon^{-1})$, the two-stage procedure replaces the overlap-dependent \textit{high-precision cost} by an overlap-dependent \textit{gap-scale cost}.
The resulting advantage is therefore expected in the spectrally resolved regime $\epsilon\ll \Delta E_0$, provided that the initial state has sufficient support on the low-lying eigenstates needed to identify the filter window.
The full proof, including the confidence allocation and the explicit constants, is given in Supplementary Note~\ref{suppl:FQPE_cost_gaussian_without_prior}.

\begin{table*}[t]
    \centering
    \begin{minipage}{0.9\textwidth}
        \centering
        \begin{tabular}{|c|c|c|c|}
            \hline
             & Total Cost & Circuit Depth &Ancilla Qubits\\
            \hline
            \textbf{FQPE [Corollary~\ref{cor:two_stage_gaussian_fqpe}]} & $\tilde{O}(\epsilon^{-1}+|\gamma_0|^{-2}\Delta E_0^{-1})$ & $\tilde{O}(\epsilon^{-1}+\Delta E_0^{-1})$& $O(\log \epsilon^{-1})$\\
            QETU~\cite[Theorem~5]{lin2022QETU}&$\tilde{O}(|\gamma_0|^{-1}\epsilon^{-1})$& $\tilde{O}(|\gamma_0|^{-1}\epsilon^{-1})$ & $O(1)$\\
            QCELS~\cite[Corollary~4]{QCELS}&$\tilde{O}(|\gamma_0|^{-4}\Delta E_0 \epsilon^{-2})$&$\tilde{O}(\Delta E_0^{-1})$&$O(1)$\\
            Fourier Filtering~\cite{LT22}&$\tilde{O}(|\gamma_0|^{-4}\epsilon^{-1})$&$\tilde{O}(\epsilon^{-1})$&$O(1)$\\
            High-confidence QPE~\cite{highc_qpe}& $\tilde{O}(|\gamma_0|^{-2}\epsilon^{-1})$&$\tilde{O}(\epsilon^{-1})$&$O(\mathrm{poly log}(|\gamma_0|^{-1}\epsilon^{-1}))$\\
            Semiclassical QPE~\cite{semicalssical_qpe}&$\tilde{O}(|\gamma_0|^{-4}\epsilon^{-1})$&$\tilde{O}(|\gamma_0|^{-2}\epsilon^{-1})$&$O(1)$\\
            \hline
        \end{tabular}
    \end{minipage}
    \caption{
        Comparison of ground state energy estimation algorithms in terms of total time cost, circuit depth, and the number of required ancilla qubits.
        Here, $|\gamma_0|=\braket{E_0|\phi_0}$ denotes the overlap between the initial state and the ground state, $\epsilon$ is the target accuracy, and $\Delta E_0$ is the spectral gap.
        QETU refers to quantum eigenvalue transformation of unitary matrix, and QCELS refers to quantum complex exponential least squares.
        Even though our algorithm uses deeper circuits, the total cost improves in the high-precision regime ($\epsilon< O(|\gamma_0|^{2}\Delta E_0)$).
        See Supplementary Note~\ref{suppl:FQPE_cost_gaussian_without_prior} for the detail.
    }
    \label{tab:prior_work_comparison}
\end{table*}

To place this end-to-end bound in context, Table~\ref{tab:prior_work_comparison} compares the asymptotic cost of two-stage Gaussian FQPE with representative ground-state energy-estimation algorithms.
The purpose of this comparison is not to claim uniform superiority over all methods, since the algorithms use different primitives and assumptions.
Rather, it shows the advantage of two-stage Gaussian FQPE in a high-precision estimation with small initial ground-state overlap and a resolvable spectral gap.

\subsection{Modified Krylov filters}
\label{subsec:krylov_filters}

The Gaussian filter analyzed above is a representative analytic band-pass filter:
its functional form is fixed in advance, and its center and width are chosen from coarse spectral information.
Other classically inspired filters, including low-pass, band-pass, and minimax filters, follow the same general philosophy of prescribing a filter shape and then bounding the resulting spectral leakage.
We summarize these additional fixed-shape filters in Supplementary Note~\ref{suppl:classically_inspired_filters}.

We now turn to a different class of filters based on Krylov subspaces.
Rather than fixing the filter function a priori, Krylov filtering constructs the filtered state from a variational subspace generated by the input state and basis functions of the Hamiltonian.
This makes the filter coefficients depend on the spectral distribution of the input state, and provides a natural setting for incorporating the cost--overlap trade-off into the state-preparation problem.

Let $\{b_k(x)\}_{k=0}^{N}$ be either a polynomial basis, such as $b_k(x)=x^k$ or $T_k(x)$, or a trigonometric basis, such as $b_k(x)=\E^{\I k\pi x}$.
For a coefficient vector $\bm{c}\in\mathbb{C}^{N+1}$, define the Krylov filter
\begin{equation}
\label{eq:krylov_filter_function}
    f_N(x;\bm{c})
    =
    \sum_{k=0}^{N} c_k b_k(x),
\end{equation}
and the corresponding Krylov-filtered state
\begin{equation}
\label{eq:krylov_filtered_state}
    \ket{\phi_{\bm{c}}}
    =
    \frac{
        f_N(\hat H;\bm{c})\ket{\phi_0}
    }{
        \|f_N(\hat H;\bm{c})\ket{\phi_0}\|
    }.
\end{equation}
This has the same structural form as the filtered state in Eq.~\eqref{eq:filtered_state}, but the coefficients are determined variationally rather than by a fixed analytic profile.

The standard Krylov subspace diagonalization (KSD) construction~\cite{saad_ksd,epperly2022theory} chooses $\bm{c}$ by minimizing the Rayleigh quotient of the Hamiltonian over this filtered-state family.
Define the Krylov Hamiltonian and overlap matrices
\begin{align}
    [\bm{H}]_{kl}
    &=
    \bra{\phi_0}
        b_k^\dagger(\hat H)\hat H b_l(\hat H)
    \ket{\phi_0},
    \\
    [\bm{S}]_{kl}
    &=
    \bra{\phi_0}
        b_k^\dagger(\hat H)b_l(\hat H)
    \ket{\phi_0}.
\end{align}
Then the variational problem is
\begin{equation}
\label{eq:KSD_minimization}
    \min_{\bm{c}\neq \bm{0}}
    \braket{\phi_{\bm{c}}|\hat H|\phi_{\bm{c}}}
    =
    \min_{\bm{c}\neq \bm{0}}
    \frac{
        \bm{c}^\dagger \bm{H}\bm{c}
    }{
        \bm{c}^\dagger \bm{S}\bm{c}
    },
\end{equation}
which leads to the generalized eigenvalue equation
\begin{equation}
\label{eq:KSD_gen_eq}
    \bm{H}\bm{c}
    =
    E^{(N)}\bm{S}\bm{c}.
\end{equation}
The lowest generalized eigenvector defines the standard Krylov filter.
For energy estimation alone, this is the usual KSD prescription.

For filtered-state preparation, however, the Rayleigh quotient is not the only relevant objective.
The state $\ket{\phi_{\bm{c}}}$ must also be prepared probabilistically as a non-unitary filtered state.
If the implemented filter is normalized by
\begin{equation}
    \alpha(\bm{c})
    =
    \max_{x\in[-1,1]} |f_N(x;\bm{c})|,
\end{equation}
then the postselection success probability is
\begin{equation}
\label{eq:krylov_pf_exact}
    p_f(\bm{c})
    =
    \frac{
        \bm{c}^\dagger \bm{S}\bm{c}
    }{
        \alpha(\bm{c})^2
    }.
\end{equation}
Thus, the KSD eigenvector can be suboptimal for FQPE even when it gives an accurate variational energy:
it may strongly suppress the excited-state components, but at the same time reduce the overall filter amplitude so much that the postselection overhead $p_f^{-1}$ dominates the total cost in Eq.~\eqref{eq:fqpe_cost}.

To incorporate this effect while retaining the simplicity of KSD, we use the bound
\begin{equation}
\label{eq:krylov_alpha_bound}
    \alpha(\bm{c})^2
    \le
    (N+1)\bm{c}^\dagger\bm{c},
\end{equation}
which follows from the Cauchy--Schwarz inequality whenever $|b_k(x)|\le 1$ on the spectral interval.
Consequently,
\begin{equation}
\label{eq:krylov_pf_bound}
    p_f(\bm{c})
    \ge
    \frac{
        \bm{c}^\dagger \bm{S}\bm{c}
    }{
        (N+1)\bm{c}^\dagger\bm{c}
    },
    \qquad
    p_f(\bm{c})^{-1}
    \le
    \frac{
        (N+1)\bm{c}^\dagger\bm{c}
    }{
        \bm{c}^\dagger \bm{S}\bm{c}
    }.
\end{equation}
This gives a tractable upper bound on the postselection penalty.

We therefore replace the purely energy-based Krylov objective by a cost-aware regularized objective,
\begin{equation}
\label{eq:modified_krylov_objective}
    \min_{\bm{c}\neq \bm{0}}
    \left[
        \frac{
            \bm{c}^\dagger \bm{H}\bm{c}
        }{
            \bm{c}^\dagger \bm{S}\bm{c}
        }
        +
        \Lambda
        \frac{
            (N+1)\bm{c}^\dagger\bm{c}
        }{
            \bm{c}^\dagger \bm{S}\bm{c}
        }
    \right],
\end{equation}
where $\Lambda\ge 0$ controls the trade-off between energy minimization and postselection success probability.
Equivalently,
\begin{equation}
\label{eq:modified_krylov_rayleigh}
    \min_{\bm{c}\neq \bm{0}}
    \frac{
        \bm{c}^\dagger
        \left[
            \bm{H}+\Lambda(N+1)\bm{I}
        \right]
        \bm{c}
    }{
        \bm{c}^\dagger\bm{S}\bm{c}
    }.
\end{equation}
The modified Krylov filter is therefore obtained from the shifted generalized eigenvalue equation
\begin{equation}
\label{eq:modified_qksd}
    \left[
        \bm{H}
        +
        \Lambda(N+1)\bm{I}
    \right]
    \bm{c}
    =
    E^{(N)}_{\Lambda}
    \bm{S}\bm{c}.
\end{equation}

The case $\Lambda=0$ recovers the standard KSD filter.
For $\Lambda>0$, coefficient vectors with small $\bm{c}^\dagger\bm{S}\bm{c}$ relative to $\bm{c}^\dagger\bm{c}$ are penalized, which increases the lower bound on $p_f$.
This generally sacrifices some spectral selectivity, and therefore may reduce $|\gamma_{f0}|^2$, but it can lower the total FQPE cost when the standard Krylov filter is dominated by postselection overhead.

A useful scale for $\Lambda$ is obtained by comparing the state-preparation depth with the QPE depth.
Using
\begin{equation}
\label{eq:optimal_Lambda}
    \Lambda
    =
    \frac{
        D_{\mathrm{sp},f}
    }{
        D_{\mathrm{QPE}}(\epsilon)
    },
\end{equation}
makes the regularization term reflect the relative contribution of filtered-state preparation to the total cost in Eq.~\eqref{eq:fqpe_cost}.
This choice is not required by the construction, but it provides a cost-aware default: the penalty becomes stronger when filtered-state preparation is expensive compared with the subsequent QPE circuit, and weaker when high-precision QPE dominates the cost.

The modified Krylov construction has two practical advantages.
First, it does not require explicit prior estimates of $E_0$ or $\Delta E_0$ to set the center and width of an analytic filter.
Second, the modification preserves the computational structure of standard KSD: after the matrices $\bm{H}$ and $\bm{S}$ are constructed, the filter coefficients are still obtained from a generalized eigenvalue problem.
The resulting Gaussian and Krylov filters are compared numerically in Sec.~\ref{subsec:properties_of_krylov_filters}.

\section{Results}\label{sec:results}
We conduct proof-of-principle numerical experiments using electronic structure Hamiltonians of Fermi-Hubbard models without chemical potential and magnetic field:
\begin{equation}
    \hat{H}_0=-t\sum_{\braket{p,q},\sigma}^{N_{\mathrm{site}}}(\hat{a}^{\dagger}_{p\sigma}\hat{a}_{q\sigma} + \hat{a}^{\dagger}_{q\sigma}\hat{a}_{p\sigma}) + U\sum_{p=1}^{N_{\mathrm{site}}}\hat{n}_{p\uparrow}\hat{n}_{p\downarrow},
\end{equation}
where $\hat{a}_{p\sigma}$ and $\hat{n}_{p\sigma}=\hat{a}^{\dagger}_{p\sigma}\hat{a}_{p\sigma}$ respectively denote the fermionic annihilation and number operators of the $p$-th site with spin $\sigma$ and $\braket{p,q}$ indicates the summation over the pairs of neighboring sites among $N_{\mathrm{site}}$ lattices.
We study systems with the onsite repulsion factor of $U/t=10$, which places the system in the strongly correlated regime.
The reference state is chosen as a Ne\'el product state, which is in antiferromagnetic phase in the hole-doped model~(i.e. less than half-filled model).
Specifically, we consider three cases of non-periodic 1-dimensional lattices with $N_{\mathrm{site}}=6$ and 7 sites with $N_{\mathrm{e}}=4$ electrons and a 2-dimensional lattice with $N_{\mathrm{site}}=2\times3$ sites with $N_{\mathrm{e}}=3$ electrons.

The Hamiltonian is mapped onto qubit operators via Bravyi-Kitaev encoding with two-qubit tapering~\cite{bravyikitaev,bravyi2017taperingqubitssimulatefermionic}.
This offers not only the reduction of the qubit count, but also focusing on the subspace that the reference state belongs to.
In each case, the resulting Hamiltonian is normalized such that its spectrum lies within $[-1,1]$.
This is done by unitary partitioning of Pauli operators~\cite{Izmaylov2020, Crawford2021efficientquantum} to represent the unnormalized Hamiltonian as a linear combination of unitary $\hat{H}_{0}=\sum_j\alpha_j\hat{U}_j$ and the normalization is done by $\hat{H}=\hat{H}_{0}/\|\bm{\alpha}\|_1$.

Through exact diagonalization, we confirmed that the initial states have poor overlaps of $|\gamma_0|^2=1.51\times10^{-2}, 3.08\times10^{-2}$ and $ 2.66\times10^{-3}$ respectively for 6, $2\times 3$, and 7 lattices, which make the problems challenging.
The spectral gaps are respectively determined as $\Delta E_0=6.20\times10^{-3}, 2.94\times 10^{-2}$ and $4.06\times10^{-3}$.
In this section, the results mainly focus on the $N_{\mathrm{site}}=7$ case with trigonometric basis only.
The results for other models and polynomial basis exhibit similar tendencies and are presented in Supplementary Note~\ref{suppl:further_numerical_results}.

\subsection{Cost Reduction in Gaussian FQPE} \label{subsec:runtime_reduction_in_gaussian_FQPE}
\begin{figure*}
    \centering
    \subfigure[~Gaussian Filters]{
        \includegraphics[width=0.435\textwidth]{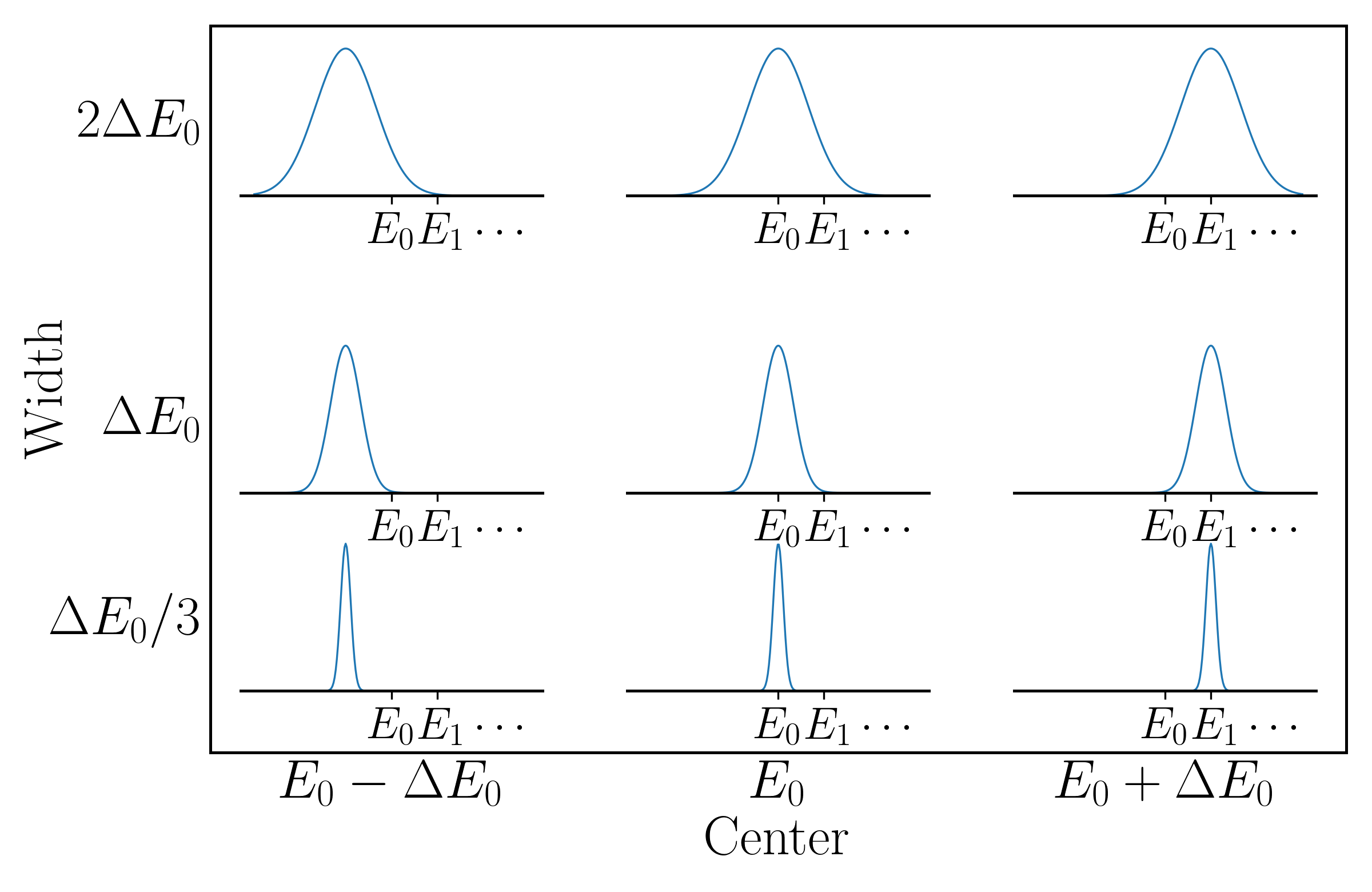}
        \label{fig:preview_pannel}
    }
    \subfigure[~$\epsilon=10^{-1}\Delta E_0$]{
        \includegraphics[width=0.525\textwidth]{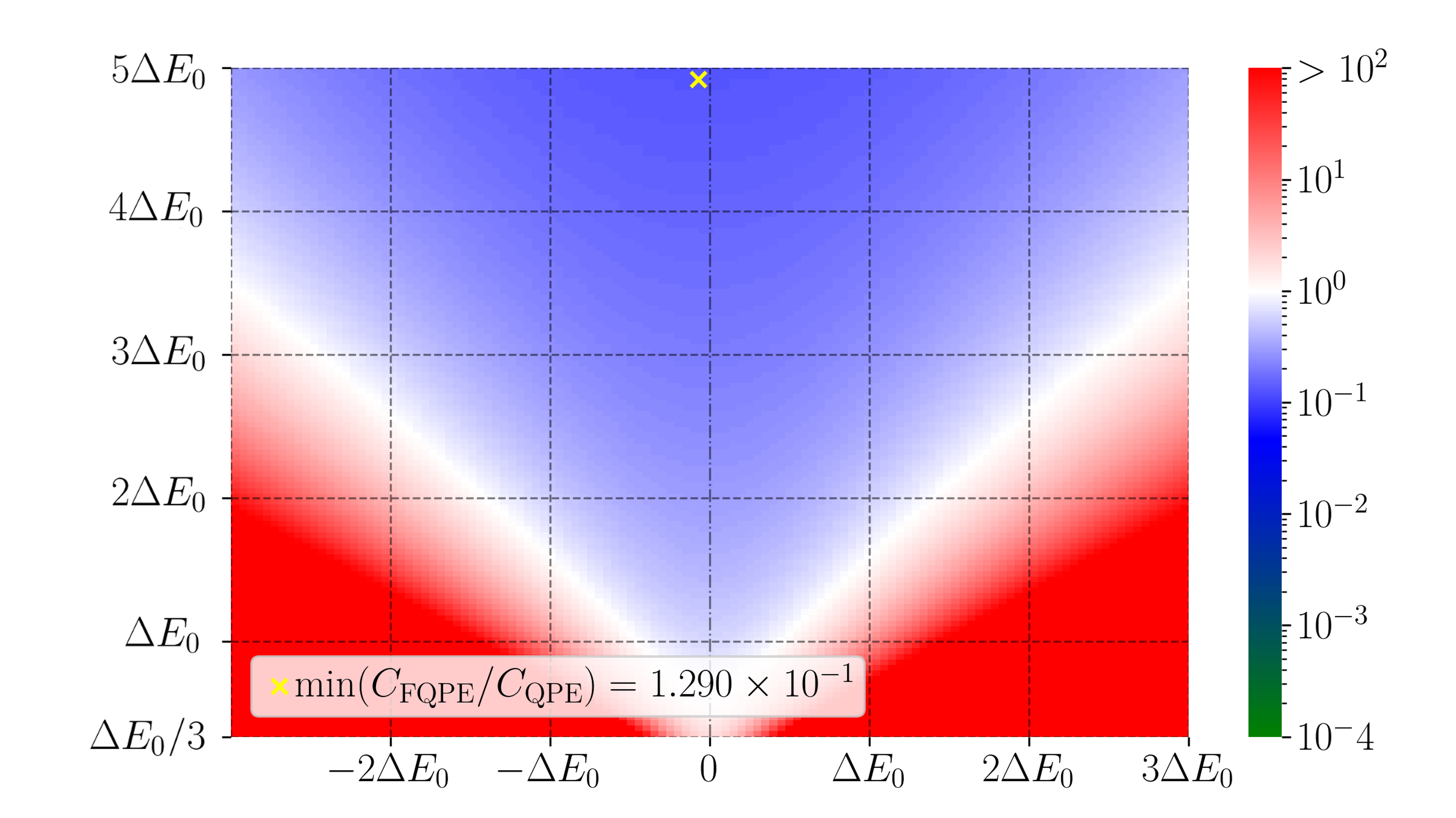}
        \label{fig:fqpe_cost_1}
    }
    \\
    \subfigure[~$\epsilon=10^{-3}\Delta E_0$]{
        \raisebox{0.75mm}{\includegraphics[width=0.435\textwidth]{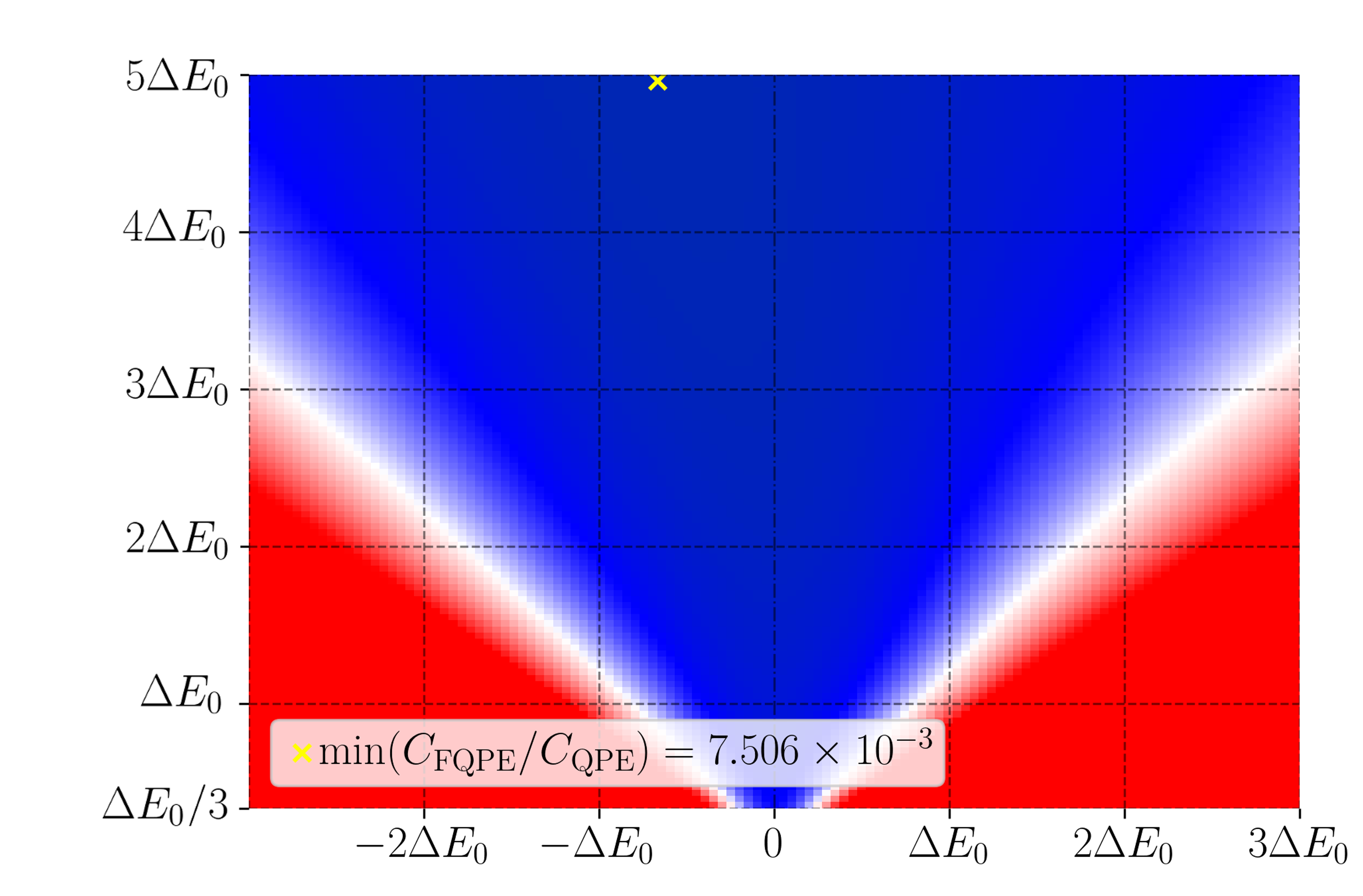}}
        \label{fig:fqpe_cost_2}
    }
    \subfigure[~$\epsilon=10^{-5}\Delta E_0$]{
        \includegraphics[width=0.525\textwidth]{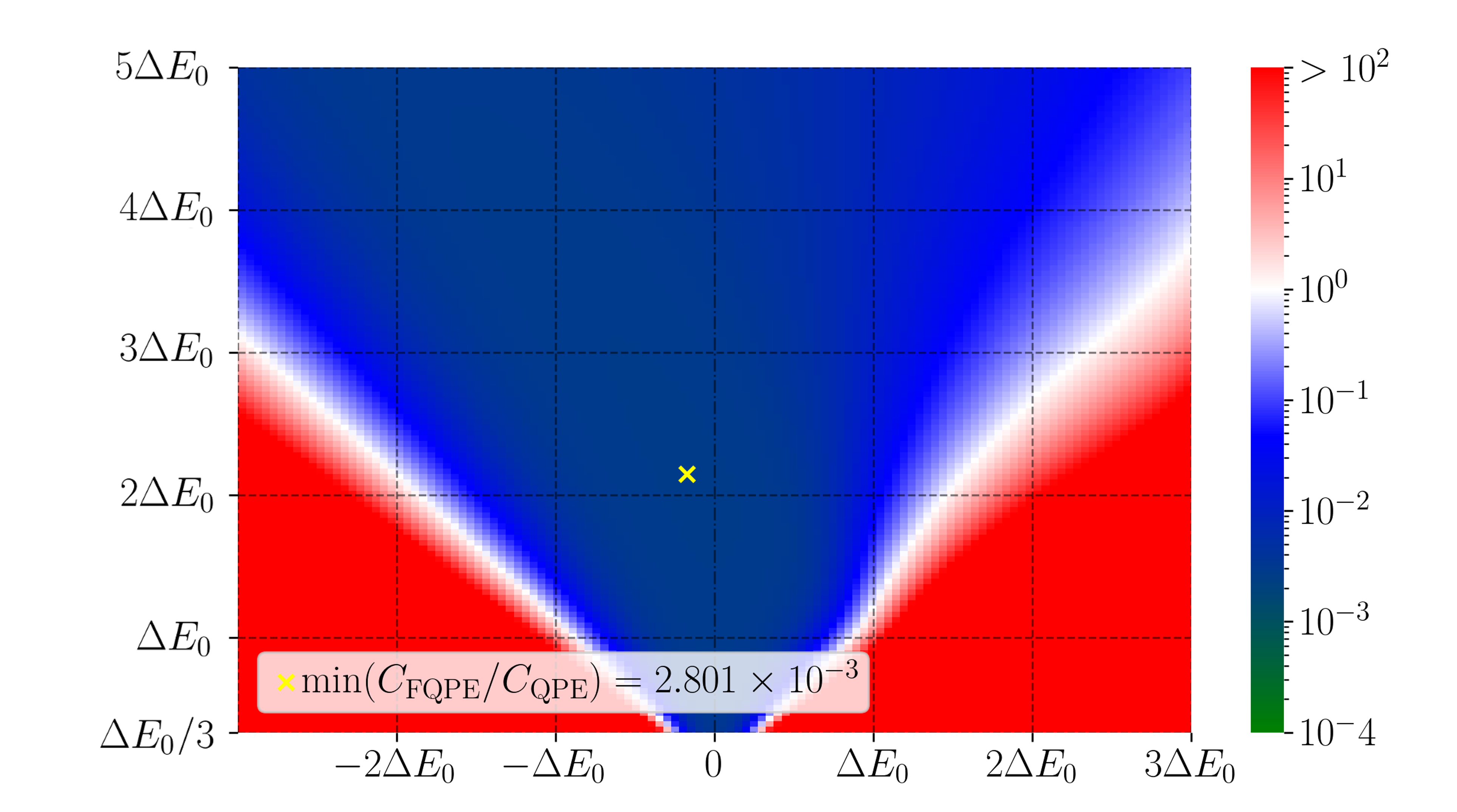}
        \label{fig:fqpe_cost_3}
    }
    \caption{
        (a): Examples of Gaussian filters with different center positions and widths, shown together with the ground- and first-excited-state energies.
        (b)-(d): Relative cost of Gaussian FQPE with target accuracies $\epsilon/\Delta E_0=10^{-1}, 10^{-3}$ and $10^{-5}$ for the Hubbard model with $N_{\mathrm{site}}=7$.
        The cost is defined as the total number of queries to $e^{-i\pi \hat{H}}$.
        In each panel, the x-axis and y-axis represent the bias of the Gaussian filter center ($\tilde{E}_0-E_0$) and the filter width ($\tilde{E}_1-\tilde{E}_0$), respectively (See Eqs.~\eqref{eq:gaussian_filter_function_approx} and \eqref{eq:gaussian_filter_params}).
        Points of minimum cost are marked and annotated with their values.
    }
    \label{fig:fqpe_cost}
\end{figure*}
% [REVISION-v2]
As illustrated in Fig.~\ref{fig:fqpe_cost}, the numerical demonstration shows that Gaussian FQPE can outperform standard QPE when $\tilde{E}_0$ and $\tilde{E}_1$ are estimated well enough for the filter peak to retain substantial amplitude at $E_0$.
This improvement holds when the main lobe of the filter is centered so that it covers the ground state energy $E_0$.
The cost reduction becomes more significant as the target accuracy becomes more stringent, because the circuit depth of the QPE part scales as $O(\epsilon^{-1})$, while the depth of the state preparation only scales as $O(\log\epsilon^{-1/2})$ (See Eq.~\eqref{eq:gaussian_filter_depth}).
Furthermore, this improvement depends on the filter width.
When the filter is too narrow, the depth for the state preparation part increases, which may prevent achieving the optimal cost reduction even if the filter center is close to $E_0$.
Conversely, an excessively large width allows more excited-state components to pass through the filter, especially when the filter center is positively biased from $E_0$.

\begin{figure*}
    \centering
    \includegraphics[width=0.6\linewidth]{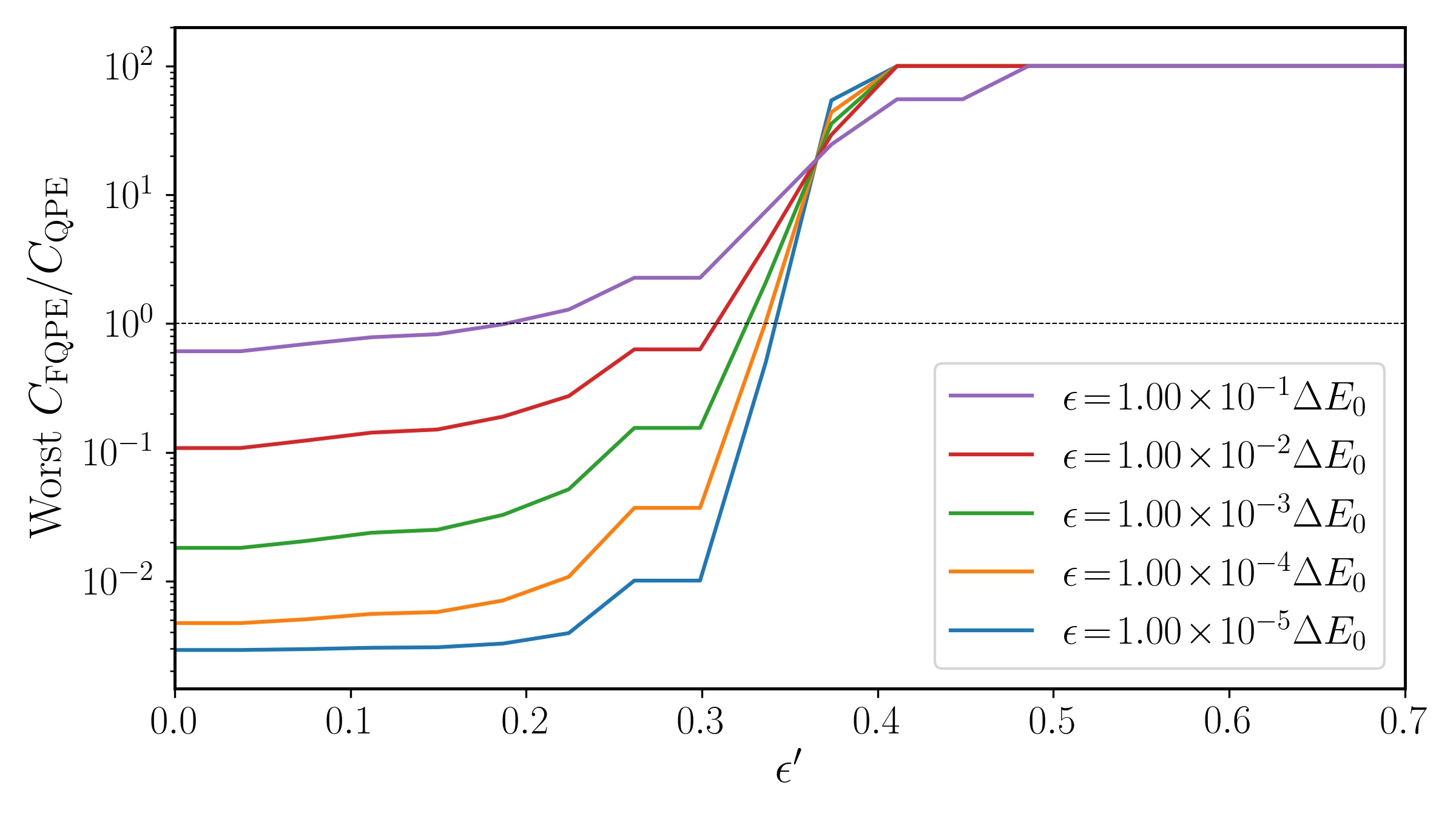}
    \caption{
        The worst-cost within the region defined by $\max(|\tilde{E}_0-E_0|,|\tilde{E}_1-E_1|)\le \epsilon'\Delta E_0$ from Fig.~\ref{fig:fqpe_cost}.
        The x-axis corresponds to the error bound $\epsilon'$ of the prior estimates $\tilde{E}_0$ and $\tilde{E}_1$.
        Colors denote the target accuracies of FQPE.}
    \label{fig:fqpe_cost_worst}
\end{figure*}

However, determining the filter parameters at the optimal points shown in Fig.~\ref{fig:fqpe_cost} is impractical, as it would require a full parameter sweep over the entire range.
Instead, Fig.~\ref{fig:fqpe_cost_worst} presents the FQPE cost under a more practical scenario, showing the worst-case cost when the parameters are chosen based on prior estimates.
Nevertheless, the cost reduction is retained when the prior estimates satisfy an error bound of $\epsilon'\le 1/5$, consistent with the cost analysis assumption in Eq.~\eqref{eq:gaussian_epsilon_prime_condition}.
We also observe that the practical upper bound of $\epsilon'$ for which cost reduction occurs is larger than the bound; up to $\epsilon'=0.3$, significant reduction is presented, particularly in the high precision regime.
Meanwhile, overly coarse estimates beyond this point eventually degrade the cost advantage.

\subsection{Properties of Krylov Filters} \label{subsec:properties_of_krylov_filters}

\begin{figure*}[!t]
    \centering
    \includegraphics[width=0.80\textwidth]{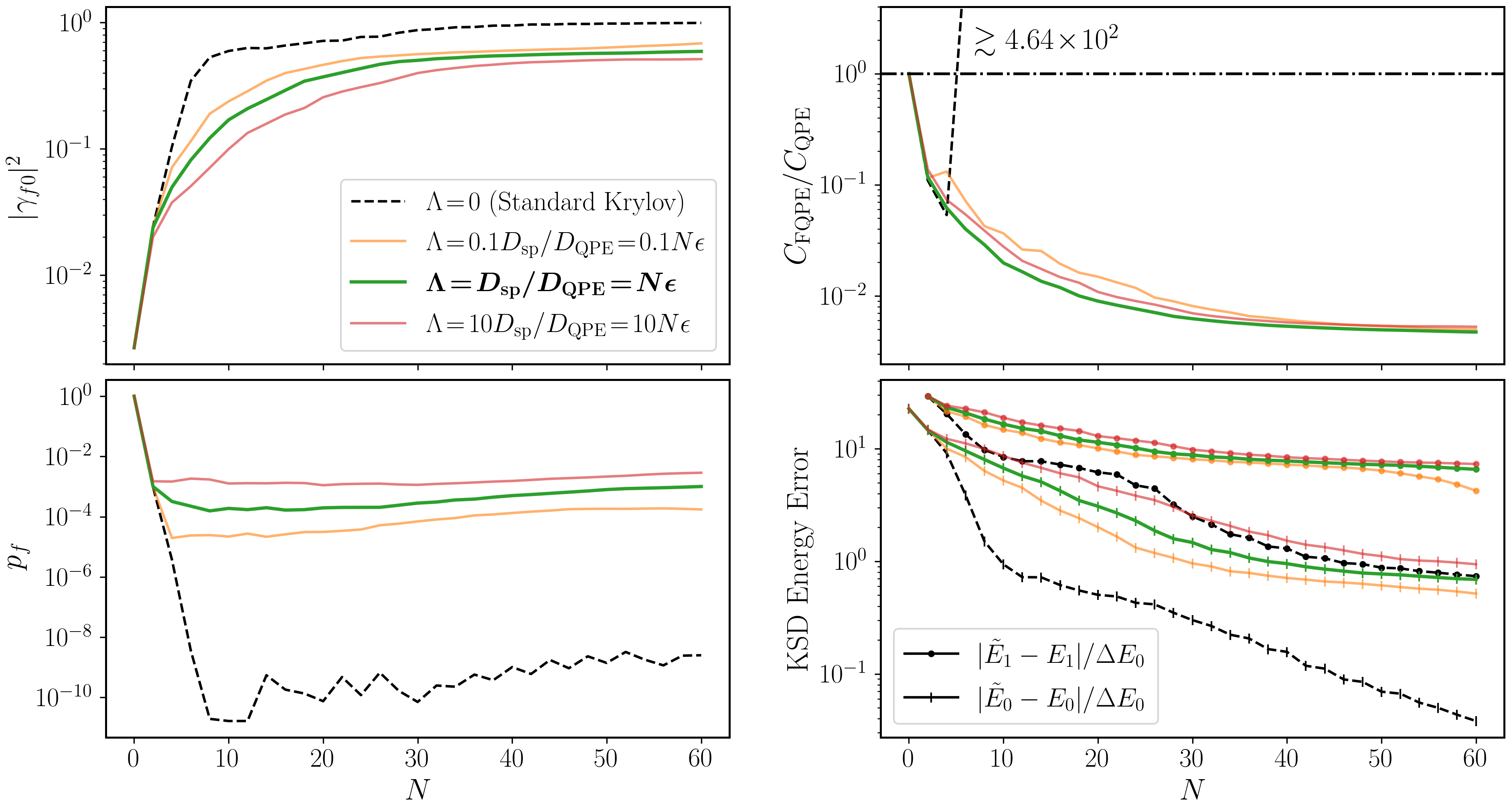}
    \caption{
        Comparison of the standard Krylov filter ($\Lambda=0$) and the modified Krylov filters ($\Lambda >0$) on the Hubbard model with $N_{\mathrm{site}}=7$.
        Curves show, as a function of the Krylov dimension $N$, the filtered-state overlap $|\gamma_{f0}|^2$, the postselection success probability $p_f$, the relative query cost $C_{\mathrm{FQPE}}/C_{\mathrm{QPE}}$, and the errors of the ground- and first-excited-state energies.
        The cost is calculated with the target accuracy of $\epsilon=10^{-4}\Delta E_0$.
    }
    \label{fig:krylov_filter}
\end{figure*}

% Krylov Filter FQPE
Numerical result in Fig.~\ref{fig:krylov_filter} contrasts the standard Krylov filter ($\Lambda=0$) and the modified Krylov filter ($\Lambda>0$).
The filters are constructed from the Krylov eigenvector (Eq.~\eqref{eq:KSD_gen_eq}) or from the modified Krylov eigenvector (Eq.~\eqref{eq:modified_qksd}).
As the Krylov dimension $N$ grows, the filtered-state overlap $|\gamma_{f0}|^2$ increases rapidly, but for $\Lambda=0$, the success probability $p_f$ collapses to a negligible level.
That collapse dominates the total query complexity, so the relative $C_{\mathrm{FQPE}}/C_{\mathrm{QPE}}$ stays above one, and FQPE becomes less efficient than the standard QPE.
Moreover, designing a Gaussian filter around the Krylov-estimated eigenvalues does not fix this failure.
It's because the first excited-state energy error saturates at $\sim \Delta E_0$, which violates the accuracy requirement $\epsilon'<0.2$ for Gaussian FQPE, as numerically verified in Fig.~\ref{fig:fqpe_cost_worst}.

\begin{figure*}[!t]
    \centering
    \subfigure[~Krylov filter and filtered state ($N=60$)]{
        \includegraphics[width=0.96\textwidth]{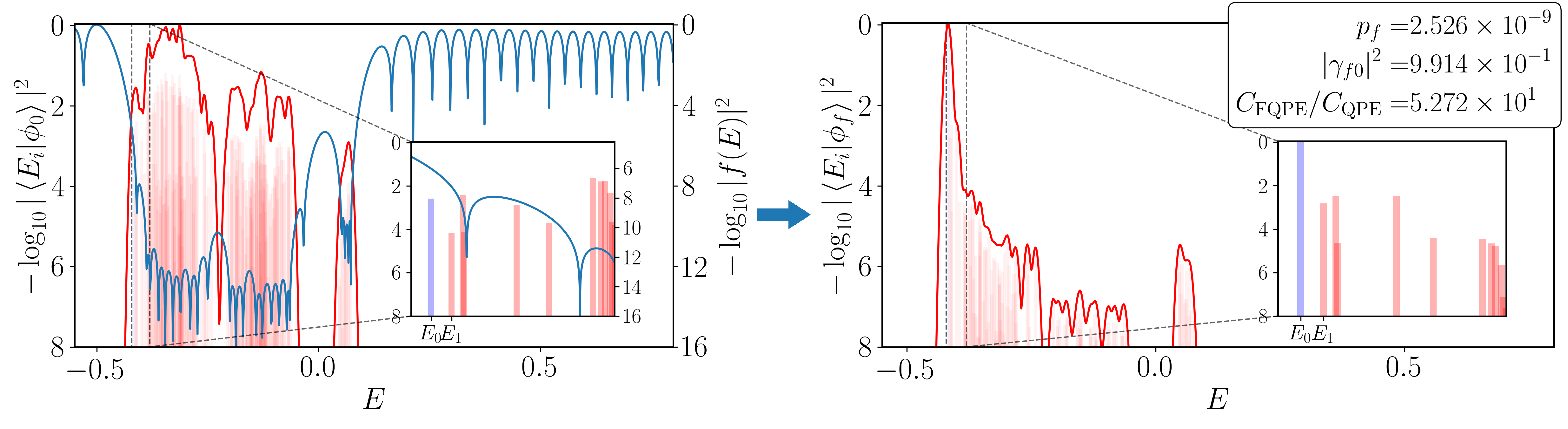}
        \label{fig:qksd_filter_func}
    }\\
    \subfigure[~Modified Krylov filter and filtered state ($N=60, \Lambda=2.3\times10^{-5}$)]{
        \includegraphics[width=0.96\textwidth]{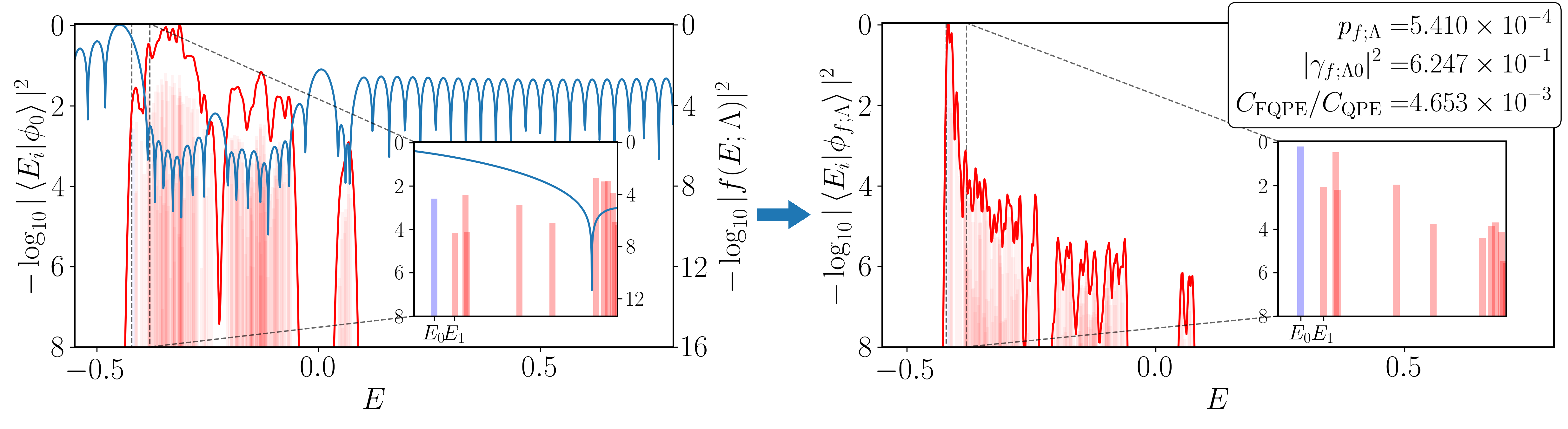}
        \label{fig:modified_qksd_filter_func}
    }\\
    \subfigure[~Gaussian filter and filtered state ($N=60$)]{
        \includegraphics[width=0.96\textwidth]{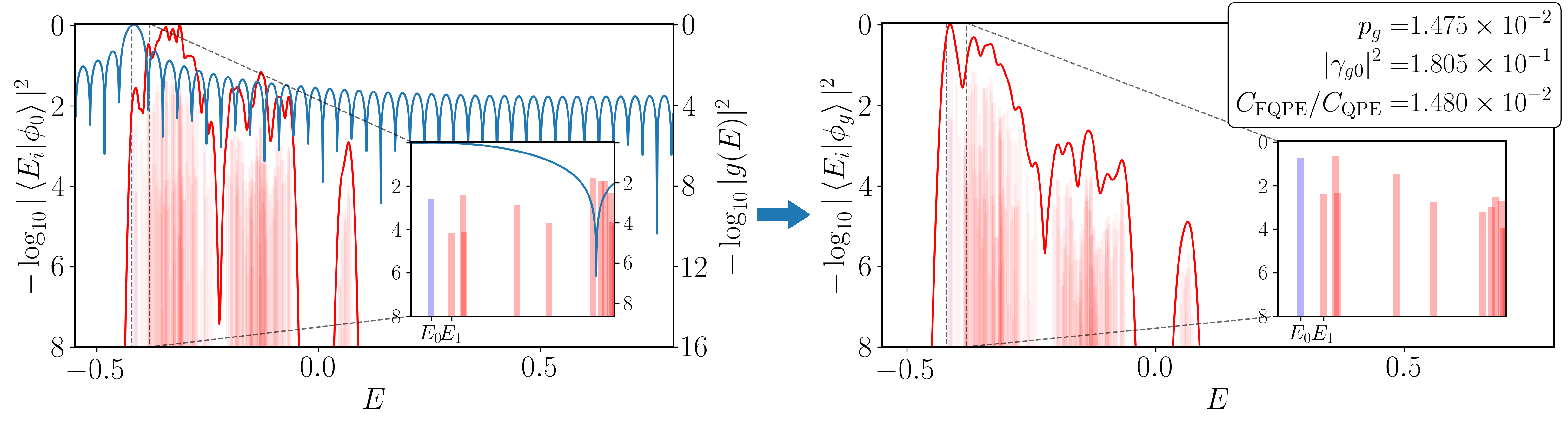}
        \label{fig:gaussian_filter_func}
    }
    \caption{
        Krylov, modified Krylov, and Gaussian filter functions, together with the energy histograms of the reference state and the filtered states, including zoomed views near the ground-state energy.
        The subfigures on the left show the filter functions and the energy histograms of the reference state, with the filter functions plotted as blue lines.
        Squared overlaps between the reference state $\ket{\phi_0}$ and the eigenstates $\ket{E_i}$ are depicted as bar histograms, with the ground-state overlap highlighted in blue, along with their kernel density estimates using a Gaussian function (red lines).
        The width of the Gaussian filter is determined by numerically fitting its main lobe to that of the modified Krylov filter, and its center is set to the Krylov ground-state energy.
        The subfigures on the right present energy histograms of the normalized filtered state, obtained by applying each filter function to the reference state.
        The filtered-state overlaps, success probabilities, and FQPE costs for $\epsilon=10^{-4}\Delta E_0$ are also annotated.
    }
    \label{fig:filter_comparison}
\end{figure*}

Fortunately, the modified Krylov filter remedies this by introducing a tunable $\Lambda$ that trades overlap for success probability at negligible classical overhead.
Increasing $\Lambda$ degrades $|\gamma_{f0}|^2$, as it departs from the proper Krylov projection, but it substantially boosts $p_f$.
Because the Krylov FQPE cost is dominated by small $p_f$, rather than by marginal gains in $|\gamma_{f0}|^2$, this trade-off is beneficial for small $\Lambda$.
We advocate the choice of Eq.~\eqref{eq:optimal_Lambda} which consistently yields the most cost-saving behavior across $N$.
In our model, this achieves a cost reduction by a factor of $3.2\times10^{-3}{\approx}1/312$ compared to the standard QPE.

\subsection{Numerical Comparison of Filter Functions}\label{subsec:numerical_comparison_of_filter_functions}

Figure~\ref{fig:filter_comparison} compares three filters: the naive Krylov filter, our proposed modified Krylov filter, and the Gaussian filter, along with their corresponding filtered states.
Both Krylov-based filters exhibit selective suppression of the excited-state components, effectively adapting to the energy distribution of the reference state, $|\gamma_i|^2=|\braket{E_i|\phi_0}|^2$.
In contrast, the Gaussian filter is non-adaptive by nature.
Note that with a limited number of basis terms ($N=60$), which is far smaller than the inverse of the spectral gap, the filters lack a sufficiently sharp spectral transition near the ground state energy.
In other words, the interval from the filter's main peak to its spectral edge remains too wide to clearly separate the ground and first-excited states.

Krylov diagonalization yields a filter resembling a band-rejection filter, where the ground state energy lies near a steep spectral edge (see Fig.~\ref{fig:qksd_filter_func}).
This results in strong \textbf{relative} suppression of excited-state components, even for small $N$, and leads to the best ground state overlap among the three filters.
However, due to the low filter amplitude at the ground state energy, $|f(E_0)|^2\approx10^{-5}$, the Krylov filter also significantly attenuates the ground state component itself. 
In other words, suppression is applied in an \textbf{absolute} sense to the ground state energy as well, reducing the success probability of preparing the filtered state.

The modified Krylov filter adopts a shape intermediate between a band-rejection and a band-pass filter, with its main peak centered near the ground state energy.
Although the slow decay of the peak results in leakage of nearby excited states into the main lobe, the filter maintains a relatively large amplitude at the ground state energy, leading to a substantially higher success probability.
As a result, the modified Krylov filter achieves the lowest FQPE cost among the three approaches.

\section{Discussion}\label{sec:discussion}

The results of this work position filtered-state preparation as a general and flexible paradigm for improving the efficiency of quantum eigenvalue estimation algorithms. 
A central bottleneck in the algorithms is the dependence of the total computational cost on the overlap between the prepared reference state and the target eigenstate.
While this limitation is well known, many existing approaches either assume a sufficiently good initial state or focus on spectral transformation techniques without explicitly quantifying how overlap amplification trades off against implementation cost.
Our framework directly addresses this issue by treating filtering as an algorithmic layer whose purpose is to reshape the spectral weight of the initial state in a controlled and analyzable manner.

From this perspective, filtered-state preparation complements and extends prior work on polynomial and Fourier-based Hamiltonian transformations, including quantum signal processing and quantum singular value transformation methods~\cite{lin2022QETU, He2022_GaussianFilter, Wang2022statepreparation, Lin2020nearoptimalground}.
Whereas these techniques are typically used to approximate functions of operators for tasks such as simulation or spectral amplification, our analysis emphasizes the role of filtering in reducing the overall cost of downstream algorithms by increasing the effective overlap with the desired eigenstate.
In particular, we provide explicit conditions under which the cost reduction obtained from overlap amplification outweighs the overhead associated with implementing a non-unitary filter, even when only coarse spectral information is available.

An important conceptual aspect of our approach is its close analogy with classical signal processing.
Viewing the Hamiltonian spectrum as a frequency domain naturally introduces notions such as bandwidth, leakage, and uncertainty, which provide intuitive guidance for filter design.
This analogy is not merely pedagogical, but also leads to concrete performance trade-offs between filter width, circuit depth, and success probability, which allows insights from classical filter theory to be translated into quantum algorithmic primitives.
In this sense, filtering serves as a bridge between continuous spectral transformations and discrete quantum circuit implementations.
For more discussion regarding this approach, refer to Supplementary Note~\ref{suppl:classically_inspired_filters}.

Within this general framework, we analyzed filtered quantum phase estimation (FQPE) using Gaussian filters as a representative example.
Compared to standard QPE, FQPE can significantly reduce the expected computational cost in regimes where high precision is required, and the initial overlap is small.
Our analysis shows that an optimal choice of filter width balances the cost of implementing the filter against the gain in overlap, leading to substantial improvements over unfiltered approaches, particularly when the target precision is much smaller than the spectral gap.
% [Revision v-2 start]
Although our present results do not support a universally practical claim, the high-precision setting $\epsilon < \Delta E_0$ considered here is not purely artificial.
Related ground-state preparation scenarios, where one seeks accuracy sufficient to separate the ground state from low-lying excitations, have also been discussed as relevant in quantum chemistry~\cite{Berry2018Improved}.
% [Revision v-2 end]

The expected scaling advantage of Gaussian FQPE is governed primarily by the initial overlap and the spectral gap.
Standard QPE has the overlap-dependent high-precision cost $\tilde{O}(|\gamma_0|^{-2}\epsilon^{-1})$, whereas two-stage Gaussian FQPE replaces this by $\tilde{O}(\epsilon^{-1}+|\gamma_{\min}|^{-2}\Delta E_0^{-1})$ up to logarithmic factors.
Thus, filtering is most useful when $\epsilon\ll \Delta E_0$ and the relevant low-energy overlaps are not too small.
When the Hamiltonian norm grows with system size, a fixed physical target accuracy corresponds to a smaller normalized accuracy $\epsilon$.
For example, for a normalized Hamiltonian $\hat H=\hat H_0/\alpha$, targeting chemical accuracy corresponds to $\epsilon=(1~\mathrm{kcal/mol})/\alpha$ in the normalized representation.
However, the normalized spectral gap is also reduced by the same factor, $\Delta E_0=\Delta E_{0,\mathrm{phys}}/\alpha$.
Therefore, the high-precision condition $\epsilon\ll \Delta E_0$ is equivalent to the physical condition $\epsilon_{\mathrm{phys}}\ll \Delta E_{0,\mathrm{phys}}$ and is not changed by normalization alone.
In this sense, the FQPE advantage can persist with increasing system size if the physical gap does not close too rapidly.
Conversely, the expected benefit can be reduced when the physical gap decreases rapidly.

We also examined Krylov-based filter constructions, which provide an alternative to analytically defined filters.
Unlike conventional Krylov subspace diagonalization methods, which are primarily designed for eigenvalue estimation, our modified Krylov ansatz is tailored for state preparation.
By adjusting the construction to trade off spectral leakage against success probability, Krylov filters can achieve strong selectivity with shallow circuits and a small number of basis states.

The cost estimates above are stated in terms of ideal block-encodings or Hamiltonian-evolution queries.
In practice, QSVT and GQSP use approximate building blocks, so the implemented filter is better viewed as $f(\hat H')$ for a perturbed Hamiltonian $\hat H'$.
Supplementary Note~\ref{suppl:filter_robustness} analyzes how such perturbations affect the filtered state.
A Lipschitz argument gives a direct sensitivity bound $\|f(\hat H')-f(\hat H)\|\le L_f\|\hat H'-\hat H\|$, while a Davis--Kahan projector argument gives a sharper gap-dependent estimate when the filter behaves approximately as a spectral projector.
For Krylov filters, perturbations in the estimated Krylov matrices $\bm H$ and $\bm S$ propagate through the generalized eigenvalue problem; Supplementary Note~\ref{suppl:krylov_state_perturbation} shows that the resulting filtered-state fidelity is controlled by the matrix perturbations and by the conditioning of $\bm S$.
These results clarify that the query-complexity advantages discussed above require sufficiently accurate block-encodings or Hamiltonian simulations and a well-conditioned Krylov overlap matrix.

Finally, numerical simulations support the theoretical analysis across both classes of filters.
In all tested cases, filtering substantially increases the effective overlap with the target eigenstate and improves the success probability of QPE.
Gaussian filters achieve robust performance with relatively few basis terms, while Krylov filters exhibit strong localization properties with shallow-depth implementations, highlighting the practical relevance of the proposed framework.

\section{Conclusion}\label{sec:conclusion}

In this work, we introduced a unified framework for quantum algorithms based on filtered-state preparation, aimed at enhancing the initial overlap with target eigenstates and thereby reducing the overall computational cost of eigenvalue estimation tasks.
By explicitly accounting for the trade-off between filter implementation cost and overlap amplification, we established general conditions under which filtering leads to a net algorithmic advantage.
This framework applies broadly across different classes of filters and algorithmic settings, and provides a systematic way to incorporate non-unitary spectral shaping into quantum algorithms.

As concrete realizations, we analyzed filtered quantum phase estimation using Gaussian filters and developed Krylov-based filters optimized for state preparation.
These examples demonstrate that filtering can be adapted to both analytic and data-driven constructions, offering flexibility in circuit depth, success probability, and spectral selectivity.
Our results show that filtered approaches can substantially outperform standard methods, particularly in regimes characterized by poor initial overlap or stringent precision requirements.

Several directions for future work naturally follow from this study.
One promising avenue is the development of iterative or adaptive filtered algorithms, in which coarse filters are applied first to obtain improved spectral estimates, followed by progressively sharper filters centered on refined energy estimates.
Such multistage filtering schemes could further reduce computational cost and are conceptually analogous to multiresolution techniques in classical signal processing.
In addition, while our focus has been on ground-state preparation, the same principles extend to excited states by appropriately shifting or shaping the filter function.
Exploring these extensions, as well as connections to hybrid and sample-based state preparation methods, represents an important direction for future research.

\section*{Data and Code Availability}
    The raw data for numerical results are available at \cite{reproduce} along with the reproducing scripts.

\section*{Funding Declaration}
    This work was partly supported by Basic Science Research Program through the National Research Foundation of Korea (NRF), funded by the Ministry of Education, Science and Technology (RS-2023-NR068116, RS-2023-NR119931, RS-2025-03532992, RS-2025-07882969). This work was also partly supported by Institute for Information \& communications Technology Promotion (IITP) grant funded by the Korea government (MSIP) (No. 2019-0-00003, Research and Development of Core technologies for Programming, Running, Implementing and Validating of Fault-Tolerant Quantum Computing System). The Ministry of Trade, Industry, and Energy (MOTIE), Korea, also partly supported this research under the Industrial Innovation Infrastructure Development Project (Project No. RS-2024-00466693). J.Huh is supported by the Yonsei University Research Fund of 2025-22-0140.

\begin{acknowledgments}
    G.L. would like to thank Youngjun Park for helpful discussions.
\end{acknowledgments}

\section*{Author Contributions}
G.L. developed the methodology, performed most of the formal analysis, implemented the software, and conducted the validation, investigation, visualization, and the majority of the manuscript writing.
M.K. and J.Hong contributed to parts of the formal analysis and assisted in writing, review, and editing of the manuscript.
S.F. contributed to the conceptualization and participated in the writing, review, and editing process.
J.Huh supervised the project, contributed to the conceptualization, and was responsible for project administration and funding acquisition.

\section*{Competing Interests}
All authors declare no financial or non-financial competing interests.

\clearpage
% -----------------------------------------------------------------------------
% Supplementary Information
% -----------------------------------------------------------------------------
\setcounter{section}{0}
\setcounter{figure}{0}
\setcounter{table}{0}
\numberwithin{equation}{section}
\renewcommand{\theequation}{\thesection.\arabic{equation}}
\renewcommand{\thefigure}{\thesection.\arabic{figure}}
\renewcommand{\thetable}{\thesection.\arabic{table}}
% Ensure that PDF destinations in the supplementary material do not collide
% with identically numbered main-text floats and equations.
\makeatletter
\renewcommand*{\theHsection}{SI.\arabic{section}}
\renewcommand*{\theHsubsection}{SI.\arabic{section}.\arabic{subsection}}
\renewcommand*{\theHsubsubsection}{SI.\arabic{section}.\arabic{subsection}.\arabic{subsubsection}}
\renewcommand*{\theHfigure}{SI.\arabic{figure}}
\renewcommand*{\theHtable}{SI.\arabic{table}}
\renewcommand*{\theHequation}{SI.\arabic{section}.\arabic{equation}}
\providecommand*{\theHsubfigure}{}
\renewcommand*{\theHsubfigure}{SI.\arabic{figure}.\arabic{subfigure}}
\makeatother
\begin{center}
\vspace*{1.5cm}
{\LARGE\bfseries Supplementary Information\par}
\vspace{0.75em}
{\large for \emph{Filtered Quantum Phase Estimation}\par}
\vspace{1em}
{\normalsize Gwonhak Lee, Minhyeok Kang, Jungsoo Hong, Stepan Fomichev, and Joonsuk Huh\par}
\vspace{1.5cm}
\end{center}
\section{Resource Analysis of Quantum Phase Estimation}\label{suppl:resource_analysis_QPE}
We provide a precise resource analysis of quantum phase estimation(QPE) for accurately determining the ground state energy, when the initial state is imperfect.

First, we recall the result for single-shot phase estimation when the exact ground state $\ket{E_0}$ is provided~\cite{Nielsen_Chuang_2010}.
The required circuit depth to ensure that the probability of the outcome $\tilde{E}_0$ estimating the true ground state energy $E_0$ within an error of $\epsilon >0$ exceeds $1-\delta_0$ for some $0<\delta_0<1$ is given by 
\begin{equation}\label{eq:single_shot_perfect_qpe_depth}
    D_{\mathrm{QPE}}(\epsilon, \delta_0)= \epsilon^{-1}\left(2+\frac{1}{2\delta_0}\right).
\end{equation}
Thus, the depth of one complete QPE trial, including input-state preparation,
is
\begin{equation}
\label{eq:single_shot_perfect_qpe_total_depth}
    D_{\phi_0}
    +
    D_{\mathrm{QPE}}(\epsilon,\delta_0).
\end{equation}

However, if an imperfect initial state $\ket{\psi}$ is provided instead, where $|\braket{\psi|E_0}|=\gamma$, the success probability with a given depth of Eq.\eqref{eq:single_shot_perfect_qpe_depth} is reduced to $\gamma^2(1-\delta_0)$ as the outcome estimate may originate from eigenvalues other than the ground state energy.

For phase estimation with $M$ trials, each using a circuit of depth
$D_{\mathrm{QPE}}(\epsilon,\delta_0)$
and an input-state preparation circuit of depth $D_{\phi_0}$, the estimated ground state energy $\tilde{E}_0^{(M)}$ is determined as the minimum among the $M$ outcomes.
Given that the target accuracy is much smaller than the spectral gap, $\epsilon\ll E_1 - E_0$, the probability that $\tilde{E}_0^{(M)}$ is $\epsilon$-accurate is
\begin{equation}\label{eq:multiple_shot_qpe_prob}
    \mathrm{Pr}\left[|\tilde{E}_0^{(M)} - E_0| < \epsilon\right] =1-(1-\gamma^2(1-\delta_0))^M,
\end{equation}
which corresponds to the complementary event of the case where none of the $M$ trials are successful.
Then, the success probability is larger than $1-\delta$ for some $0<\delta<1$ whenever
\begin{equation}
    M(\gamma, \delta, \delta_0) = \lceil\gamma^{-2}(1-\delta_0)^{-1}\log \delta^{-1}\rceil.
\end{equation}
The expected total circuit depth of standard QPE is therefore
\begin{equation}
\label{eq:standard_qpe_total_depth_delta0}
    C_{\mathrm{QPE}}
    (\gamma,\epsilon,\delta,\delta_0)
    =
    M(\gamma,\delta,\delta_0)
    \left[
        D_{\phi_0}
        +
        D_{\mathrm{QPE}}(\epsilon,\delta_0)
    \right].
\end{equation}

We optimize the parameter $\delta_0$ by minimizing the part of the total
depth that depends on the single-shot phase-estimation accuracy:
\begin{equation}
\label{eq:optimal_singleshot_succprob}
\begin{split}
    \delta_0^{\star}
    &=
    \argmin_{0<\delta_0<1}
    M(\gamma,\delta,\delta_0)
    D_{\mathrm{QPE}}(\epsilon,\delta_0)
    \\
    &=
    \argmin_{0<\delta_0<1}
    \gamma^{-2}
    \epsilon^{-1}
    \log \delta^{-1}
    \left(
        2
        +
        \frac{1}{2\delta_0}
    \right)
    (1-\delta_0)^{-1}
    \\
    &\approx
    0.309017 .
\end{split}
\end{equation}
This is the same choice as in the state-preparation-free resource estimate,
because $D_{\phi_0}$ is independent of $\delta_0$.

With this choice, we use the shorthand
\begin{align}
\label{eq:optimal_singleshot_cost}
    D_{\mathrm{QPE}}(\epsilon)
    &\approx
    3.61803\,\epsilon^{-1},
    \\
    M(\gamma,\delta)
    &\approx
    \left\lceil
        1.44721\,
        \gamma^{-2}
        \log \delta^{-1}
    \right\rceil .
\end{align}
Hence, the total depth of standard QPE initialized with $\ket{\phi_0}$ is
\begin{equation}
\label{eq:standard_qpe_total_depth}
    C_{\mathrm{QPE}}
    (\gamma,\epsilon,\delta)
    =
    M(\gamma,\delta)
    \left[
        D_{\phi_0}
        +
        D_{\mathrm{QPE}}(\epsilon)
    \right].
\end{equation}

\section{Hamiltonian Function Implementation}\label{suppl:hamiltonian_funciton_implementation}

We review the foundational techniques on which quantum filter functions are built: qubitization~\cite{Low2019hamiltonian}, QSVT~\cite{gilyen2019quantum, tang2023}, quantum eigenvalue transformation of unitaries (QETU)~\cite{lin2022QETU}, and GQSP~\cite{motlagh2024GQSP}, which efficiently implement functions of a Hamiltonian by quantum circuit.

\subsection{Block Encoding}
We begin by defining an $(\alpha,m,\epsilon_{\mathrm{BE}})$-block-encoding of a normal operator $\hat{A}\in\mathbb{C}^{2^n\times2^n}$ with $\|\hat{A}\|\le\alpha$ and $\epsilon_{\mathrm{BE}}\in[0,1/2)$.
This encoding corresponds to a unitary operator $\hat{U}\in\mathbb{C}^{2^{n+m}\times2^{n+m}}$ that satisfies
\begin{equation}\label{eq:block-encoding}
    \|\hat{A}-\alpha(\bra{0^{m}}\otimes\hat{I})\hat{U}(\ket{0^m}\otimes\hat{I})\|\le\epsilon_{\mathrm{BE}},
\end{equation}
meaning that $\hat{U}$ encodes $\hat{A}/\alpha$ with $\epsilon_{\mathrm{BE}}$-accuracy within its $\ket{0^m}\bra{0^{m}}$ subspace.

Notably, while $\hat{U}$ itself is unitary, the operator encoded in the subspace may be non-unitary.
Thus, applying the encoded operator $\hat{A}$ can fail with a finite probability, where the resulting state lies outside the encoded subspace.
Consider applying $\hat{U}$ with $\epsilon_{\mathrm{BE}}=0$ to an $(n+m)$-qubit state $\ket{\phi}\ket{0^m}$:
\begin{equation}
    \hat{U}\ket{\phi}\ket{0^m}=\frac{\hat{A}}{\alpha}\ket{\phi}\ket{0^m}+\ket{\perp},
\end{equation}
where $\ket{\perp}$ is an unnormalized state orthogonal to $\ket{0^m}$.
To successfully apply $\hat{A}/\alpha$ to $\ket{\phi}$, post-selection on the $\ket{0^m}$ ancilla state is required, which occurs with probability $\tfrac{1}{\alpha^2}\braket{\phi|\hat{A}^{\dagger}\hat{A}|\phi}$.

\subsection{Polynomial Implementation}
Qubitization of $\hat{H}$, denoted as $\hat{Q}(\hat{H})$, provides a specific implementation of $(1, m, 0)$-block-encoding of $\hat{H}$ that exploits an $\mathrm{SU}(2)$-invariant subspace allowing $\hat{Q}(\hat{H})$ to be interpreted as a single-qubit operator~\cite{Low2019hamiltonian}.
When combined with quantum signal processing, which enables polynomial amplitude transformations of qubit states, one can construct bounded real polynomials of $\hat{H}$.

This technique, known as QSVT, facilitates the polynomial approximation of Hamiltonian functions $f(\hat{H})$ bounded as $\|f(\hat{H})\|\le 1$~\cite[Theorem~4]{gilyen2019quantum}:
\begin{equation}\label{eq:polynomial_approx}
    f(\hat{H})\approx p_{\bm{c}}(\hat{H})=\sum_{k=0}^N c_k T_k(\hat{H}),
\end{equation}
where $\bm{c}\in\mathbb{R}^{N+1}$ that satisfies $\max_{x\in[-1,1]}|p_{\bm{c}}(x)|\le 1$ and $T_k$ is the $k$-th Chebyshev polynomial.
Here, both $f(x)$ and $p_{\bm{c}}(x)$ are further assumed to have definite parity.

Specifically, QSVT allows implementation of a quantum circuit $\hat{U}_{p_{\bm{c}}}^{(\mathrm{P})}\in\mathbb{C}^{2^{n+m+1}\times2^{n+m+1}}$, which is a ($1, m+1, \epsilon_f$)-block-encoding of the function $f(\hat{H})$:
\begin{equation}\label{eq:polynomial_blockencoding}
    \|f(\hat{H})-\bra{0^{m+1}}\hat{U}_{p_{\bm{c}}}^{(\mathrm{P})}\ket{0^{m+1}}\|\le \epsilon_f,
\end{equation}
where $\epsilon_f=\max_{x\in[-1,1]}|f(x)-p_{\bm{c}}(x)|$ is an additive error of $N$-order polynomial approximation.
This implementation queries the qubitization operators and its inverse $N$ times, along with a classical precomputation of complexity $O(\mathrm{poly}(N))$.
Furthermore, its success probability is given as 
\begin{equation}
    p_f=\braket{\phi_0|f(\hat{H})^\dagger f(\hat{H})|\phi_0}-O(\epsilon_f),
\end{equation}
if $\hat{U}_{p_{\bm{c}}}^{(\mathrm{P})}$ is applied to $\ket{\phi_0}$ along with $\ket{0^{m+2}}$ ancilla qubits.

However, the functions employed in this work often exhibit a peak at a certain value $\mu\in[-1,1]$, and thus do not have definite parity, which conflicts with the assumption in the QSVT.
In such case, we adopt qubitization operator $\hat{Q}\left(\frac{\hat{H}-\mu}{1+|\mu|}\right)$ to implement $p_{\bm{c}}\left(\frac{\hat{H}-\mu}{1+|\mu|}\right)$ by QSVT.
Alternatively, a modified QSVT method~\cite[Theorem~31]{gilyen2019quantum} presents the $(2,m+2,\epsilon_f)$-block-encoding of $f(\hat{H})$ with indefinite parity.
Although it offers implementation of more general Hamiltonian functions, the success probability is reduced by the factor of $1/4$.

QSVT has been further extended to generalized QSVT (GQSVT)~\cite{sunderhauf2023GQSVT}, which achieves a ($\beta, m+1, \epsilon_f$)-block-encoding of bounded complex functions with half the circuit depth compared to QSVT.
However, the normalization prefactor $\beta=O(\log N)$ may increase arbitrarily and thus lower the overall success probability.
Therefore, in this work, we adopt the standard QSVT in Eqs.~\eqref{eq:polynomial_approx} and \eqref{eq:polynomial_blockencoding} to implement polynomial approximations of Hamiltonian functions.

Notably, QSVT achieves optimal Hamiltonian simulation, attaining the minimal cost permitted by the no-fast-forwarding theorem~\cite[Corollary~32]{gilyen2019quantum}.
Specifically, a $(1, m+2, \epsilon_{\mathrm{HE}})$-block-encoding of Hamiltonian evolution operator $\E^{-\I\hat{H}t}$ can be implemented by querying the controlled-$\hat{Q}(\hat{H})$ and its inverse $N_{\mathrm{sim}}=O(|t|+\log(\epsilon_{\mathrm{HE}}^{-1}))$ times.
This relies on the Jacobi-Anger expansion:
\begin{equation}\label{eq:trunc_jacobi_anger}
    \E^{-\I\hat{H}t}= J_0(t)+2\sum_{k=1}^{N_{\mathrm{sim}}}\I^k J_k( t)T_k(\hat{H})+O(\epsilon_{\mathrm{HE}}),
\end{equation}
where $J_k$ is the $k$-th Bessel function of the first kind.

\subsection{Trigonometric Series Implementation}
In contrast, QETU is a technique for block encoding of Laurent polynomials of the Hamiltonian evolution operator $\E^{\I\pi\hat{H}}$~\cite{lin2022QETU}.
It was motivated by applications in early fault-tolerant quantum computing~(EFTQC), aiming to circumvent block encoding of $\hat{H}$ and the use of QSVT, which typically require a large number of ancilla qubits and costly multi-qubit Toffoli gates.
Thus, QETU disregards the possibility of implementing the Hamiltonian evolution operator using QSVT, as in Eq.~\eqref{eq:trunc_jacobi_anger}.
Moreover, if the evolution operator is approximated as a polynomial in $\hat{H}$, then the resulting QETU effectively reduces to a polynomial transformation of $\hat{H}$, aligning it with QSVT.
QETU was originally designed to use more practical approximations, such as the first-order Trotter product formula, which requires $O(\epsilon_{\mathrm{HE}}^{-1})$ Trotter steps.
Although this causes a higher asymptotic cost compared to the logarithmic scaling in QSVT-based methods, it avoids the use of ancilla qubits and is more practical in the EFTQC regime.

QETU has since been generalized to GQSP~\cite[Theorem~6]{motlagh2024GQSP}, which supports complex polynomials without parity constraints.
This enables flexible trigonometric series representations of the Hamiltonian function:
\begin{equation}\label{eq:trigonometric_approx}
    f(\hat{H})\approx P_{\bm{c}}(\E^{\I\pi\hat{H}})=\sum_{k=-N/2}^{N/2}c_{k} \E^{\I\pi k\hat{H}},
\end{equation}
where $\bm{c}\in\mathbb{C}^{N+1}$ satisfies $\max_{|z|=1}|P_{\bm{c}}(z)|\le 1$.
The use of complex coefficients allows convenient spectral shifting via simple phase shifts: $c_k\rightarrow c_k\E^{-\I\pi k \mu}$.
In this work, we primarily consider functions that have a main peak at $\mu\in[-1,1]$, while the trigonometric series are inherently periodic.
To prevent the appearance of peaks in the neighboring periods within $[-1,1]$, a period of at least $1+|\mu|\le2$ is sufficient.
For simplicity, we fix the period to $2$ for the trigonometric basis functions.
Specifically, the block-encoding operator of $f(\hat{H})$ is written as
\begin{equation}\label{eq:trigonometric_blocencoding}
    \|f(\hat{H})-\braket{0^{m+1}|\hat{U}_{P_{\bm{c}}}^{(\mathrm{T})}|0^{m+1}}\|\le\epsilon_f,%L(P_{\bm{c}})\epsilon_{\mathrm{HE}},
\end{equation}
where $\hat{U}_{P_{\bm{c}}}^{(\mathrm{T})}\in\mathbb{C}^{(n+m+1)\times(n+m+1)}$ encodes the approximation in Eq.\eqref{eq:trigonometric_approx} with error $\epsilon_f$.

\section{Filtered Quantum Phase Estimation (FQPE) with a Generic Filter}\label{suppl:generic_FQPE}
In this section, we restate and prove Theorem~\ref{thm:generic_fqpe} of the main text in a more detailed form.

\begin{theorem}[Generic FQPE, formal version of Theorem~\ref{thm:generic_fqpe}]\label{thm:_generic_FQPE}
    Suppose a generic filter function $f(\hat{H})$, satisfying $|f(x)|\le 1~\forall x\in[-1,1]$, can be block-encoded with the circuit depth of $D_{\mathrm{sp},f}$.
    Then, for FQPE to solve the ground-state energy estimation problem with an accuracy $\epsilon$ and confidence at least $1-\delta$, the required number of state-preparation attempts satisfies
    \begin{equation}\label{eq:thm_Mf_bound}
      M_f = \Theta(|\gamma_{f0}|^{-2})\,p_f^{-1}\,\log(\delta^{-1})
          = |f(E_0)|^{-2}\,M_{\mathrm{QPE}}(\gamma_0,\delta).
    \end{equation}
    Furthermore, the expected total circuit depth and its standard deviation are
    \begin{gather}
    \begin{split}
    \mathbb{E}[C_{\mathrm{FQPE}}(M;M_f)]
      =& M_f\!\bigl(D_{\phi_0}+D_{\mathrm{sp},f}+p_fD_{\mathrm{QPE}}(\epsilon)\bigr)\\
      =& M_{\mathrm{QPE}}(\gamma_0,\delta)D_{\mathrm{QPE}}(\epsilon)
         \!\left(|f(E_0)|^{-2}\frac{D_{\phi_0}+D_{\mathrm{sp},f}}{D_{\mathrm{QPE}}(\epsilon)}
         + \left|\frac{\gamma_0}{\gamma_{f0}}\right|^2\right),
    \end{split}\label{eq:thm_exact_expected_cost}\\
        \sigma_{C_{\mathrm{FQPE}}} = D_{\mathrm{QPE}}(\epsilon)\sqrt{M_{\mathrm{QPE}}(\gamma_{f0},\delta) (1-p_f)}.\label{eq:thm_exact_standard_deviation}
    \end{gather}
\end{theorem}
\begin{proof}
    Let $M$ denote the number of successful filtered-state preparations out of $M_f$ independent attempts.
    Then $M\sim\mathrm{Binomial}(M_f,p_f)$, giving
    \begin{equation*}
        P_{\mathrm{SP}}(M;M_f)=\binom{M_f}{M}p_f^M(1-p_f)^{M_f-M}.
    \end{equation*}
    Conditioned on a successful preparation, a QPE run projects onto the ground state with probability $|\gamma_{f0}|^2(1-\delta_0)=\Theta(|\gamma_{f0}|^2)$ (See Supplementary Note.~\ref{suppl:resource_analysis_QPE}). Thus, the probability that at least one of $M$ QPE runs succeeds is
    \begin{equation*}
        P_{\mathrm{QPE}}(M,\gamma_{f0}) = 1 - (1-\Theta(|\gamma_{f0}|^2))^M.        
    \end{equation*}
    Averaging over $M$ yields the total success probability
    \begin{equation*}
        P_{\mathrm{FQPE}}(M_{f}) = \sum_{M=0}^{M_{f}}P_{\mathrm{QPE}}(M,\gamma_{f0})P_{\mathrm{SP}}(M;M_{f})=1-(1-p_{f}\Theta(|\gamma_{f0}|^2))^{M_{f}}.
    \end{equation*}
    Imposing $P_{\mathrm{FQPE}}(M_f)\ge 1-\delta$ gives Eq.~\eqref{eq:thm_Mf_bound}.
    
    The total cost for fixed $M$ and $M_f$ is given by
    \begin{equation*}
        C_{\mathrm{FQPE}}(M;M_{f})=M_{f}(D_{\phi_0}+D_{\mathrm{sp},f})+MD_{\mathrm{QPE}}(\epsilon).
    \end{equation*}
    Taking the expectation gives 
    \begin{equation*}
      \mathbb{E}[C_{\mathrm{FQPE}}(M;M_f)]
       = M_f(D_{\phi_0}+D_{\mathrm{sp},f})+\mathbb{E}[M]D_{\mathrm{QPE}}(\epsilon)
       = M_f(D_{\phi_0}+D_{\mathrm{sp},f}+p_fD_{\mathrm{QPE}}(\epsilon)),
    \end{equation*}
    leading directly to Eq.~\eqref{eq:thm_exact_expected_cost} using Eq.~\eqref{eq:prob_over_tradeoff}.
    Similarly, the variance becomes
    \begin{align*}
        \mathrm{Var}_M[C_{\mathrm{FQPE}}(M;M_f)]
        =&D_{\mathrm{QPE}}(\epsilon)^2M_fp_f(1-p_f)\\
        =&D_{\mathrm{QPE}}(\epsilon)^2\Theta(|\gamma_{f0}|^{-2})\log(\delta^{-1})(1-p_f)\quad\text{(By Eq.~\eqref{eq:thm_Mf_bound})}\\
        =&D_{\mathrm{QPE}}(\epsilon)^2M_{\mathrm{QPE}}(\gamma_{f0},\delta)(1-p_f),
    \end{align*}
    which results in Eq.~\eqref{eq:thm_exact_standard_deviation}.
\end{proof}

\section{Classically Inspired Filters and Rejection Bounds}
\label{suppl:classically_inspired_filters}

\subsection{Rejection ratio of a filtered state}

For a filtered state
\begin{equation}
    \ket{\phi_f}
    =
    \frac{f(\hat H)\ket{\phi_0}}
    {\|f(\hat H)\ket{\phi_0}\|},
\end{equation}
the squared overlap with the ground state can be written as
\begin{equation}
    |\gamma_{f0}|^2
    =
    \frac{1}{1+R_f},
    \qquad
    R_f
    :=
    \sum_{i>0}
    \left|
        \frac{\gamma_i f(E_i)}
        {\gamma_0 f(E_0)}
    \right|^2.
    \label{eq:supp_rejection_ratio}
\end{equation}
Thus, filter design can be viewed as the problem of suppressing the relative excited-state leakage $R_f$ while maintaining a non-negligible postselection probability.

A simple upper bound is
\begin{equation}
    R_f
    \le
    \left(|\gamma_0|^{-2}-1\right)
    \max_{i>0}
    \left|
        \frac{f(E_i)}{f(E_0)}
    \right|^2.
    \label{eq:supp_rejection_bound_max}
\end{equation}
This bound motivates minimax filter design, where one minimizes the largest filter amplitude in the rejection region subject to normalization at the target energy.

\subsection{Low-pass and band-pass filters}

\begin{figure*}[t]
    \centering
    \subfigure[Low-pass filter]{
        \includegraphics[width=0.3\textwidth]{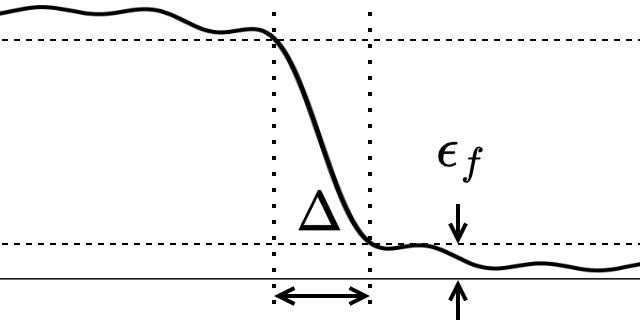}
        \label{fig:supp_low_pass_filter}
    }
    \subfigure[Band-pass filter]{
        \includegraphics[width=0.3\textwidth]{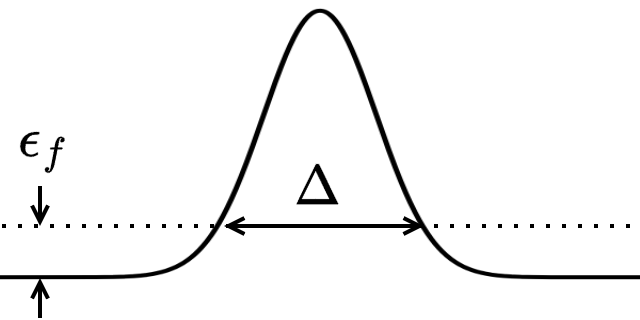}
        \label{fig:supp_band_pass_filter}
    }
    \caption{
    Illustrations of representative classically inspired filter profiles for state preparation.
    (a) A low-pass filter accepts eigenstates with energies below a threshold and suppresses higher-energy components.
    The threshold is specified by $\mu$, the transition width by $\Delta$, and the maximum rejection-band fluctuation by $\epsilon_f$.
    (b) A band-pass filter targets an energy window $[\mu-\Delta/2,\mu+\Delta/2]$ around an estimated eigenvalue $\mu$, while suppressing components outside the window.
    The parameter $\Delta$ defines the passband width, and $\epsilon_f$ bounds the maximum filter amplitude in the rejection region.
    }
    \label{fig:supp_low_band_pass_filters}
\end{figure*}

Classically inspired filters can be broadly divided into low-pass and band-pass designs.
A low-pass filter accepts eigenvalues below a threshold and suppresses those above it, while a band-pass filter targets a finite energy window around an estimated eigenvalue.
The Gaussian filter analyzed in the main text is a smooth band-pass example~\cite{He2022_GaussianFilter,low2017hamiltoniansimulationuniformspectral}.
Figure~\ref{fig:supp_low_band_pass_filters} illustrates these two representative profiles.

A low-pass filter with threshold $\mu$ and transition width $\Delta$ satisfies, for a fluctuation parameter $0<\epsilon_f<1/2$,
\begin{equation}
    f(x)\in
    \begin{cases}
        [1-\epsilon_f,1],
        & x\in[-1,\mu-\Delta/2],
        \\
        [-\epsilon_f,\epsilon_f],
        & x\in[\mu+\Delta/2,1].
    \end{cases}
\end{equation}
If the ground state lies in the pass region and all excited states lie in the rejection region, then
\begin{equation}
    R_f
    \le
    \left(|\gamma_0|^{-2}-1\right)
    \left(\epsilon_f^{-1}-1\right)^{-2}
    =
    O(|\gamma_0|^{-2}\epsilon_f^2).
\end{equation}

An ideal low-pass filter is the shifted Heaviside function, but its discontinuity makes it difficult to approximate by a finite polynomial or Fourier series without oscillatory artifacts.
Smooth approximations, such as error-function filters, avoid this issue and can be implemented using polynomial or trigonometric approximations~\cite{Lin2020nearoptimalground,lin2022QETU}.
For both polynomial and trigonometric series representations, the number of required basis functions scales as $N=O(\Delta^{-1}\log\epsilon_f^{-1})$~\cite[Lemma~14]{gilyen2019quantum}, \cite{lin2022QETU}.

Band-pass filters instead target an energy window $[\mu-\Delta/2,\mu+\Delta/2]$ and suppress components outside this region.
They are useful when an approximate target eigenvalue is available, but their performance depends on the accuracy of the center $\mu$ and on the separation between the target state and nearby eigenvalues.

\subsection{Connection to minimax and spectral-concentration filters}

Equation~\eqref{eq:supp_rejection_bound_max} naturally leads to a minimax design problem:
one seeks a filter that minimizes the maximum relative fluctuation in the rejection region.
This gives Chebyshev-type filters.
For trigonometric series, the corresponding construction is closely related to the Dolph--Chebyshev filter~\cite{epperly2022theory}, while the polynomial case yields an analogous Chebyshev construction.

Alternatively, one may minimize an integrated leakage measure,
\begin{equation}
    R_f
    \lessapprox
    \Delta E_{\min}^{-1}
    \left(
        \max_{i>0}
        \left|
            \frac{\gamma_i}{\gamma_0}
        \right|^2
    \right)
    \int_{E_1}^{E_{d-1}}
    \left|
        \frac{f(x)}{f(E_0)}
    \right|^2 dx,
\end{equation}
where $\Delta E_{\min}=\min_{0<i<d-1}(E_{i+1}-E_i)$.
This viewpoint connects filter design to spectral-concentration problems such as Slepian or DPSS filters~\cite{slepian1983} and their practical approximations, including Kaiser filters~\cite{kaiser1980}.

In the next note, we give the minimax construction used to bound the suppression-band leakage.

\section{Minimax Filter Design in the Suppression Band}\label{suppl:minimax_filter_design}
In this section, we derive functions that minimize the maximum fluctuations within the suppression band, $\Delta\le|x|\le1$ for some $0<\Delta<1$.

\subsection{Minimax Problem with General Basis}
The problem is explicitly formulated as:
\begin{equation}\label{eq:minimax_fluct}
    f^{\star}_{\Delta}(x)=\argmin_{f\in\mathcal{F}_N;f(0)=1}\max_{\Delta\le|x|\le1}|f(x)|,
\end{equation}
where $\mathcal{F}_N=\{f(x)=\sum_{k=0}^{N}c_kb_k(x):\bm{c}\in\mathbb{C}^{N+1}, |f(x)|\le1~\forall|x|\le1\}$, and $\{b_k(x)\}_{k=0}^{N}$ is a chosen function basis.

Consider decomposing $f(x)=f_\mathrm{e}(x)+f_\mathrm{o}(x)$ into its even and odd components, defined by $2f_\mathrm{e}(x)=f(x)+f(-x)$ and $2f_\mathrm{o}(x)=f(x)-f(-x)$.
Because of the symmetry of the domain and the conditions $f(0)=f_\mathrm{e}(0)=1$ and $f_\mathrm{o}(0)=0$, it follows that the odd component must vanish, since:
\begin{align*}
    \max_{\Delta\le|x|\le1}|f(x)|=&\max_{\Delta\le x\le1}\max\left\{ |f_\mathrm{e}(x)+f_\mathrm{o}(x)|, |f_\mathrm{e}(x)-f_\mathrm{o}(x)| \right\}\\
    =&\max_{\Delta\le x\le1}|f_\mathrm{e}(x)|+|f_\mathrm{o}(x)|\\
    \ge&\max_{\Delta\le x\le1}|f_\mathrm{e}(x)|.
\end{align*}
Therefore, it suffices to consider even basis functions in $\mathcal{F}_N$.
For instance, in the polynomial case, we take $\mathcal{F}_N=\mathcal{P}_{N}^{\mathrm{even}}$ with basis functions $\{b_k(x)=T_{2k}(x)\}_{k=0}^{N}$, while in the trigonometric case, we consider $\mathcal{F}_N=\mathcal{T}_N^{\mathrm{even}}$ with $\{b_k(x)=\cos k\pi x\}_{k=0}^{N}$.

To solve Eq.\eqref{eq:minimax_fluct}, we begin with Chebyshev's equioscillation theorem \cite{golomb1962lectures}:
\begin{equation}\label{eq:chebyshev_equioscillation}
    \argmin_{f\in\mathcal{P}_N'}\max_{|x|\le1}|f(x)|=\frac{T_N(x)}{2^{N-1}},
\end{equation}
where $\mathcal{P}_N'$ is the set of degree-$N$ monic polynomials.
The minimized fluctuation in $|x|\le1$ is $|f(x)|\le\epsilon_N=2^{-N+1}$.
Now, for any interval $[a,b]$ with $c\notin[a,b]$, a linear mapping from $[a,b]\rightarrow[-1,1]$ and appropriate scaling yields:
\begin{equation}\label{eq:shifted_chebyshev_equioscillation}
    \argmin_{f\in\mathcal{P}_N;f(c)=1}\max_{x\in[a,b]}|f(x)|=\epsilon_{N} T_N\left(\frac{a+b-2x}{b-a}\right),
\end{equation}
where $\mathcal{P}_N$ is the set of degree-$N$ polynomials, and $\epsilon_N=T_N(z)^{-1}$ with $z=1+2\tfrac{a-c}{b-a}$ ensures $f(c)=1$.
The achieved minimum value is $\epsilon_{N}=T_N(|z|)^{-1}$, which decays exponentially in $N$ because $|z|>1$ always holds.

To determine the required $N$ for a target error $\epsilon_f$, let $|z|=1+2\tan^2\theta$ with $0<\theta<\pi/2$, then:
\begin{equation*}
    \epsilon_N=T_N(1+2\tan^2\theta)^{-1}
    <2\left(\tan\theta+\sec\theta\right)^{-2N}
\end{equation*}
using the inequality $T_N(z)>\frac{1}{2}\left(z+\sqrt{z^2-1}\right)^N$ for $z>1$.
Therefore, to ensure $\epsilon_{N}<\epsilon_f$, it suffices that:
\begin{equation*}
    N>\frac{\log(2\epsilon_N^{-1})}{0.9\log |z|}.
\end{equation*}

\subsection{Polynomial Minimax Problem}
First, the polynomial case ($\mathcal{F}_N=\mathcal{P}^{\mathrm{even}}_N$) is solved by applying a quadratic transformation $x\rightarrow x^2$ to Eq.\eqref{eq:shifted_chebyshev_equioscillation} with $c=0$, $a=\Delta^2$, and $b=1$.
Therefore, we obtain:
\begin{gather*}
    \argmin_{f\in\mathcal{P}_N^{\mathrm{even}};f(0)=1}\max_{\Delta^2\le x^2\le1}|f(x)|=\epsilon_N^{(\mathrm{P})}T_N\left(\frac{\Delta^2+1-2x^2}{1-\Delta^2}\right),
    \epsilon_N^{(\mathrm{P})}=T_N(1+\tfrac{2}{\Delta^{-2}-1})^{-1}\lesssim2\E^{-2N\Delta}.
\end{gather*}
To ensure $\epsilon_N^{(\mathrm{P})}\le\epsilon_f$, it suffices that:
\begin{equation*}
    N^{(\mathrm{P})}>\frac{\log(2\epsilon_f^{-1})}{0.9\log\left(1+\frac{2}{\Delta^{-2}-1}\right)}=O\left(\Delta^{-2}\log\epsilon_f^{-1}\right).
\end{equation*}

\subsection{Trigonometric Minimax Problem}
Meanwhile, the trigonometric case ($\mathcal{F}_N=\mathcal{T}_N^{\mathrm{even}}$) has been studied in \cite{epperly2022theory}, where the transformation $x\rightarrow\cos \pi x$ leads to:
\begin{gather*}
    \argmin_{f\in\mathcal{T}_N^{\mathrm{even}};f(0)=1}\max_{-1\le\cos \pi x\le \cos \pi \Delta}|f(x)|=\epsilon_{N}^{(\mathrm{T})}T_N\left(1+2\frac{\cos \pi x - \cos \pi \Delta}{1+\cos \pi \Delta}\right),\\
    \epsilon_N^{(\mathrm{T})}=T_N(1+2\tan^2\tfrac{\pi\Delta}{2})^{-1}\lesssim 2\E^{-\pi N\Delta}.\nonumber
\end{gather*}
The order needed to achieve $\epsilon_N^{(\mathrm{T})}\le\epsilon_f$ is:
\begin{equation*}
    N^{(\mathrm{T})}>\frac{\log(2\epsilon_f^{-1})}{2\log\left(1+\pi\Delta/2\right)}=O(\Delta^{-1}\log\epsilon_f^{-1}).
\end{equation*}

\subsection{Shifted Minimax Filter Function}
In practice, filter functions are often centered at $\mu\in[-1,1]$ to cover a specific energy window $[\mu-\Delta, \mu+\Delta]$.
Thus, a more practical formulation than Eq.\eqref{eq:minimax_fluct} is:
\begin{equation}\label{eq:minimax_fluct_mu}
    f^{\star}_{\mu,\Delta}(x)=\argmin_{f\in\mathcal{F}_N;f(\mu)=1}\max_{|x-\mu|\ge\Delta\wedge|x|\le1}|f(x)|,
\end{equation}
which is generally asymmetric with respect to $x=\mu$.
Unfortunately, no closed-form solution is currently known for this case.
As a practical alternative, we consider a relaxed version that expands the fluctuation region:
\begin{equation}\label{eq:minimax_fluct_mu_subopt}
    \tilde{f}^{\star}_{\mu,\Delta}(x)=\argmin_{f\in\mathcal{F}_N;f(\mu)=1}\max_{\Delta\le|x-\mu|\le1+|\mu|}|f(x)|,
\end{equation}
which is solvable using the methods discussed for Eq.\eqref{eq:minimax_fluct}:
Specifically, we have an optimal solution for Eq.\eqref{eq:minimax_fluct_mu_subopt} as
\begin{equation*}
    \tilde{f}^{\star}_{\mu,\Delta}(x)=f^{\star}_{\Delta'}\left(\frac{x-\mu}{1+|\mu|}\right),
\end{equation*}
where $\Delta'=\Delta/(1+|\mu|)$.
This function $\tilde{f}^{\star}_{\mu,\Delta}$ serves as a suboptimal but tractable solution to the problem in Eq.\eqref{eq:minimax_fluct_mu}.

\section{Fourier Series Analysis of Gaussian Function}\label{suppl:gaussian_fourier}
In this section, we review the error and determine the number of trigonometric basis functions required to approximate a Gaussian function centered at $\mu\in[-1,1]$ with the width $\sigma\in(0,1/2]$ by using the well-established Fourier series method.

We begin by considering the Gaussian filter function $f(x)= \E^{-\frac{x^2}{2\sigma^2}}$, which we approximate over the domain $|x|\le L=1+|\mu|$ using a finite Fourier series.
We then consider a shifting $x\rightarrow (x-\mu)$ so that the shifted domain includes $|x|\le1$.
Let the maximum approximation error be bounded by $\epsilon_f$:
\begin{equation}\label{eq:gaussian_fourier_error}
\max_{|x|\le L}\left| f(x) - \sum_{k=0}^{N}{}^{'} c_k \cos(2\pi kx/L) \right| \le \epsilon_f,
\end{equation}
where $c_k=\tfrac{1}{L}\int_{-L}^{L}f(x)\cos(\pi kx/L)dx$, and $\sum'$ indicates that the first term $(k=0)$ is taken with half weight.

Assuming $L\gg\sigma$, the Fourier coefficients can be approximated as:
\begin{align*}
    c_k=&\sqrt{2\pi}\frac{\sigma}{L}\E^{-\frac{\pi^2k^2\sigma^2}{2L^2}}\mathrm{Re}\left[\mathrm{erf}\left(\frac{1}{\sqrt{2}}\left(\frac{L}{\sigma}+\I\pi k\frac{\sigma}{L}\right)\right)\right]\\
    \approx&\sqrt{2\pi}\frac{\sigma}{L}\E^{-\frac{\pi^2k^2\sigma^2}{2L^2}},
\end{align*}
where the approximation holds due to the rapid decay of the imaginary part of the error function in the complex plane for large $L/\sigma$.

The residual error in Eq.\eqref{eq:gaussian_fourier_error} is then bounded by:
\begin{align*}
    \sum_{k>N}|c_k|\lesssim\int_{N\sigma/L}^{\infty}\E^{-\frac{\pi^2x^2}{2}}dx = \mathrm{erfc}\left(\frac{\pi N \sigma}{\sqrt{2}L}\right)< \E^{-\frac{\pi^2N^2\sigma^2}{2L^2}}.
\end{align*}
To ensure the error is less than $\epsilon_f$, the minimum number of Fourier terms required satisfies:
\begin{equation}
    N>\frac{\sqrt{2}L}{\pi\sigma}\sqrt{\log \epsilon_f^{-1}} = O\left(\sigma^{-1}\sqrt{\log \epsilon_f^{-1}}\right).
\end{equation}

For a fair comparison with the Chebyshev filter that satisfies $f(\Delta)\le\epsilon_f$, we need rescaling of the width to $\sigma=\Delta/\sqrt{2\log\epsilon_f^{-1}}$, which results in the number of basis similar to that of Chebyshev filter:
\begin{equation}\label{eq:gaussian_num_basis}
    N> O\left(\Delta^{-1}\log \epsilon_f^{-1}\right).
\end{equation}

\section{Gaussian FQPE with Coarse Spectral Estimates}\label{suppl:gaussian_FQPE}
\begin{figure*}
    \centering
    \subfigure[]{
        \includegraphics[width=0.4\linewidth]{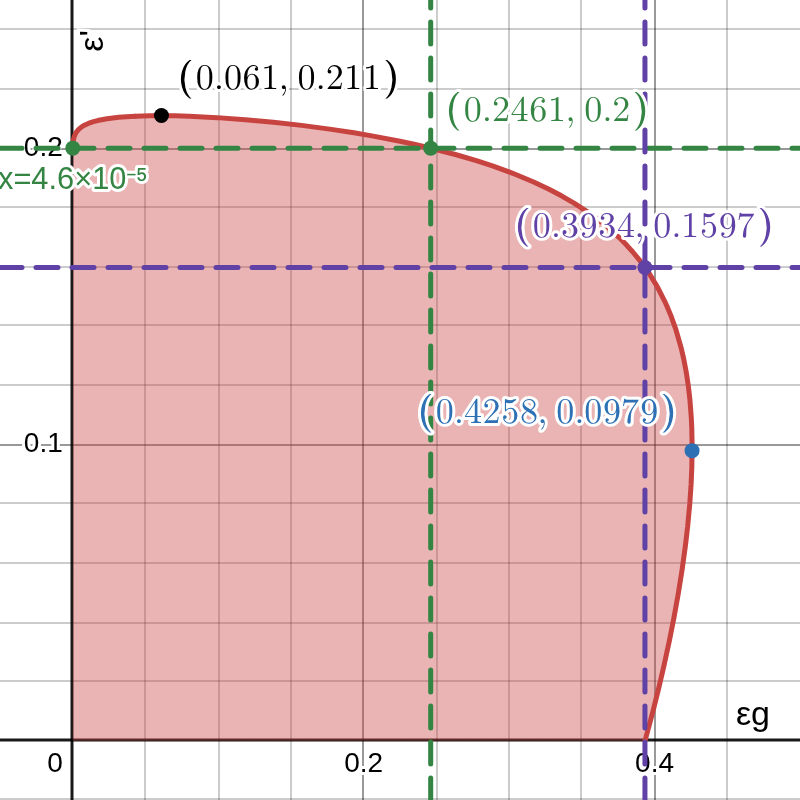}
        \label{fig:epsilon_prime}
    }
    \caption{The range of $(\epsilon_g,\epsilon')$ that satisfies the inequality \eqref{eq:thm_gauss_fqpe_g2} and thus the filter value at the ground state energy $|\tilde{g}(E_0)|$ is ensured to be larger than $\E^{-1}\epsilon_g^{2\epsilon'}$.
    In the figure, some important points and lines are denoted: maximally allowed $\epsilon'$(black), maximally allowed $\epsilon_g$(blue), the region used in Theorems~\ref{thm:gaussianfqpe} and \ref{thm:gaussianfqpe_poly}~(green) and the another region that allows loose bound for $\epsilon_g$ and tighter $\epsilon'$(purple).
    }
    \label{fig:bound_epsilons}
\end{figure*}

In this section, we derive an upper bound on the cost of FQPE using the approximated Gaussian filter both for trigonometric and polynomial bases, as stated in the following theorems.

\begin{theorem}[Gaussian FQPE with Trigonometric Approximation, formal version of Theorem~\ref{thm:gaussian_fqpe}]\label{thm:gaussianfqpe}
    Suppose that prior estimates $\tilde{E}_0$ and $\tilde{E}_1$ of $E_0$ and $E_1$ are given with accuracy of $\epsilon'\Delta E_0$, where 
    \begin{equation}\label{eq:epsilonprime_bound}
        0<\epsilon'\le1/5.        
    \end{equation}
    For the accuracy of 
    \begin{equation}\label{eq:epsilon_bound}
        1.92\times10^{-8}< \frac{\epsilon}{\Delta E_0}< \epsilon',
    \end{equation}
    there exists a filtered QPE algorithm that estimates the ground state energy with confidence at least $1-\delta$, whose total cost is upper-bounded by:
    \begin{equation}\label{eq:thm_gauss_fqpe_result}
    \begin{split}
        &M_{\mathrm{QPE}}(\gamma_0,\delta)D_{\mathrm{QPE}}(\epsilon)\times 4\E\left( \frac{\kappa\epsilon}{\Delta E_0} \right)^{1-\epsilon'}\left(1+\log \frac{\Delta E_0}{\kappa\epsilon}\right)+M_{\mathrm{QPE}}(1,\delta)D_{\mathrm{QPE}}(\epsilon)\\
        =&M_{\mathrm{QPE}}(\gamma_0,\delta)D_{\mathrm{QPE}}(\epsilon)\times O\left(\left(\frac{\epsilon}{\Delta E_0}\right)^{1-\epsilon'}\log \frac{\Delta E_0}{\epsilon}\right),
    \end{split}
    \end{equation}
    where a global constant is defined as
    \begin{equation*}
        \kappa=\frac{5}{4\pi}\left(2+\frac{1}{2\delta_0}\right)^{-1}\approx 0.110.
    \end{equation*}
\end{theorem}

\begin{proof}
Consider the following approximate Gaussian function of the Hamiltonian $\tilde{g}(\hat{H})$:
\begin{equation*}
    \max_{x\in[-1,1]}|\tilde{g}(x)-g(x)|\le\epsilon_g,\quad
    g(x)=\exp\left(\left(\frac{x-\tilde{E}_0}{(1-\epsilon')(\tilde{E}_1-\tilde{E}_0)}\right)^2\log\epsilon_g \right),
\end{equation*}
where $g(x)$ represents an ideal Gaussian filter constructed using coarse estimates $\tilde{E}_0$ and $\tilde{E}_1$, satisfying $|\tilde{E}_{0(1)}-E_{0(1)}|\le\epsilon'\Delta E_0$ and some $0<\epsilon_g<1$, which will be determined later.
Using the analysis presented in Supplementary Note~\ref{suppl:gaussian_fourier}, we can implement such an approximation using a Fourier series of order
\begin{equation*}
    N=\left\lceil\frac{4\log \epsilon_g^{-1}}{\pi(1-\epsilon')(\tilde{E}_1-\tilde{E}_0)}\right\rceil\le\left\lceil\frac{10\log \epsilon_g^{-1}}{\pi\Delta E_0}\right\rceil,
\end{equation*}
since $(1-\epsilon')(\tilde{E}_1-\tilde{E}_0) \ge (1-3\epsilon'+2\epsilon'^2)\Delta E_0>0.4\Delta E_0$ under the assumed $\epsilon'<1/5$.
This results in the a GQSP circuit depth of
\begin{equation}\label{eq:thm_gauss_fqpe_g0}
    D_{\mathrm{sp},\tilde{g}}=N\le c_0\log\epsilon_g^{-1}D_{\mathrm{QPE}}(\Delta E_0),
\end{equation}
where
\begin{equation*}
    c_0=10\pi^{-1}\left(2+\frac{1}{2\delta_0}\right)^{-1}\approx 0.8796
\end{equation*}

We now aim to bound the cost factor in Eq.~\eqref{eq:thm_exact_expected_cost} corresponding to $f(\hat{H})=\tilde{g}(\hat{H})$:
\begin{equation}\label{eq:thm_gauss_fpqe_g1}
    |\tilde{g}(E_0)|^{-2}\frac{D_{\mathrm{sp},\tilde{g}}}{D_{\mathrm{QPE}}(\epsilon)}+\left|\frac{\gamma_0}{\gamma_{\tilde{g}0}}\right|^2\le|\tilde{g}(E_0)|^{-2}\left(\frac{D_{\mathrm{sp},\tilde{g}}}{D_{\mathrm{QPE}}(\epsilon)}+|\tilde{g}(E_1)|^2\right)+|\gamma_0|^2,
\end{equation}
since $|\gamma_{\tilde{g}0}|^{-2}=1+R_{\tilde{g}}\le1+\left|\frac{\tilde{g}(E_1)}{\tilde{g}(E_0)}\right|^2(|\gamma_0|^{-2}-1)$ from Eq.~\eqref{eq:supp_rejection_bound_max}.
Furthermore, because the main lobe of $g(x)$ never covers $E_1$ ($E_1>\tilde{E_1}+\epsilon'\Delta E_0$), we can bound $|\tilde{g}(E_1)|^2$ as
\begin{equation}\label{eq:thm_gauss_fqpe_g2}
    |\tilde{g}(E_1)|^2\le(g(E_1)+\epsilon_g)^2\le\left[\exp\left(\left(\frac{1+\epsilon'}{(1-\epsilon')(1-2\epsilon')}\right)^2\log\epsilon_g\right)+\epsilon_g\right]^2\le4\epsilon_g^{2}.
\end{equation}
The factor in Eq.~\eqref{eq:thm_gauss_fpqe_g1}, combined with Eqs.~\eqref{eq:thm_gauss_fqpe_g0} and \eqref{eq:thm_gauss_fqpe_g2}, is bounded as
\begin{equation*}
    \frac{D_{\mathrm{sp},\tilde{g}}}{D_{\mathrm{QPE}}(\epsilon)}+|\tilde{g}(E_1)|^2\le \frac{c_0\epsilon}{\Delta E_0} \log \epsilon_g^{-1}+4\epsilon_g^{2}=:\alpha(\epsilon_g),
\end{equation*}
which is minimized by 
\begin{equation}\label{eq:opt_epsilon_g}
    \epsilon_g^\star=\sqrt{\frac{c_0\epsilon}{8\Delta E_0}},
\end{equation}
and its minimum is 
\begin{equation}
    \alpha(\epsilon_g^{\star})=\frac{1}{2}\frac{c_0\epsilon}{\Delta E_0}\left(1+\log \frac{8\Delta E_0}{c_0\epsilon}\right).
\end{equation}

To bound the other factor $|\tilde{g}(E_0)|^{-2}$, observe:
\begin{align}
    |\tilde{g}(E_0)|^{-2}\le&|{g}(E_0)-\epsilon_g|^{-2} \nonumber\\
    =& \left(\exp\left(\left(\frac{E_0-\tilde{E}_0}{\tilde{E}_1-\tilde{E}_0-\epsilon'\Delta E_0}\right)^2\log\epsilon_g\right)-\epsilon_g\right)^{-2}\nonumber\\
    \le& \left(\exp\left(\left(\frac{\epsilon'}{1-3\epsilon'}\right)^2\log\epsilon_g\right)-\epsilon_g\right)^{-2}\nonumber\\
    %\le& (\epsilon^{\epsilon'^2+O(\epsilon'^3)}-\epsilon)^{-2}\\
    <& \E\epsilon_g^{-2\epsilon'}\label{eq:thm_gauss_fqpe_g3}.
    %\le& 4\epsilon^{-2\epsilon'^2},
\end{align}
Numerical validation with 128-bit arithmetics shows that the bound of Eq.~\eqref{eq:thm_gauss_fqpe_g2} holds whenever $0<\epsilon'<1/5$ and $4.6\times10^{-5}<\epsilon_g<0.24$, which leads to 
\begin{equation}\label{eq:epsilon_bound_proof}
        1.92\times10^{-8} < \frac{\epsilon}{\Delta E_0} <0.5238,
\end{equation}
by Eq.~\eqref{eq:opt_epsilon_g}.
However, for $\epsilon/\Delta E_0>\epsilon'$, FQPE doesn't improve the accuracy of ground state energy from the prior estimate.
Thus we take the lower bound of Eq.~\eqref{eq:epsilon_bound_proof}, which results in the bound of Eq.~\eqref{eq:epsilon_bound}.

Substituting $\kappa=c_0/8$, the overall bounds of Eq.~\eqref{eq:thm_gauss_fpqe_g1} is given by
\begin{equation*}
     |\tilde{g}(E_0)|^{-2}\frac{D_{\mathrm{sp},\tilde{g}}}{D_{\mathrm{QPE}}(\epsilon)}+\left|\frac{\gamma_0}{\gamma_{\tilde{g}0}}\right|^2<4\E\left( \frac{\kappa\epsilon}{\Delta E_0} \right)^{1-\epsilon'}\left(1+\log \frac{\Delta E_0}{\kappa\epsilon}\right)+|\gamma_0|^2,
\end{equation*}
which leads to Eq.~\eqref{eq:thm_gauss_fqpe_result}.
\end{proof}

\begin{remark}
    Beyond the bound on $\epsilon'$ given in Eq.~\eqref{eq:epsilonprime_bound}, a more general admissible region for $(\epsilon_g,\epsilon')$ and the corresponding values of $\epsilon/\Delta E_0$ is illustrated in Fig.~\ref{fig:bound_epsilons}, where Eq.~\eqref{eq:thm_gauss_fqpe_g3} is satisfied.
\end{remark}

\begin{remark}
    It is in principle possible to minimize the bound in Eq.~\eqref{eq:thm_gauss_fpqe_g1} after combining Eqs.~\eqref{eq:thm_gauss_fqpe_g2} and \eqref{eq:thm_gauss_fqpe_g3}, yielding a slightly tighter cost estimate. 
    However, the improvement is only of order $O(\epsilon'^2)$, while the resulting expression for $\epsilon_g$ and final cost bound become substantially more complicated.
\end{remark}

\begin{remark}[Sufficient condition for Gaussian FQPE advantage]
\label{rem:suppl_gaussian_fqpe_advantage}
Substituting the explicit Gaussian FQPE bound gives the sufficient condition
\begin{equation}
\label{eq:suppl_gaussian_fqpe_advantage_condition}
    \frac{\bar{C}_{\mathrm{FQPE},g}}{C_{\mathrm{QPE}}}=4\E
    \left(
        \kappa\frac{\epsilon}{\Delta E_0}
    \right)^{1-\epsilon'}
    \left(
        1+
        \log
        \frac{\Delta E_0}{\kappa\epsilon}
    \right)
    +
    \frac{
        M_{\mathrm{QPE}}(1,\delta)
    }{
        M_{\mathrm{QPE}}(\gamma_0,\delta)
    }
    <1 .
\end{equation}
Since the second term in Eq.~\eqref{eq:suppl_gaussian_fqpe_advantage_condition} scales as
\[
    \frac{
        M_{\mathrm{QPE}}(1,\delta)
    }{
        M_{\mathrm{QPE}}(\gamma_0,\delta)
    }
    =
    \Theta(|\gamma_0|^2).
\]
Thus, up to constants in the repetition bound, Eq.~\eqref{eq:suppl_gaussian_fqpe_advantage_condition} reduces to
\begin{equation}
\label{eq:suppl_gaussian_fqpe_advantage_condition_simple}
    4\E
    \left(
        \kappa\frac{\epsilon}{\Delta E_0}
    \right)^{1-\epsilon'}
    \left(
        1+
        \log
        \frac{\Delta E_0}{\kappa\epsilon}
    \right)
    +
    |\gamma_0|^2
    <1 .
\end{equation}
Equivalently, for fixed $\epsilon'$ and $|\gamma_0|^2$, the sufficient advantage regime can be written as
\begin{equation}
\label{eq:suppl_gaussian_fqpe_advantage_rstar}
    \frac{\epsilon}{\Delta E_0}
    <
    r_\ast(\epsilon',|\gamma_0|^2),
\end{equation}
where $r_\ast(\epsilon',|\gamma_0|^2)$ is the positive solution obtained by replacing the inequality in
Eq.~\eqref{eq:suppl_gaussian_fqpe_advantage_condition_simple} with equality.
This form makes clear that the advantage is enhanced when the initial overlap is small, when the target precision is much smaller than the spectral gap, and when the coarse spectral estimates are accurate.
In particular, for $\epsilon'\ll 1$, the filtering-dominated part of the Gaussian FQPE cost scales approximately as
\[
    \tilde O\!\left(
        |\gamma_0|^{-2}\Delta E_0^{-1}
    \right),
\]
rather than
\[
    \tilde O\!\left(
        |\gamma_0|^{-2}\epsilon^{-1}
    \right),
\]
so the dominant precision dependence is shifted from the final target precision $\epsilon$ to the spectral gap $\Delta E_0$.
\end{remark}

\begin{theorem}[Gaussian FQPE with Polynomial Approximation]\label{thm:gaussianfqpe_poly}
    Import the notation and assumptions from Theorem~\ref{thm:gaussianfqpe}, except for Eq.~\eqref{eq:epsilon_bound}.
    Let $\epsilon_{\mathrm{HE}}\in(0,\E^{-\E\pi/2})$. Suppose the phase estimation uses an approximate time evolution operator $\hat{U}_{\hat{H}}$ satisfying    
    \begin{equation}\label{eq:thm_gauss_fqpe_poly_assum_simerr}
        \|\hat{U}_{\hat{H}}-\E^{-\I\pi\hat{H}}\|\le \epsilon_{\mathrm{HE}}<\E^{-\E\pi/2}\approx1.40\times10^{-2},
    \end{equation}
    implemented via QSVT based on truncated Jacobi-Anger expansion (Eq.~\ref{eq:trunc_jacobi_anger}).
    For the accuracy satisfying
    \begin{equation}\label{eq:thm_gauss_fqpe_poly_assum_epsilon}
        1.80\times 10^{-8}<\frac{\epsilon}{\Delta E_0 \log\epsilon_{\mathrm{HE}}^{-1}}\le \frac{\epsilon'}{\log \epsilon_{\mathrm{HE}}^{-1}},
    \end{equation}
    there exists a state preparation such that the total cost is reduced by
    \begin{equation}\label{eq:thm_gauss_fqpe_poly_result}
    \begin{split}
        &M_{\mathrm{QPE}}(\gamma_0,\delta)D_{\mathrm{QPE}}(\epsilon)\times 4\E\left( \frac{c\epsilon}{\Delta E_0 \log \epsilon_{\mathrm{HE}}^{-1}}\right)^{1-\epsilon'}\left(1+\log\frac{16\Delta E_0\log \epsilon_{\mathrm{HE}}^{-1}}{c\epsilon}\right)+M_{\mathrm{QPE}}(1,\delta)D_{\mathrm{QPE}}(\epsilon)\\
        =&M_{\mathrm{QPE}}(\gamma_0,\delta)D_{\mathrm{QPE}}(\epsilon)\times O\left(\left(\frac{\epsilon}{\Delta E_0\log \epsilon_{\mathrm{HE}}^{-1}}\right)^{1-\epsilon'}\log \frac{\Delta E_0\log \epsilon_{\mathrm{HE}}^{-1}}{\epsilon}\right),
    \end{split}
    \end{equation}
    with a constant is defined as
    \begin{equation*}
        c = \frac{5\E}{32}\left(2+\frac{1}{2\delta_0}\right)^{-1}\approx 0.1174.
    \end{equation*}
\end{theorem}

\begin{proof}
    The proof is identical to that of Theorem~\ref{thm:gaussianfqpe} except for $D_{\mathrm{QPE}}$ and $D_{\mathrm{sp},f}$.
    The truncation order of the Hamiltonian simulation in Eq.~\ref{eq:trunc_jacobi_anger} is replaced by a simple bound based on the assumption in Eq.~\eqref{eq:thm_gauss_fqpe_poly_assum_simerr}:
    \begin{equation*}
        N_{\mathrm{sim}}=\frac{e\pi}{2}+\log\epsilon_{\mathrm{HE}}^{-1}\le 2\log\epsilon_{\mathrm{HE}}^{-1},    
    \end{equation*}
    This results in the QPE depth of
    \begin{equation*}
        D_{\mathrm{QPE}}(\epsilon)=\epsilon^{-1}\left(2+\frac{1}{2\delta_0}\right)\cdot 2\log\epsilon_{\mathrm{HE}}^{-1},
    \end{equation*}
    from Eq.~\eqref{eq:optimal_singleshot_cost}.
    
    Furthermore, using \cite[Corollary~3]{low2017hamiltoniansimulationuniformspectral}, the polynomial approximation to the Gaussian filter satisfies:
    \begin{equation*}
        D_{\mathrm{sp},\tilde{g}}\le \sqrt{\max\left[\frac{\E^2\log1/\epsilon_g}{(1-\epsilon')^2(\tilde{E}_1-\tilde{E}_0)^2},\log(2/\epsilon_g)\right]\log(4/\epsilon_g)},
    \end{equation*}
    For a sufficiently small $\epsilon_g$, we can take the first term as the maximum value since $\tilde{E}_1-\tilde{E}_0\le 2 <\E/(1-\epsilon')$.
    Furthermore, since
    \begin{align*}
        (1-\epsilon')(\tilde{E}_1-\tilde{E}_0)\ge& (1-\epsilon')(1-2\epsilon')\Delta E_0\\
        \ge& (1-3\epsilon')\Delta E_0\\
        \ge& \frac{2}{5}\Delta E_0,
    \end{align*}
    we can bound $D_{\mathrm{sp},\tilde{g}}$ by
    \begin{equation*}
        D_{\mathrm{sp},\tilde{g}} < \frac{5\E}{2\Delta E_0}\log 4\epsilon_g^{-1}.
    \end{equation*}
    
    Then, analogous to the proof of Theorem~\ref{thm:gaussianfqpe}, we have the following bound of
    \begin{equation*}
        \frac{D_{\mathrm{sp},\tilde{g}}}{D_{\mathrm{QPE}}(\epsilon)} + |\tilde{g}(E_1)|^2\le \frac{c_0 \epsilon}{\Delta E_0 \log \epsilon_{\mathrm{HE}}^{-1}}\log4\epsilon_g^{-1}+4\epsilon_g^2=:\alpha(\epsilon_g),
    \end{equation*}
    where
    \begin{equation*}
        c_0=\frac{5\E}{4}\left(2+\frac{1}{2\delta_0}\right)^{-1}\approx 0.9391.
    \end{equation*}
    The bound $\alpha(\epsilon_g)$ is minimized by
    \begin{equation}\label{eq:epsilon_g_in_poly}
        \epsilon_g^{\star}=\sqrt{\frac{c_0\epsilon}{8\Delta E_0\log\epsilon_{\mathrm{HE}}^{-1}}},
    \end{equation}
    and the corresponding minimum value is
    \begin{equation*}
        \alpha(\epsilon_g^{\star})= \frac{c_0\epsilon}{2\Delta E_0 \log \epsilon_{\mathrm{HE}}^{-1}}\left( 1+ \log\frac{128\Delta E_0  \log\epsilon_{\mathrm{HE}}^{-1}}{c_0\epsilon} \right).
    \end{equation*}

    The bound of $|\tilde{g}(E_0)|^{-2}<\E \epsilon_g^{-2\epsilon'}$ is also similarly achieved, with the same condition for $\epsilon'$ and $\epsilon_g$.
    However, since $\epsilon_g$ in Eq.~\eqref{eq:epsilon_g_in_poly} is different, we have the bound of
    \begin{equation*}
       1.80\times 10^{-8} <\frac{\epsilon}{\Delta E_0 \log \epsilon_{\mathrm{HE}}^{-1}} < 0.49.
    \end{equation*}
    Here, the conditions of $\epsilon'<1/5$ and $\epsilon_{\mathrm{HE}}<\E^{-\E\pi/2}$ makes the upper bound trivial, leading to Eq.~\eqref{eq:thm_gauss_fqpe_poly_assum_epsilon}.

    Finally, the total cost factor becomes
    \begin{equation*}
        |\tilde{g}(E_0)|^{-2}\frac{D_{\mathrm{sp},\tilde{g}}}{D_{\mathrm{QPE}}(\epsilon)}+\left|\frac{\gamma_0}{\gamma_{\tilde{g}0}}\right|^2<4\E\left( \frac{c\epsilon}{\Delta E_0 \log \epsilon_{\mathrm{HE}}^{-1}} \right)^{1-\epsilon'}\left(1+\log \frac{16\Delta E_0 \log \epsilon_{\mathrm{HE}}^{-1}}{c\epsilon}\right)+|\gamma_0|^2,
    \end{equation*}
    which establishes the bound in the theorem.
\end{proof}

\section{Gaussian FQPE without Prior Estimations}\label{suppl:FQPE_cost_gaussian_without_prior}
In Theorem~\ref{thm:gaussianfqpe}, we assumed that coarse prior estimates $\tilde{E}_0$ and $\tilde{E}_1$ of $E_0$ and $E_1$ are available with accuracy $\epsilon'\Delta E_0$.
To allow a fair comparison with algorithms that do not make this assumption, the following theorem gives a total cost bound that explicitly includes the cost of obtaining these prior estimates via standard QPE, in addition to the FQPE stage.
It also specifies a choice of $\epsilon'$ that minimizes the total cost.

\begin{corollary}[thm:gaussianfqpe][Gaussian FQPE without Prior Estimates, formal version of Corollary \ref{cor:two_stage_gaussian_fqpe}]\label{corr:gaussian_fqpe_wo_prior}
    For the accuracy range
    \begin{equation}\label{eq:cor_interior_condition}
        1.92\times10^{-8}<\frac{\epsilon}{\Delta E_0}<\frac{1}{5}
    \end{equation}
    the two-stage algorithm estimates $E_0$ with confidence at least $1-\delta$, and the total cost is bounded by
    \begin{equation}\label{eq:cor_simple_cost_bound_final}
        C_{\mathrm{2stage}}
        \le
        \left(
            \frac{c_1}{|\gamma_{\mathrm{min}}|^2\Delta E_0}\log \frac{\Delta E_0}{\kappa \epsilon}
            +
            \frac{c_2}{\epsilon}
        \right)
        \left[\log\frac{3}{\delta}
        +\log\frac{|\gamma_0|^2\Delta E_0}{c_3\epsilon}\right],
    \end{equation}
    where $|\gamma_{\mathrm{min}}|^2:=\min\{|\gamma_0|^2,|\gamma_1|^2\}$, and the universal constants are $c_1 \approx 45.420$, $c_2 \approx 3.618$, $c_3\approx 5.192$, and $\kappa\approx 0.110$.
    In particular, hiding polylogarithmic factors and assuming $|\gamma_0|\approx |\gamma_1|$,
    \begin{equation}
        C_{\mathrm{2stage}}=\tilde{O}\!\left(\epsilon^{-1}+|\gamma_0|^{-2}\,\Delta E_0^{-1}\right).        
    \end{equation}
\end{corollary}

\begin{proof}
    We define the confidence of the prior (coarse) estimates as $1-\delta_1$:
    \begin{equation*}
        P_{\mathrm{prior}}(\epsilon'):=\Pr\left[|\tilde{E}_j-E_j|\le \epsilon'\Delta E_0, j=0,1\right]\ge1-\delta_1,
    \end{equation*}
    and the conditional confidence $1-\delta_2$ of FQPE (given the prior estimates are accurate within $\epsilon'\Delta E_0$):
    \begin{equation*}
        P_{\mathrm{FQPE|prior}}(\epsilon')=\Pr\left[|\tilde{E}_{0}^{(\mathrm{FQPE})}-E_0|\le\epsilon~\big\rvert~|\tilde{E}_j-E_j|\le\epsilon'\Delta E_0, j=0, 1\right]\ge1-\delta_2.
    \end{equation*}
    Then the total success probability satisfies
    \begin{equation}\label{eq:pf4_total_confidence}
         \Pr\left[|\tilde{E}_{0}^{(\mathrm{FQPE})}-E_0|\le\epsilon\right]\ge P_{\mathrm{FQPE}}(\epsilon'):=P_{\mathrm{prior}}P_{\mathrm{FQPE|prior}}\ge(1-\delta_1)(1-\delta_2)\ge 1-\delta.
    \end{equation}

    \textbf{QPE for prior estimates (Stage 1).}
    Recall Eq.~\eqref{eq:optimal_singleshot_succprob} and we set $p_j:=(1-\delta_0^\star)|\gamma_j|^2$ for $j=0,1$.
    Over $M_1$ independent QPE trials at target accuracy $\epsilon'\Delta E_0$, the probability of finding both $\tilde{E}_0$ and $\tilde{E}_1$ is
    \begin{equation*}
        P_{\mathrm{prior}}(\epsilon')=1-(1-p_0)^{M_1}-(1-p_1)^{M_1}+(1-p_0-p_1)^{M_1}.
    \end{equation*}
    For 
    \begin{equation*}
       M_1=\left\lceil p_{\mathrm{min}}^{-1}\log (2\delta_1^{-1})\right\rceil,\quad p_{\mathrm{min}}:=\min\{p_0,p_1\},
    \end{equation*}
    we have 
    \begin{equation*}
        P_{\mathrm{prior}}(\epsilon')\ge1-2(1-p_{\mathrm{min}})^{M_1}+(1-2p_{\mathrm{min}})^{M_1}\ge 1-2\E^{-M_1 p_{\mathrm{min}}}\ge 1-\delta_1.
    \end{equation*}
    Thus the Stage~1 cost is
    \begin{align*}
        C_{\mathrm{prior}}(\epsilon', \delta_1)=D_{\mathrm{QPE}}(\epsilon'\Delta E_0)\left\lceil p_{\mathrm{min}}^{-1}\log (2\delta_1^{-1})\right\rceil
        =\frac{A_0}{\epsilon'}\log \frac{2}{\delta_1},
    \end{align*}
    where 
    \begin{equation*}
        A_0:=\frac{c_D}{\Delta E_0 p_{\mathrm{min}}},     
    \end{equation*}
    and $c_D=2+(2\delta_0)^{-1}\approx3.61803$ is from Eq.~\eqref{eq:optimal_singleshot_cost}.

    \textbf{FQPE (Stage 2).}
    From Theorem~\ref{thm:gaussianfqpe}, we have the FQPE part cost of
    \begin{equation*}
        C_{\mathrm{FQPE}}(\epsilon', \delta_2)
        \le \left(B_0 C_0^{1-\epsilon'}+\frac{c_D}{\epsilon}\right)\log \delta_2^{-1},
    \end{equation*}
    where
    \begin{equation*}
        B_0:=\frac{4\E c_D}{p_0\epsilon}\left(1+\log \frac{\Delta E_0}{\kappa\epsilon}\right),\quad
        C_0:=\frac{\kappa \epsilon}{\Delta E_0},
    \end{equation*}
    and $\kappa:=\frac{5}{4\pi}\left(2+\frac{1}{2\delta_0}\right)^{-1}\approx 0.110$.

    \textbf{Splitting success probabilities.}
    The total cost becomes:
    \begin{align*}
        C_{\mathrm{2stage}}(\epsilon',\delta_1, \delta_2)\le&\frac{A_0}{\epsilon'}\log\frac{2}{\delta_1}+\left(B_0C_0^{1-\epsilon'}+\frac{c_D}{\epsilon}\right)\log \frac{1}{\delta_2}\\
        =&\alpha(\epsilon')\log\frac{2}{\delta_1}+\beta(\epsilon')\log\frac{1}{\delta_2},
    \end{align*}
    with
    \begin{equation*}
        \alpha(\epsilon'):=\frac{A_0}{\epsilon'},\quad \beta(\epsilon'):=B_0C_0^{1-\epsilon'}+\frac{c_D}{\epsilon}.
    \end{equation*}
    Under Eq.~\eqref{eq:pf4_total_confidence} and small $\delta$ (so $\delta_1\delta_2\approx0$), a near-optimal split is
    \begin{equation*}
        \delta_1^{\star}=\frac{\alpha(\epsilon')}{\alpha(\epsilon')+\beta(\epsilon')}\delta,\quad \delta_2^{\star}=\frac{\beta(\epsilon')}{\alpha(\epsilon')+\beta(\epsilon')}\delta,
    \end{equation*}
    giving
    \begin{equation}\label{eq:pf4_intermediate_cost}
        C_{\mathrm{2stage}}(\epsilon')\le \alpha(\epsilon')\log \frac{2(1+\beta(\epsilon')/\alpha(\epsilon'))}{\delta}+\beta(\epsilon')\log \frac{1+\alpha(\epsilon')/\beta(\epsilon')}{\delta}.
    \end{equation}

    \textbf{Balanced choice of the prior accuracy $\epsilon'$.}
    Choose 
    \begin{equation*}
        \epsilon'^\star=\frac{1}{L}W(a),\quad L:=\log C_0^{-1}=\log\frac{\Delta E_0}{\kappa \epsilon}>1,
    \end{equation*}
    which ensures
    \begin{equation*}
        \beta(\epsilon'^\star)=\frac{1}{2}\alpha(\epsilon'^\star)+\frac{c_D}{\epsilon}.
    \end{equation*}
    Here 
    \begin{gather*}
        a:=\frac{A_0L}{2B_0C_0}=\frac{p_0}{8\E \kappa (1+L^{-1})p_\mathrm{min}},\\
        L:=\log C_0^{-1} >0,\quad \alpha^{\star}:=\alpha(\epsilon'^{\star}),\quad \beta:=\beta(\epsilon'^{\star}),
    \end{gather*}
    and $W(\cdot)$ is the Lambert $W$ function.
    Since $\beta^\star/\alpha^\star>\tfrac{1}{2}$, we have $2(1+\beta^\star/\alpha^\star)>3>1+\alpha^\star/\beta^\star$; hence from Eq.~\eqref{eq:pf4_intermediate_cost},
    \begin{align*}
        C_{\mathrm{2stage}}(\epsilon'^\star)\le& (\alpha^\star+\beta^\star)\log\left[\frac{2}{\delta}\left(1+\frac{\beta^\star}{\alpha^\star}\right)\right]\\
        =&\left(\frac{3}{2}\alpha^\star+\frac{c_D}{\epsilon}\right)\log\left[\frac{1}{\delta}\left(3+\frac{2c_D}{\epsilon\alpha^{\star}}\right)\right].
    \end{align*}
    Using $a/(1+a)\le W(a)\le a$ and $(1+L^{-1})^{-1} \le 8\E \kappa a\le p_0/p_{\mathrm{min}}$
    we finally obtain
    \begin{equation*}
        C_\mathrm{2stage}(\epsilon'^\star)\le \frac{c_D}{2}\left(\frac{3(1+16\E \kappa)}{p_{\mathrm{min}}\Delta E_0}\log\frac{\Delta E_0}{\kappa \epsilon}+\frac{2}{\epsilon} \right) \left(\log \frac{3}{\delta}+\log \frac{p_0\Delta E_0}{12\E \kappa\epsilon}\right),
    \end{equation*}
    which yields Eq.~\eqref{eq:cor_simple_cost_bound_final} with
    \begin{align*}
        c_1&=\frac{3c_D(1+16\E\kappa)}{2(1-\delta_0^{\star})}\approx45.420,\\
        c_2&=c_D\approx3.618,\\
        c_3&=\frac{12\E \kappa}{1-\delta_0^{\star}}\approx 5.192.
    \end{align*}
\end{proof}

\begin{remark}
    The true optimal $\delta_1$ and $\delta_2$ differ from the near optimal splits $\delta_1^{\star}$ and $\delta_2^{\star}$ by at most $O(\delta^2)$, which has a negligible effect on the total cost bound.
    Furthermore, Eq.~\eqref{eq:pf4_intermediate_cost} can be further reduced by selecting the optimal $\epsilon'$, rather than the algebraically convenient $\epsilon'^{\star}$.
\end{remark}

\section{Filter Robustness Analysis}
In the QSVT implementation, an imperfect qubitization operator may be viewed as a block-encoding of a perturbed Hamiltonian $\hat H'$ satisfying $\|\hat H'-\hat H\|\le \epsilon_{\hat H}$.
Such perturbations can arise, for example, from truncating or omitting small Hamiltonian terms in the construction of the block encoding.
In the GQSP implementation, the Hamiltonian evolution operator is typically approximated by a product formula or another Hamiltonian-simulation routine, which can likewise be represented by an effective Hamiltonian $\hat H'$.
For instance, for a first-order Trotter formula with $N_{\mathrm{Trot}}$ steps over evolution time $\pi$, the effective Hamiltonian error scales as $\epsilon_{\hat H}=O(N_{\mathrm{Trot}}^{-1})$ under standard assumptions~\cite{Yi2022, PhysRevX.11.011020}.

In this note, we analyze the robustness of filter functions under practical imperfections that arise during their implementation.
In realistic settings, the building blocks used to implement polynomial or trigonometric functions, such as qubitization or time evolution operators, are often approximated due to limited circuit depth.
These approximations lead to perturbations in the encoded Hamiltonian, which in turn affect the fidelity of the filtered state.
We quantify this effect using both Lipschitz continuity arguments and perturbation theory, including a bound based on the Davis–Kahan theorem, to establish error tolerance criteria for the reliable implementation of filters.

Furthermore, in Krylov based filters, imperfect circuit realizations or sampling noise induce perturbations to the subspace matrices, thereby distorting the coefficients of the filter and altering the filtered state.
To assess this sensitivity, we analyze how such perturbations propagate through the generalized eigenvalue problem and quantify their influence on the fidelity of the resulting Krylov-filtered state.
This analysis extends the discussion to subspace-projected filters and establishes explicit error bounds via the Davis–Kahan theorem and matrix-condition arguments.

\subsection{Davis-Kahan Theorem}
First, we provide a modified version of Davis-Kahan theorem for eigenspace perturbation, which will be used in filter function robustness and Krylov filter perturbation.

\begin{lemma}[Modified Davis-Kahan theorem{~\cite[Theorem~V.3.6]{Stewart90}}]\label{thm:davis_kahan}
    Let $\bm{A}, \tilde{\bm{A}}\in\mathbb{C}^{d\times d}$ be Hermitian matrices with ordered eigenvalues
    \begin{equation*}
        \lambda_0\le\cdots\le\lambda_{d-1}, \qquad\tilde{\lambda}_0\le\cdots\le\tilde{\lambda}_{d-1},
    \end{equation*}
    and corresponding eigenvectors $\{\bm{\lambda_i}\}$ and $\{\tilde{\bm{\lambda_i}}\}$.
    Fix integers $0\le l \le r\le d-1$.
    Let
    \begin{equation*}
        \bm{\Pi} = \sum_{i=l}^{r}\bm{\lambda_i}\bm{\lambda_i}^{\dagger},\qquad \tilde{\bm{\Pi}} = \sum_{i=l}^{r}\tilde{\bm{\lambda_i}}\tilde{\bm{\lambda_i}}^{\dagger}
    \end{equation*}
    be the projectors onto the eigenspaces of $\bm{A}$ and $\tilde{\bm{A}}$ corresponding to the selected eigenvalue indices.
    Define the eigengap
    \begin{equation*}
        \Delta = \min(\lambda_l-\lambda_{l-1}, \lambda_{r+1}-\lambda_r),
    \end{equation*}
    where $\lambda_{-1}=-\infty$ and $\lambda_{d}=\infty$ are defined.
    If
    \begin{equation}\label{eq:DKT_gap_assumption}
        \Delta > \|\tilde{\bm{A}}-\bm{A}\|,
    \end{equation}
    the following inequality holds:
    \begin{equation}\label{eq:DKT_result}
        \|\tilde{\bm{\Pi}}-\bm{\Pi}\|\le \frac{1}{\|\tilde{\bm{A}}-\bm{A}\|^{-1}\Delta-1}.
    \end{equation}
\end{lemma}
\begin{proof}
    Set $n:=r-l+1$ and choose principal vectors to form semi-unitary matrices $\bm{U}=(\bm{u_1},\cdots, \bm{u_n}), \tilde{\bm{U}}=(\tilde{\bm{u}}_{\bm{1}},\cdots, \tilde{\bm{u}}_{\bm{n}})\in\mathbb{C}^{d\times n}$, i.e. 
    \begin{equation*}
        \bm{U}^{\dagger}\bm{U}=\tilde{\bm{U}}^{\dagger}\tilde{\bm{U}}=I_n, \quad \bm{U}^{\dagger}\tilde{\bm{U}}=\operatorname{diag}(\cos\theta_1,\cdots,\cos\theta_n), 
    \end{equation*}
    so that $\bm{\Pi}=\bm{U}\bm{U}^\dagger, ~\tilde{\bm{\Pi}}=\tilde{\bm{U}}\tilde{\bm{U}}^{\dagger}$.
    
    In the orthogonal basis of $\{\bm{u_i} ,\bm{w_i}\}_{i=1}^{n}$ with $\tilde{\bm{u}}_{\bm{i}}=\cos\theta_i\bm{u_i}+\sin\theta_i{\bm{w}}_{\bm{i}}$, the two projectors take the block form of
    \begin{equation*}
        \bm{\Pi}=\bigoplus_{i=1}^{n}\begin{bmatrix}
            1 & 0 \\
            0 & 0
        \end{bmatrix}_{\{\bm{u_i} ,\bm{w_i}\}}\oplus~\bm{0}_{d-2n},
        \quad
        \tilde{\bm{\Pi}}=\bigoplus_{i=1}^{n}\begin{bmatrix}
            \cos^2 \theta_i & \cos \theta_i \sin \theta_i \\
            \cos \theta_i \sin \theta_i & \sin^2 \theta_i 
        \end{bmatrix}_{\{\bm{u_i} ,\bm{w_i}\}}\oplus~\bm{0}_{d-2n}.
    \end{equation*}
    Conjugating each $2\times2$ block by the rotation $\bm{R_i}:=\bigl[\!\begin{smallmatrix}
        \cos \alpha_i & -\sin \alpha_i\\ \sin\alpha_i &\phantom{-}\cos\alpha_i
    \end{smallmatrix}\!\bigr]$ with $\alpha_i=\tfrac{\theta_i}{2}-\tfrac{\pi}{4}$ diagonalizes the difference:
    \begin{equation*}
        \bm{R_i}^{\dagger}
        \begin{bmatrix}
            \sin^2\theta_i & \cos\theta_i\sin\theta_i\\
            -\cos\theta_i\sin\theta_i & -\sin^2\theta_i
        \end{bmatrix}
        \bm{R_i}=\begin{bmatrix}
            \sin \theta_i & 0 \\
            0 & -\sin \theta_i
        \end{bmatrix}.
    \end{equation*}
    Putting the $\bm{R_i}$'s together we obtain a global unitary $\bm{W}=\bigoplus_{i=1}^n \bm{R_i}\oplus \bm{I}_{d-2n}$ such that
    {\setstretch{1.0}
    \begin{equation*}
        \bm{W}^\dagger (\bm{\Pi}-\tilde{\bm{\Pi}})\bm{W}=\bigoplus_{i=1}^{n}\begin{bmatrix}
            \sin \theta_i & 0\\
            0 & -\sin \theta_i
        \end{bmatrix}_{\{\bm{u_i} ,\bm{w_i}\}}\oplus~\bm{0}_{d-2n},
     \end{equation*}
     }
    hence
    \begin{equation*}
        \|\tilde{\bm{\Pi}}-\bm{\Pi}\|=\max_{1\le i\le n}|\sin\theta_i|.
    \end{equation*}
    
    The sin $\bm{\Theta}$ version of the Davis-Kahan theorem (see, eg., \cite[Theorem V.3.6]{Stewart90} or \cite[Theorem 1]{davis_kahan_stat}) gives
    \begin{equation*}
        \max_{1\le i\le n}|\sin\theta_i| \le \frac{\|\tilde{\bm{A}}-\bm{A}\|}{\delta},
    \end{equation*}
    where
    \begin{equation*}
        \delta := \inf_{\substack{l\le i\le r\\-1\le j<l\text{~or~}r<j\le d}}|\tilde{\lambda}_j-\lambda_i|>0
    \end{equation*}
    is defined.
    Furthermore, it can be shown that $\delta^{-1}\le(\Delta-\|\tilde{\bm{A}}-\bm{A}\|)^{-1}$ based on the Weyl's inequality ($|\tilde{\lambda}_j-\lambda_j|\le\|\tilde{\bm{A}}-\bm{A}\|$) and the assumption in Eq.\eqref{eq:DKT_gap_assumption}, which exactly leads to Eq.\eqref{eq:DKT_result}.
\end{proof}

This lemma differs from the original Davis-Kahan theorem in two respects.
First, while the classical theorem bounds $\max \sin \theta_i$; we explicitly show that this equals the spectral norm of the projector difference, $\|\tilde {\bm{\Pi}}-\bm{\Pi}\|$.
Second, the original gap parameter $\delta=\inf_{i\in[l,r],\, j\notin[l,r]}|\tilde{\lambda}_j-\lambda_i|$ is replaced by the unperturbed eigengap $\Delta=\min(\lambda_l-\lambda_{l-1},\,\lambda_{r+1}-\lambda_r)$.
Using Weyl's inequality, the usual bound $\|\tilde{\Pi}-\Pi\|\le \|A-\tilde{A}\|/\delta$ is converted into the explicit form $\|\tilde{\Pi}-\Pi\|\le(\|A-\tilde{A}\|^{-1}\Delta-1)^{-1}$, which no longer depends on the perturbed eigenvalues.

\subsection{Filter Robustness against Basis Perturbation}\label{suppl:filter_robustness}
Here, we analyze the robustness of filter functions under practical imperfections that arise during their implementation.
We quantify this effect using both Lipschitz continuity arguments and perturbation theory, including a bound based on the Davis–Kahan theorem, to establish error tolerance criteria for the reliable implementation of filters.

\subsubsection{Analysis Based on Lipschitz Continuity}
An analysis can be carried out from the Lipschitz continuity of a target filter function $f$ and thus the sensitivity is chosen as the Lipschitz constant:
\begin{equation*}
    \|f(\hat{H}')-f(\hat{H})\|\le L_{f} \epsilon_{\hat{H}},\quad\kappa_f\le L_f,
\end{equation*}
where $L_{f}$ is the Lipschitz constant of $f$, which amplifies the Hamiltonian perturbation. 
A bounds for the Lipschitz constant can be determined by the mean value theorem:
\begin{equation*}
    L_f = \sup_{x\in[-1,1]}\left|f'(x)\right|\le 
        \pi\sum_{k=0}^{n-1}|kc_k|\le \pi n\|\bm{c}\|_1.
\end{equation*}

For the Chebyshev filter function analyzed in Supplementary Note~\ref{suppl:minimax_filter_design}, we have the Lipschitz constant of 
\begin{equation*}
    L_{\mathrm{cheby}}\le\frac{2\pi n^2\E^{-\pi n \Delta}}{\cos^2\pi\Delta/2}=O(\Delta^{-2}\epsilon_f(\log\epsilon_f^{-1})^2).
\end{equation*}
In order to sufficiently suppress the error to achieve $L_{\mathrm{cheby}}\epsilon_{\hat{H}}\le\epsilon_f$ with Trotterization, the total circuit depth should be 
\begin{equation*}
    D_{\mathrm{cheby}}=\Omega\left(n\epsilon_{\hat{H}}^{-1}\right)=\Omega\left(nL_{\mathrm{cheby}}\epsilon_f^{-1}\right)=\Omega\left(\Delta^{-3}(\log\epsilon_f^{-1})^3\right).    
\end{equation*}

For the Gaussian filter function, the Lipschitz constant is slightly more significant, $L_{\mathrm{gauss}}=O(\Delta^{-2}\log\epsilon_f^{-1})$, which requires circuit depth of
\begin{equation*}
    D_{\mathrm{gauss}}=\Omega\left(\Delta^{-3}\epsilon_f^{-1}(\log\epsilon_f^{-1})^2\right).
\end{equation*}

\subsubsection{Analysis Based on Eigenspace Perturbation}
However, the above analysis with the Lipschitz continuity often overestimates the circuit depth for filter functions.
Instead, we propose another analysis based on eigenspace perturbation after approximating the filter function to a projector.
Especially for a filter function that sharply discriminates between the accepted and rejected eigenvalues, it can be approximated as a projector onto a subspace that series of eigenstates spans:
\begin{equation*}
    f(\hat{H})=\sum_{i=0}^{d-1}f(E_i)\ket{E_i}\bra{E_i}\approx \sum_{l\le i\le r}\ket{E_i}\bra{E_i}=:\Pi_f(\hat{H}),
\end{equation*}
for some integers $0\le l\le r\le d-1$.
Let us denote the approximation error as
\begin{equation*}
    \epsilon_{\Pi_f}=\|f(\hat{H})-\Pi_f(\hat{H})\|=\max_{i\in[0,d-1]}|f(E_i)-\delta_{i\in[l,r]}|,
\end{equation*}
where $\delta_{i\in[l,r]}$ equals to 1 if $l\le i\le r$ and 0 otherwise.
Since we focus on the ground state, we assign $l=r=0$, resulting in $\Pi_f(\hat{H})=\ket{E_0}\bra{E_0}$ and
\begin{equation*}
    \epsilon_{\Pi_f}\le\max\{|1-f(E_0)|, \epsilon_f\}
\end{equation*}

Now, consider the effective Hamiltonian,
\begin{equation*}
    \hat{H}'=\sum_{i=0}^{d-1}E'_i\ket{E'_i}\bra{E'_i},    
\end{equation*}
where $(E'_i, \ket{E'_{i}})$ is the $i$-th eigenpair of $\hat{H}'$.
Assume that $\epsilon_{\hat{H}}<\Delta$ is small enough for $\Pi_f(\hat{H}')$ to project to its ground state, $\Pi_f(\hat{H}')=\ket{E_0}\bra{E_0}$.
The eigenspace perturbation of $\Pi_f(\hat{H}')$ can be presented by simply substituting $\bm{A}=\hat{H}$ and $\tilde{\bm{A}}=\tilde{H}'$ in Theorem \ref{thm:davis_kahan}.
Then, we have the bound of
\begin{equation*}
    \|\Pi_{f}(\hat{H}')-\Pi_f(\hat{H})\|\le\frac{1}{\epsilon_{\hat{H}}^{-1}\Delta-1}.
\end{equation*}
Then, putting all together provides
\begin{equation*}
\begin{split}
    \|f(\hat{H}')-f(\hat{H})\|\le& \|f(\hat{H}')-\Pi_f(\hat{H}')\|+ \|\Pi_f(\hat{H}')-\Pi_f(\hat{H})\|+\|f(\hat{H})-\Pi_f(\hat{H})\|\\
    \lesssim&2\max\{|1-f(E_0)|, \epsilon_f\}+\frac{1}{\epsilon_{\hat{H}}^{-1}\Delta-1}.
\end{split}
\end{equation*}
Once the ideal filter clearly separates between the ground state and first excited state, the approximation error $\epsilon_{\Pi_f}$ becomes small enough.
In order to suppress the perturbation error below the fluctuation level $\Delta^{-1}\epsilon_{\mathrm{HE}}<\epsilon_f$, the circuit depth should be 
\begin{equation*}
    D=O(n\epsilon_{\mathrm{HE}}^{-1})=O(\Delta^{-2}\epsilon_f^{-1}\log\epsilon_f^{-1}),    
\end{equation*}
which shows much smaller than the result of Lipschitz analysis.

\subsection{Krylov State Perturbation}\label{suppl:krylov_state_perturbation}
Here, we analyze how the perturbations in the Krylov matrices affect the Krylov state fidelity.
\begin{theorem}[Subspace Ground State Stability]
    For $(N+1)$ fixed normalized basis states $\{\ket{\phi_k}\}_{k=0}^{N}$, define the following two normalized states:
    \begin{equation*}
        \ket{\phi_f}:=\frac{1}{\sqrt{\bm{c}^{\dagger}\bm{S}\bm{c}}}\sum_{k=0}^{N} c_k \ket{\phi_k},\quad \ket{\phi_{\tilde{f}}}:=\frac{1}{\sqrt{\tilde{\bm{c}}^{\dagger}\bm{S}\tilde{\bm{c}}}}\sum_{k=0}^{N} \tilde{c}_k \ket{\phi_k},
    \end{equation*}
    where complex vectors $\bm{c}$ and $\tilde{\bm{c}}$ are respectively obtained from the eigenvalue problems with the lowest eigenvalues:
    \begin{align}
        \bm{Hc}=E^{(N)}\bm{S}\bm{c},\label{eq:subspace_perturbation_original_gevp}\\
        \tilde{\bm{H}}\tilde{\bm{c}}=\tilde{E}^{(N)}\tilde{\bm{S}}\tilde{\bm{c}},
    \end{align}
    Here, we define the matrices as $H_{kl}=\braket{\phi_k|\hat{H}|\phi_l}$ and $S_{kl}=\braket{\phi_k|\phi_l}$.
    Suppose that the matrices $\tilde{\bm{H}}$ and $\tilde{\bm{S}}$ satisfies
    \begin{equation}\label{eq:subspace_perturbation_assumption}
        \|\Delta(\bm{S}^{-1/2}\bm{H}\bm{S}^{-1/2})\|:=\|\tilde{\bm{S}}^{-1/2}\tilde{\bm{H}}\tilde{\bm{S}}^{-1/2}-\bm{S}^{-1/2}\bm{H}\bm{S}^{-1/2}\|\le g:=E_1^{(N)}-E_0^{(N)},
    \end{equation}
    where $E_0^{(N)}$ and $E_1^{(N)}$ are the two lowest and second-lowest eigenvalues of Eq.~\eqref{eq:subspace_perturbation_original_gevp}.
    Then, the fidelity between two states is bounded by
    \begin{equation}\label{eq:subspace_perturbation_result}
        |\braket{\phi_f|\phi_{\tilde{f}}}|\ge 1-\frac{\|\Delta(\bm{S}^{-1/2}\bm{H}\bm{S}^{-1/2})\|}{E_1^{(N)}-E_0^{(N)}-\|\Delta(\bm{S}^{-1/2}\bm{H}\bm{S}^{-1/2})\|},
    \end{equation}
    where the condition number of $\bm{S}$ is defined as $\kappa(\bm{S})=\|\bm{S}^{-1}\|\|\bm{S}\|$.
    Moreover, we have the following bound:
    \begin{equation*}
        \|\Delta(\bm{S}^{-1/2}\bm{H}\bm{S}^{-1/2})\|\le \frac{\kappa(\bm{S})}{\|\bm{S}\|}\frac{1}{1-\|\bm{S}^{-1}\|\|\bm{\Delta_S}\|}\|\bm{\Delta_H}\|+\frac{\kappa(\bm{S})^2}{\|\bm{S}\|^2}\frac{\|\bm{H}\|}{(1-\|\bm{S}^{-1}\|\|\bm{\Delta_S}\|)^2}\|\bm{\Delta_S}\|,
    \end{equation*}
    with $\|\bm{S}^{-1}\|\|\bm{\Delta_S}\|<1$.
\end{theorem}
\begin{remark}
    For a tighter bound of $\|\bm{S}^{-1}\|\|\bm{\Delta_S}\|<1/2$, we have
    \begin{equation*}
        \|\Delta(\bm{S}^{-1/2}\bm{H}\bm{S}^{-1/2})\|\le2\left( \|\bm{S}^{-1}\|\|\bm{\Delta_H}\|+\|\bm{S}\|\|\bm{S}^{-1}\|^2 \|\bm{\Delta_S}\| \right),
    \end{equation*}
    since $\|\bm{H}\|\le \|\bm{S}^{-1/2}\bm{H}\bm{S}^{-1/2}\|\|\bm{S}^{1/2}\|^2\le \|\hat{H}\|\|\bm{S}\|$.
    Furthermore, Given that small perturbation of $\|\Delta(\bm{S}^{-1/2}\bm{H}\bm{S}^{-1/2})\|<g/2$, we can replace the the bound as
    \begin{equation*}
        |\braket{\phi_f|\phi_{\tilde{f}}}|\ge 1-2\frac{\|\Delta(\bm{S}^{-1/2}\bm{H}\bm{S}^{-1/2})\|}{E_1^{(N)}-E_0^{(N)}}\ge 1-4\frac{\|\bm{S}^{-1}\|\|\bm{\Delta_H}\|+\|\bm{S}\|\|\bm{S}^{-1}\|^2 \|\bm{\Delta_S}\|}{E_1^{(N)}-E_0^{(N)}}.
    \end{equation*}
\end{remark}
\begin{remark}
    In addition to the assumption in the previous remark, we apply this theorem to Krylov subspace, where $E_0^{(N)}$ and $E_1^{(N)}$ is close to the spectral gap of $\hat{H}$.
    Then the bound leads to
    \begin{equation*}
        |\braket{\phi_f|\phi_{\tilde{f}}}|\gtrsim 1-4\frac{\|\bm{S}^{-1}\|\|\bm{\Delta_H}\|+\|\bm{S}\|\|\bm{S}^{-1}\|^2 \|\bm{\Delta_S}\|}{\Delta}.
    \end{equation*}
\end{remark}
\begin{proof}
    Consider the standard Hermitian reductions
    \begin{equation*}
        \bm{A}=\bm{S}^{-1/2}\bm{H}\bm{S}^{-1/2},\qquad 
        \tilde {\bm{A}}=\tilde {\bm{S}}^{-1/2}\tilde {\bm{H}}\,\tilde {\bm{S}}^{-1/2}.
    \end{equation*}
    Let $\bm{u}:=\bm{S}^{1/2}\bm{c}$ and $\tilde{\bm{u}}:=\tilde {\bm{S}}^{1/2}\tilde {\bm{c}}$. 
    Without loss of generality, we can assume $\bm{c}^\dagger \bm{S} \bm{c}=\tilde {\bm{c}}^\dagger \bm{S} \tilde {\bm{c}}=1$, so we have $\|\bm{u}\|_2=\|\tilde {\bm{u}}\|_2=1$ and
    \begin{equation*}
        \bm{A}\bm{u}=E_0^{(N)}\bm{u},\qquad \tilde {\bm{A}}\,\tilde {\bm{u}}=\tilde E_0^{(N)}\tilde {\bm{u}}.
    \end{equation*}
    Let $\theta$ be the (acute) principal angle between the 1-D eigenspaces $\operatorname{span}\{u\}$ and $\operatorname{span}\{\tilde u\}$; then 
    \begin{equation*}
        \bigl|\braket{\phi_f|\phi_{\tilde f}}\bigr|=\cos\theta,
    \end{equation*}
    since 
    \begin{equation*}
      \braket{\phi_f|\phi_{\tilde f}}=
      \bm{c}^\dagger \bm{S} \tilde {\bm{c}} = \bm{u}^\dagger \tilde {\bm{u}}
    \end{equation*}
    under the unit-norm choices above.

    By Theorem~\ref{thm:davis_kahan} applied to $\bm{A}$ and $\tilde {\bm{A}}$ with eigen-gap $g=E_1^{(N)}-E_0^{(N)}$, one has
    \begin{equation*}
        \sin\theta \;\le\; \frac{\ \|\tilde {\bm{A}}-\bm{A}\|\ }{\,g-\|\tilde {\bm{A}}-\bm{A}\|\,}
        \quad\text{whenever }\|\tilde {\bm{A}}-\bm{A}\|<g.
    \end{equation*}
    Observe that
    \begin{equation*}
        \tilde {\bm{A}}-\bm{A}
        = \tilde{\bm{S}}^{-1/2}\bm{\Delta_H}\tilde{\bm{S}}^{-1/2}+(\tilde{\bm{S}}^{-1/2}-\bm{S}^{-1/2})\bm{H}\tilde{\bm{S}}^{-1/2}+\bm{S}^{-1/2}\bm{H}(\tilde{\bm{S}}^{-1/2}-\bm{S}^{-1/2}).
    \end{equation*}
    So,
    \begin{equation*}
        \|\tilde{\bm{A}}-\bm{A}\|\le \|\tilde{\bm{S}}^{-1/2}\|^2\|\bm{\Delta_H}\|+\|\tilde{\bm{S}}^{-1/2}-\bm{S}^{-1/2}\|\|\bm{H}\|(\|\tilde{\bm{S}}^{-1/2}\|+\|\bm{S}^{-1/2}\|).
    \end{equation*}

    Weyl's inequality gives
    \begin{equation*}
        \sigma_{\mathrm{min}}(\tilde{\bm{S}})\ge \sigma_{\mathrm{min}}(\bm{S})-\|\bm{\Delta_S}\|\ge (1-\eta)\sigma_{\mathrm{min}}(\bm{S}),
    \end{equation*}
    where $\eta:=\|\bm{S}^{-1}\|\|\bm{\Delta_S}\|<1$ is defined.
    Hence,
    \begin{equation*}
        \|\tilde{\bm{S}}^{-1/2}\|\le(1-\eta)^{-1/2}\sqrt{\|\bm{S}^{-1}\|}.
    \end{equation*}

    For the inverse-square-root difference, we can use operator Lipschitz bound for $f(x)=x^{-1/2}$ on the positive definite cone (i.e. $|f(x)-f(y)|\le \sup_{t\in[x,y]}|f'(t)|x-y|$ by the mean-value theorem):
    \begin{equation*}
        \|\tilde{\bm{S}}^{-1/2}-\bm{S}^{-1/2}\|\le\frac{1}{2}(1-\eta)^{-3/2}\|\bm{S}^{-1}\|^{3/2}\|\bm{\Delta_S}\|.
    \end{equation*}

    Put these into the split bound, we yield Eq.~\eqref{eq:subspace_perturbation_result}
\end{proof}

\newpage
\section{Further Numerical Results}\label{suppl:further_numerical_results}
In this section, we present numerical results beyond those shown in the main manuscript.
\begin{figure*}[ht]
    \centering
    \subfigure[~$N_{\mathrm{site}}=6, \epsilon=10^{-1}\Delta E_0$, Trig.]{
        \includegraphics[width=0.295\textwidth]{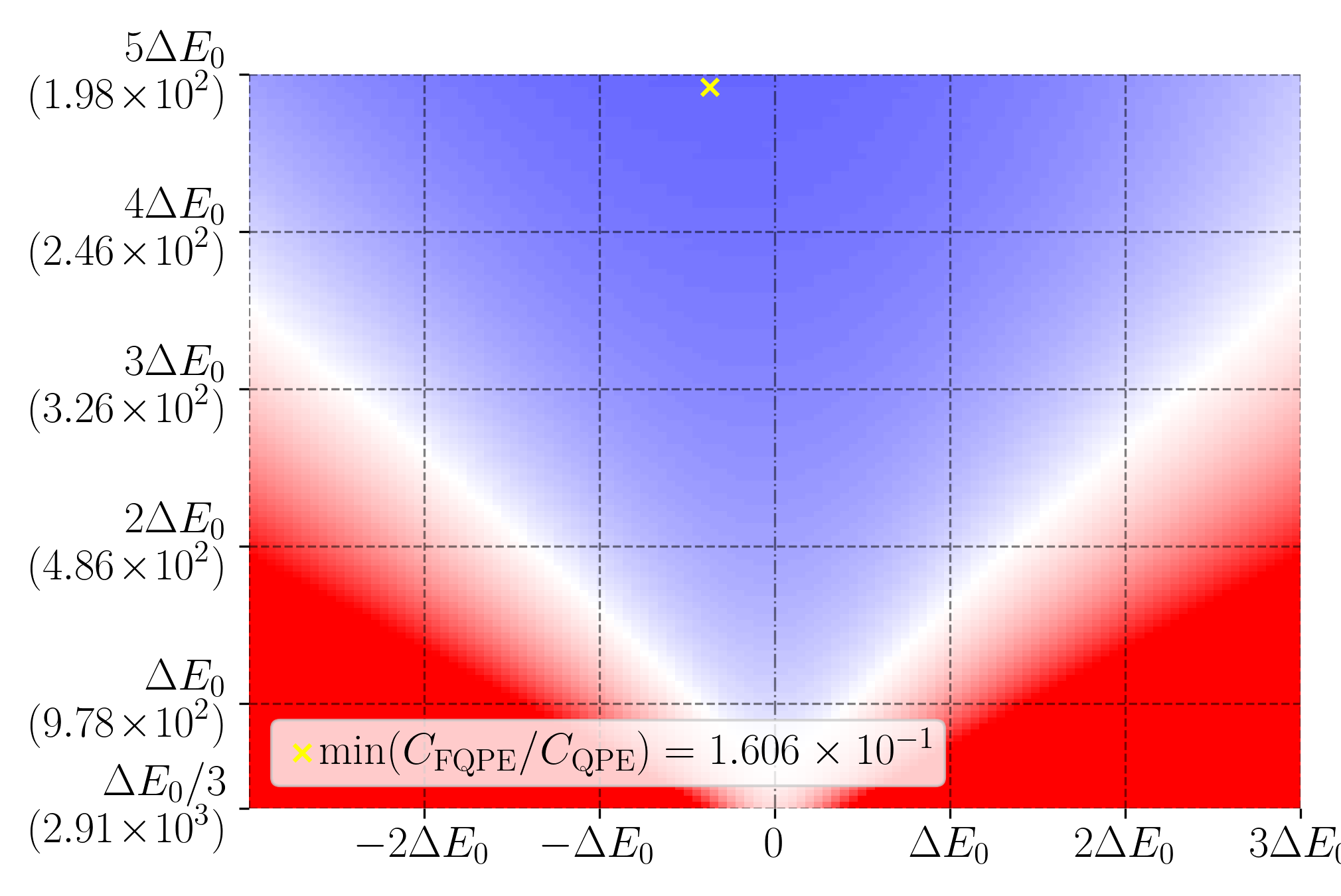}
    }
    \subfigure[~$N_{\mathrm{site}}=6, \epsilon=10^{-3}\Delta E_0$, Trig.]{
        \includegraphics[width=0.295\textwidth]{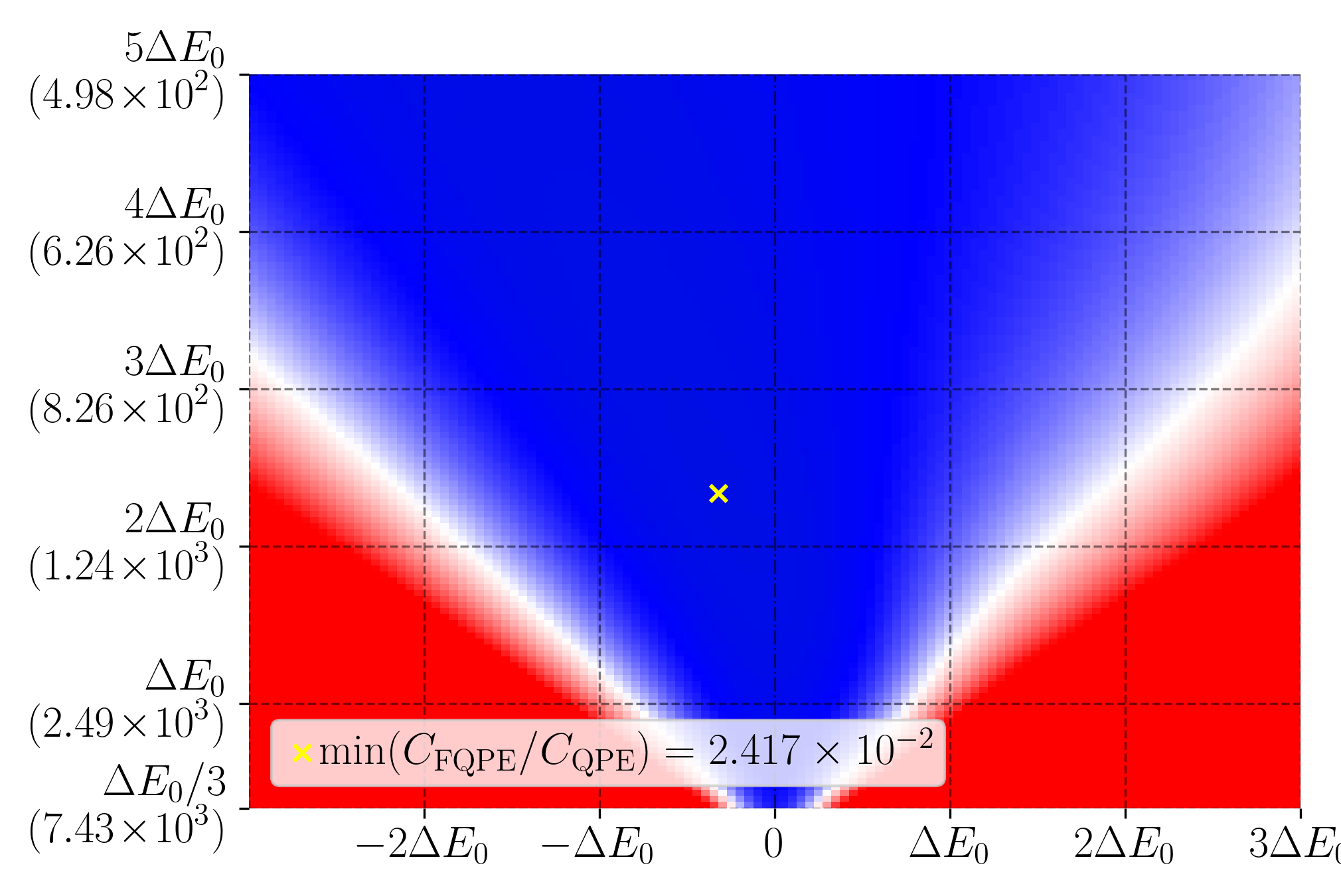}
    }
    \subfigure[~$N_{\mathrm{site}}=6, \epsilon=10^{-5}\Delta E_0$, Trig.]{
        \includegraphics[width=0.35\textwidth]{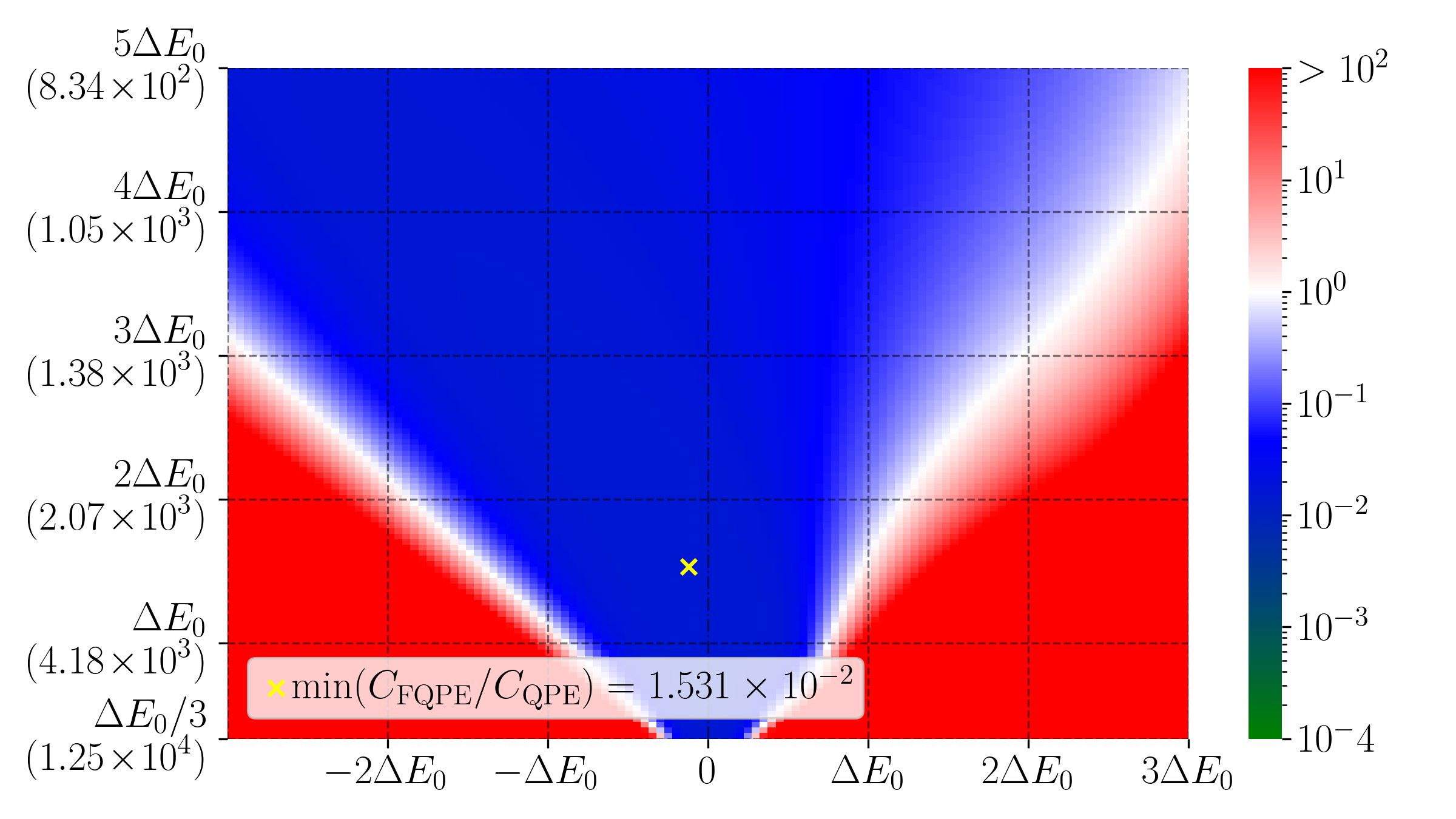}
    }
    \\
    \subfigure[~$N_{\mathrm{site}}=2\times3, \epsilon=10^{-1}\Delta E_0$, Trig.]{
        \includegraphics[width=0.295\textwidth]{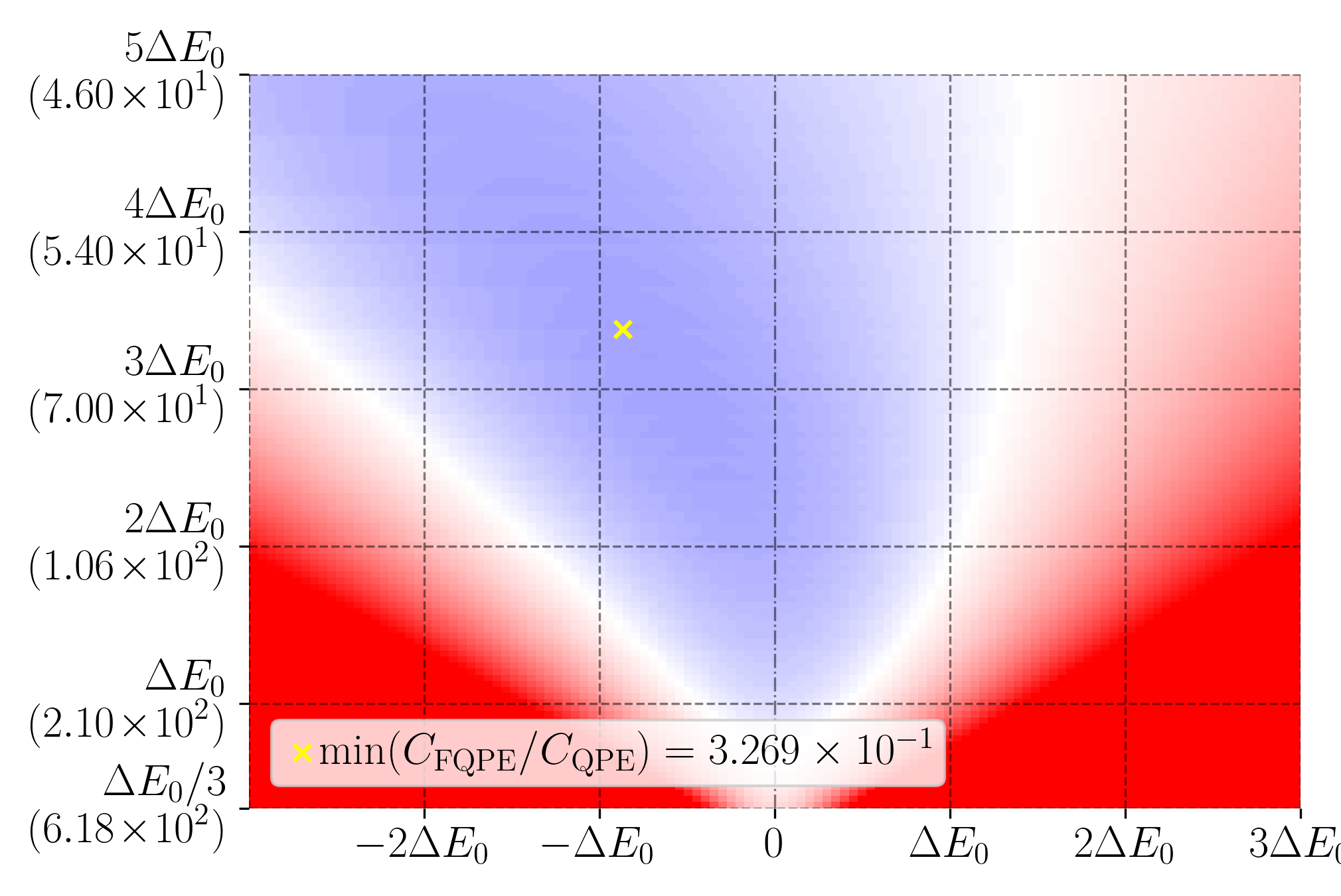}
    }
    \subfigure[~$N_{\mathrm{site}}=2\times3, \epsilon=10^{-3}\Delta E_0$, Trig.]{
        \includegraphics[width=0.295\textwidth]{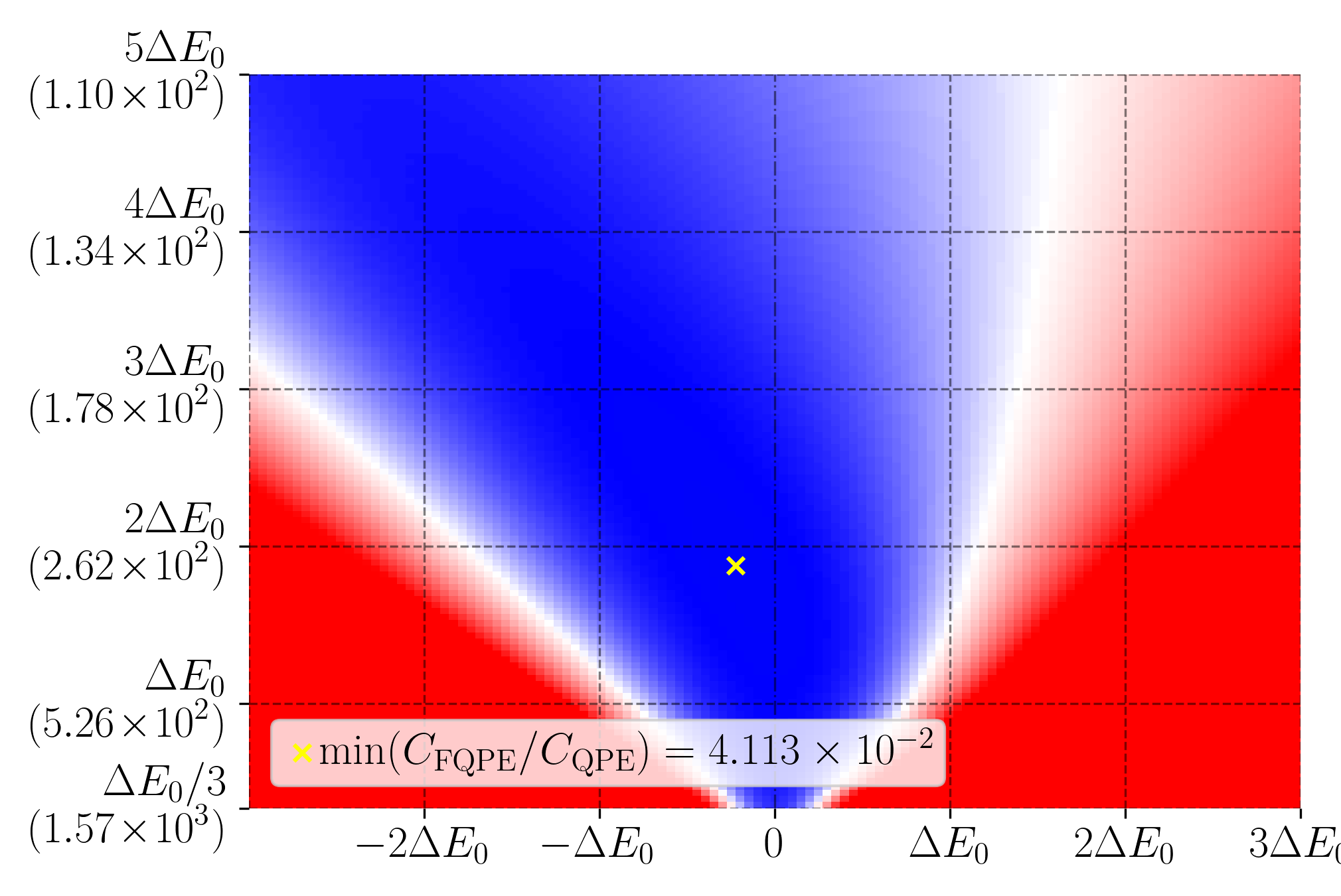}
    }
    \subfigure[~$N_{\mathrm{site}}=2\times3, \epsilon=10^{-5}\Delta E_0$, Trig.]{
        \includegraphics[width=0.35\textwidth]{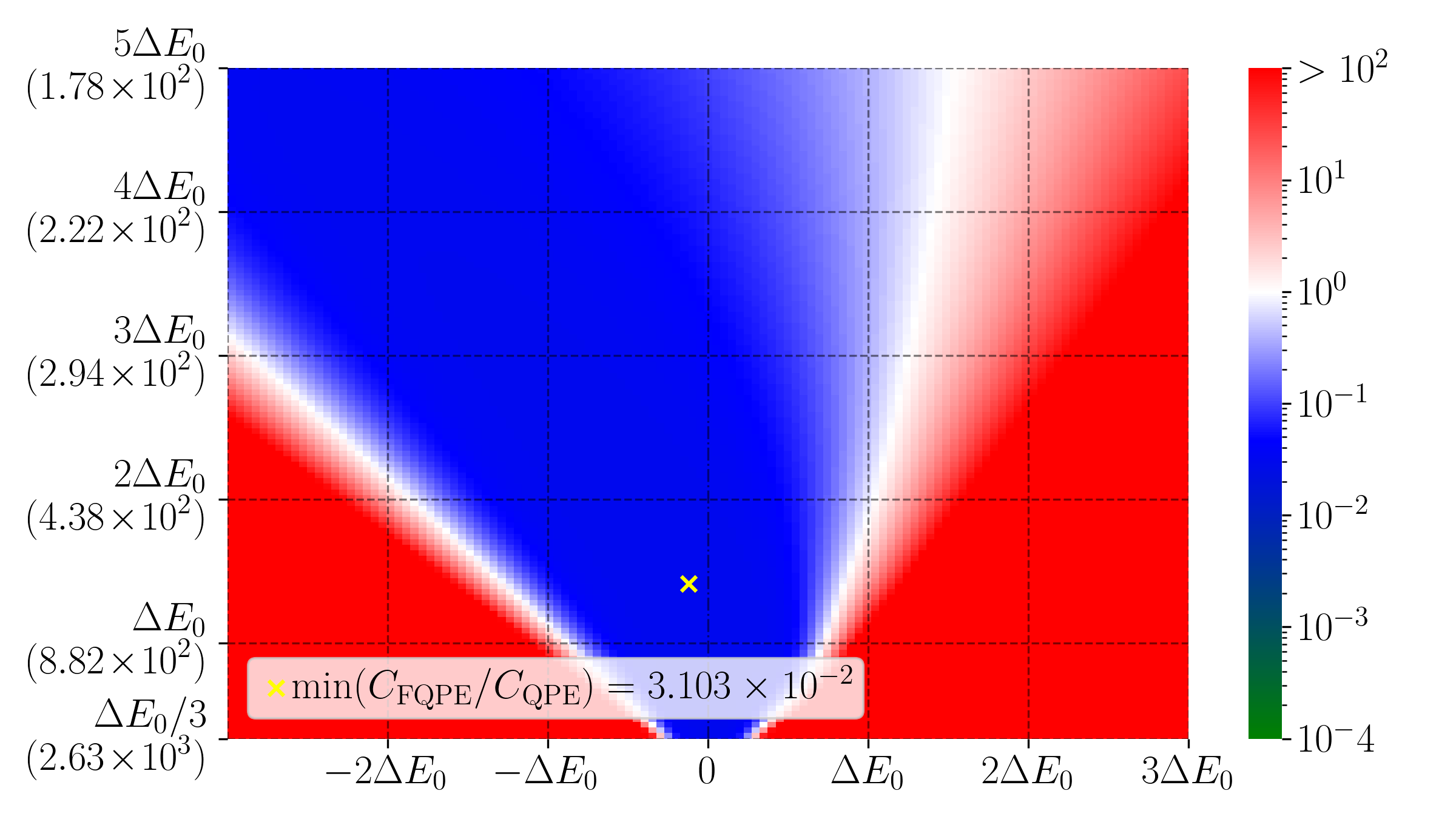}
    }
    \\
    \subfigure[~$N_{\mathrm{site}}=6, \epsilon=10^{-1}\Delta E_0$, Poly.]{
        \includegraphics[width=0.295\textwidth]{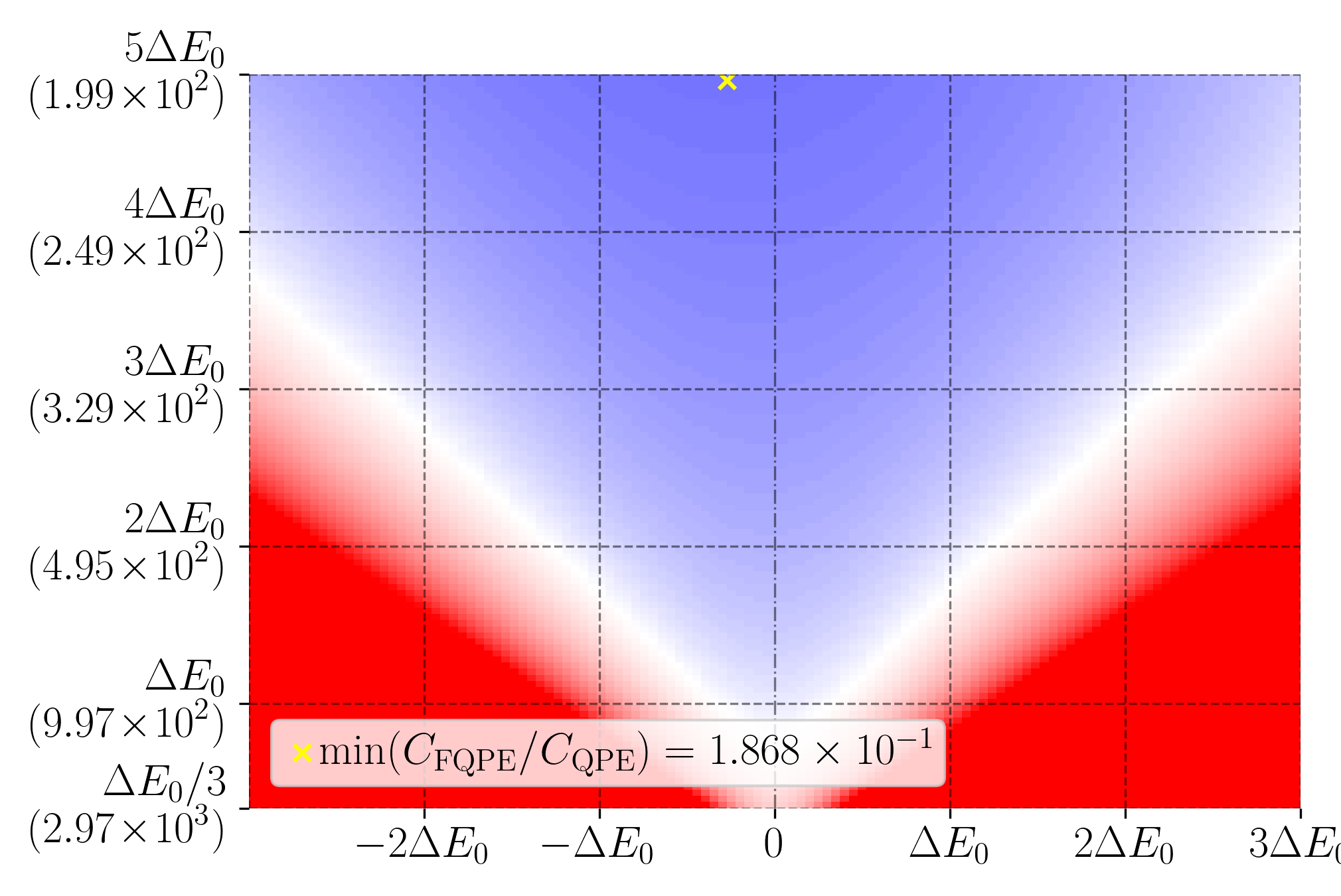}
    }
    \subfigure[~$N_{\mathrm{site}}=6, \epsilon=10^{-3}\Delta E_0$, Poly.]{
        \includegraphics[width=0.295\textwidth]{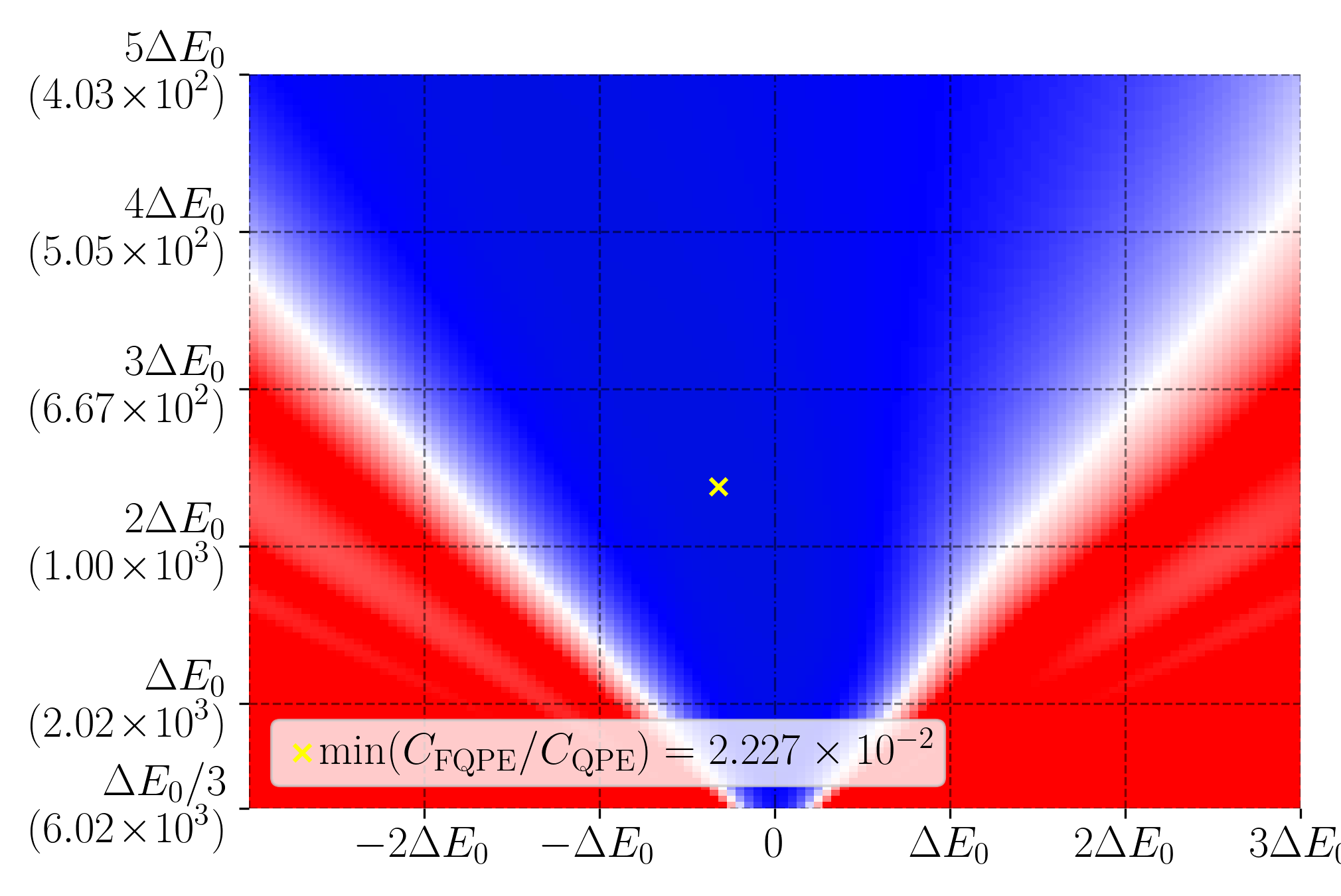}
    }
    \subfigure[~$N_{\mathrm{site}}=6, \epsilon=10^{-5}\Delta E_0$, Poly.]{
        \includegraphics[width=0.35\textwidth]{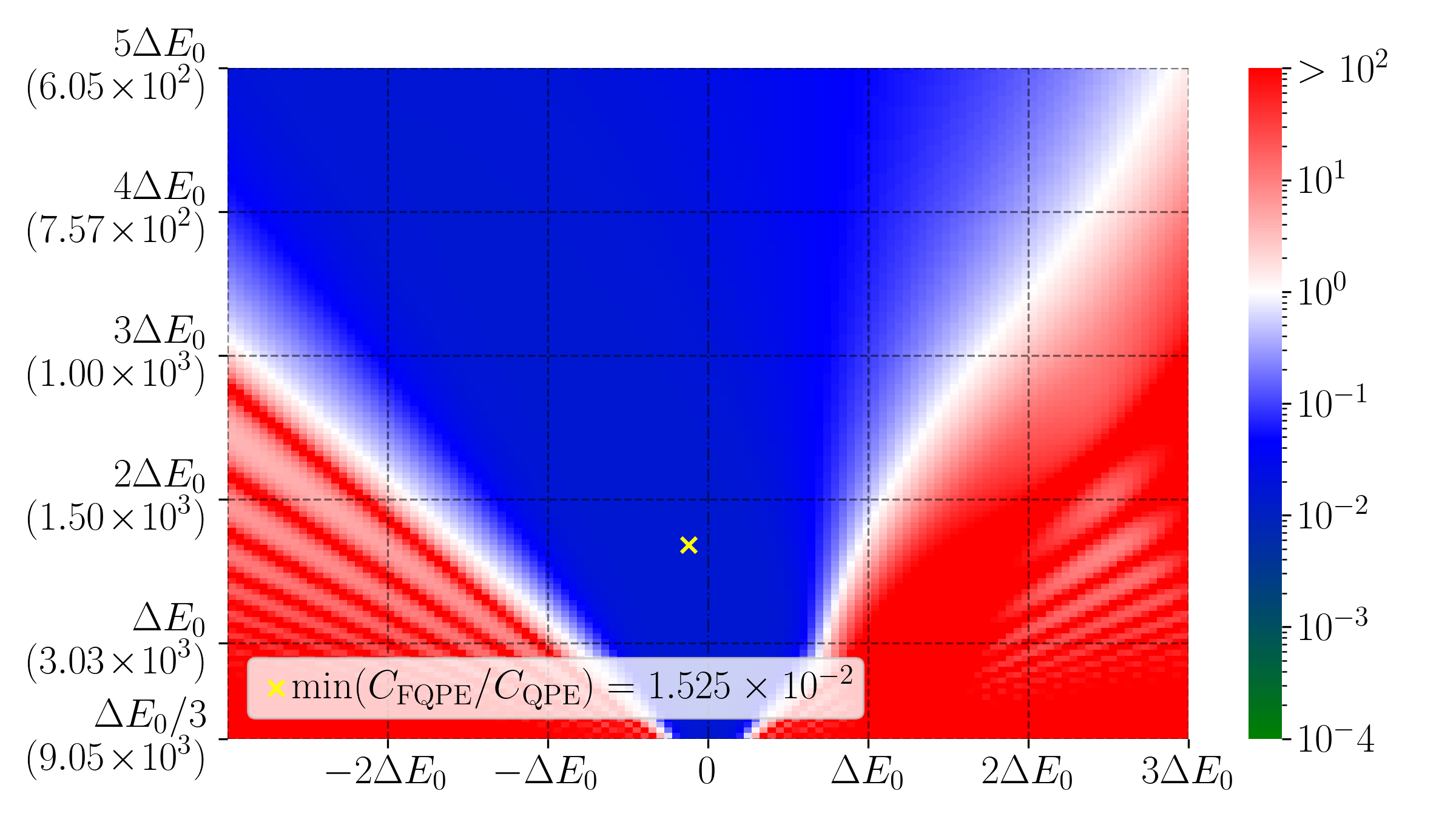}
    }
    \\
    \subfigure[~$N_{\mathrm{site}}=2\times3, \epsilon=10^{-1}\Delta E_0$, Poly.]{
        \includegraphics[width=0.295\textwidth]{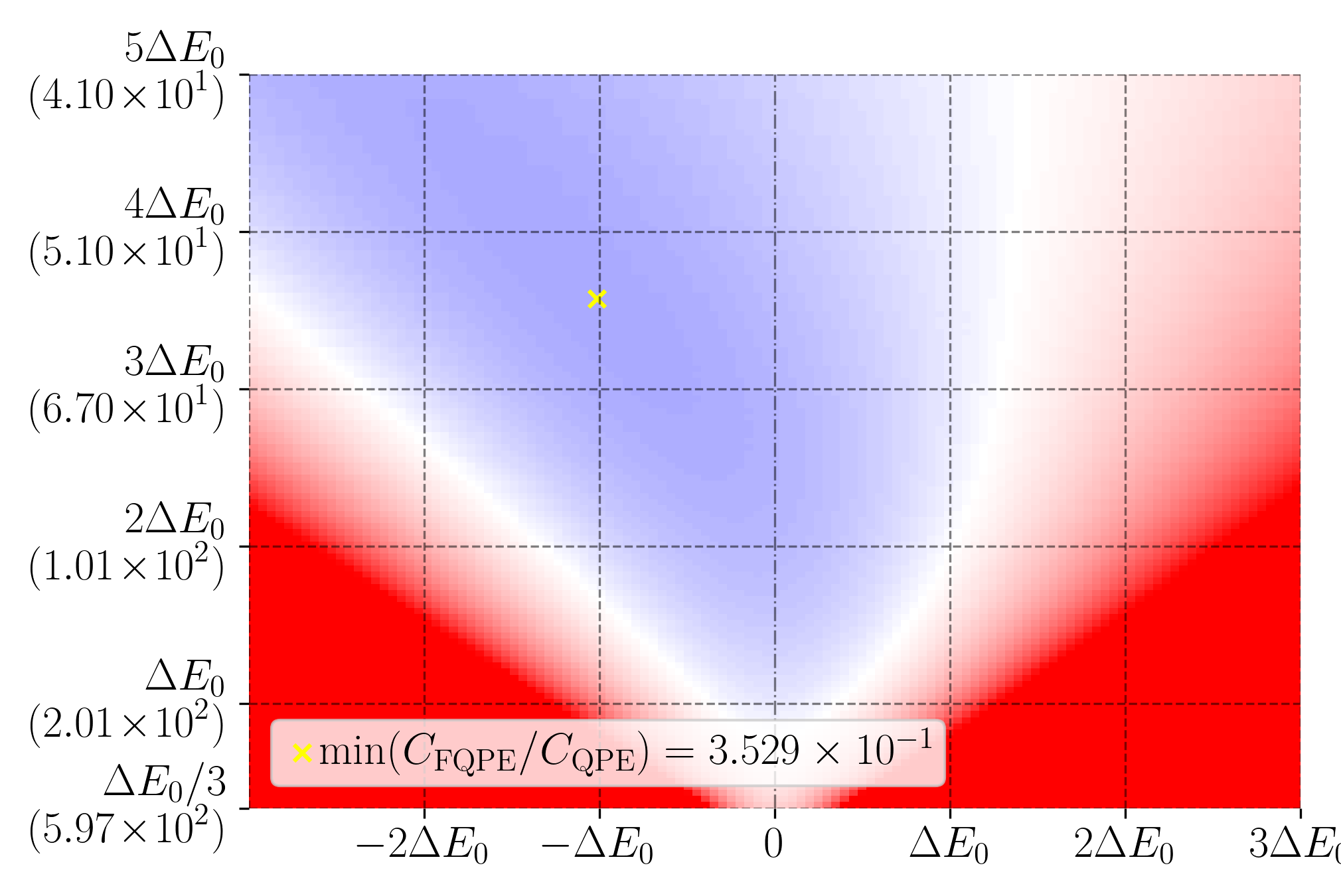}
    }
    \subfigure[~$N_{\mathrm{site}}=2\times3, \epsilon=10^{-3}\Delta E_0$, Poly.]{
        \includegraphics[width=0.295\textwidth]{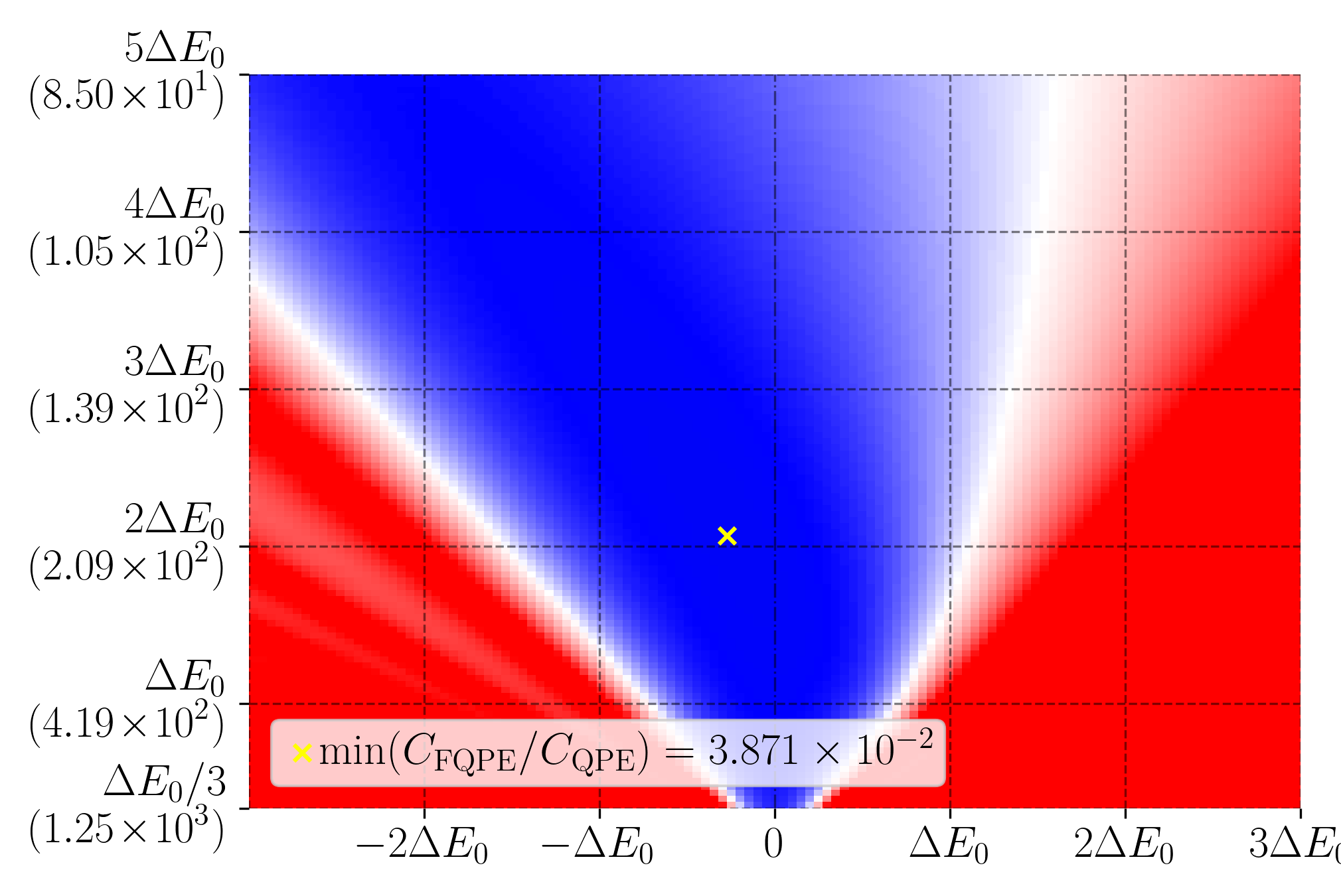}
    }
    \subfigure[~$N_{\mathrm{site}}=2\times3, \epsilon=10^{-5}\Delta E_0$, Poly.]{
        \includegraphics[width=0.35\textwidth]{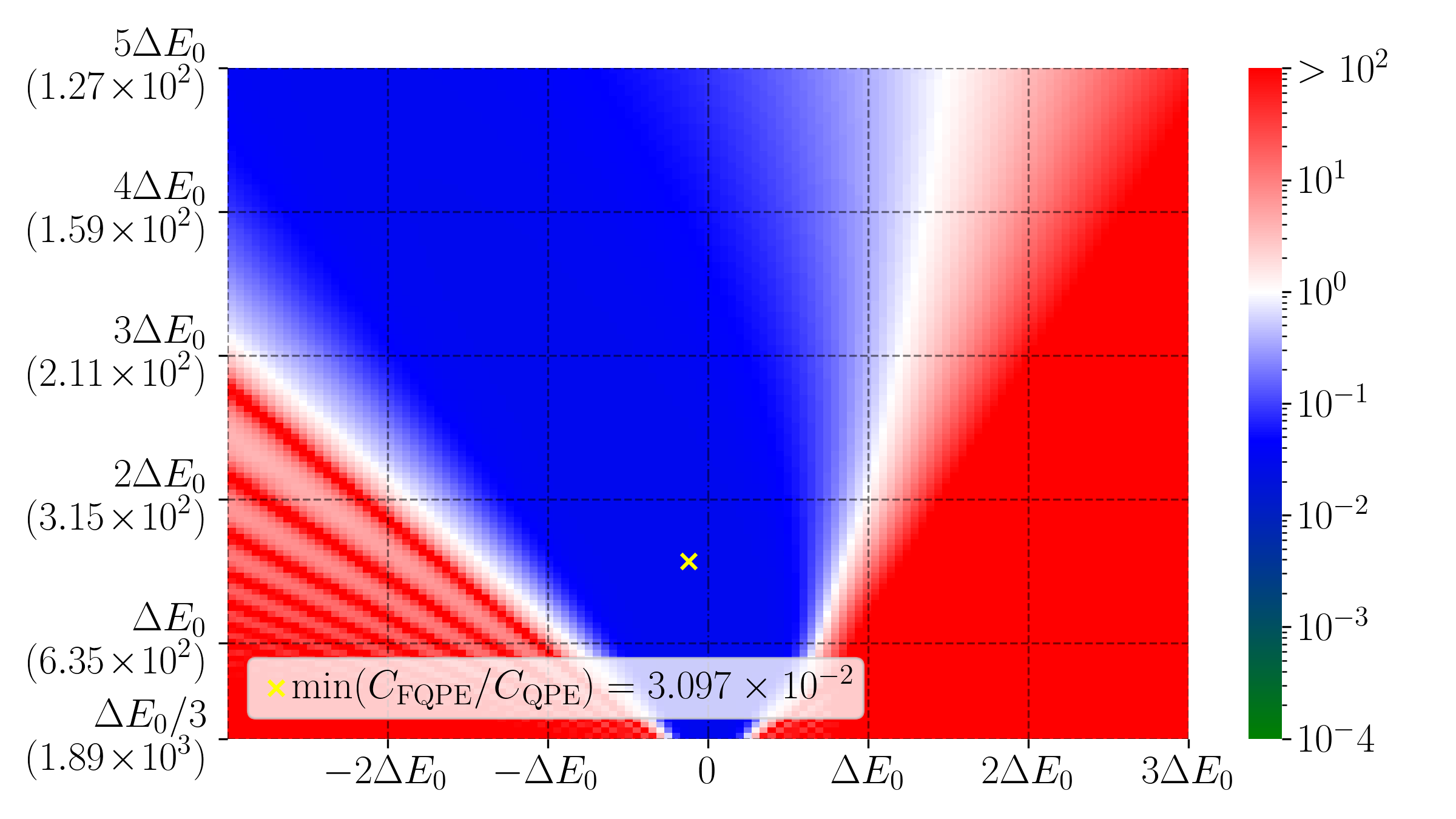}
    }
    \caption{
        Relative Gaussian FQPE cost for the Hubbard model with $N_{\mathrm{site}}=6$ (1D) and $2\times3$ (2D), using trigonometric (Trig.) and polynomial (Poly.) Gaussian filters.
        Further description is analogous to the caption of Fig.~\ref{fig:fqpe_cost}.
    }
    \label{fig:fqpe_cost_extend}
\end{figure*}

\newpage

\begin{figure*}
    \centering
    \subfigure[~$N_{\mathrm{site}}=6$, Trigonometric]{
        \includegraphics[width=0.47\textwidth]{
            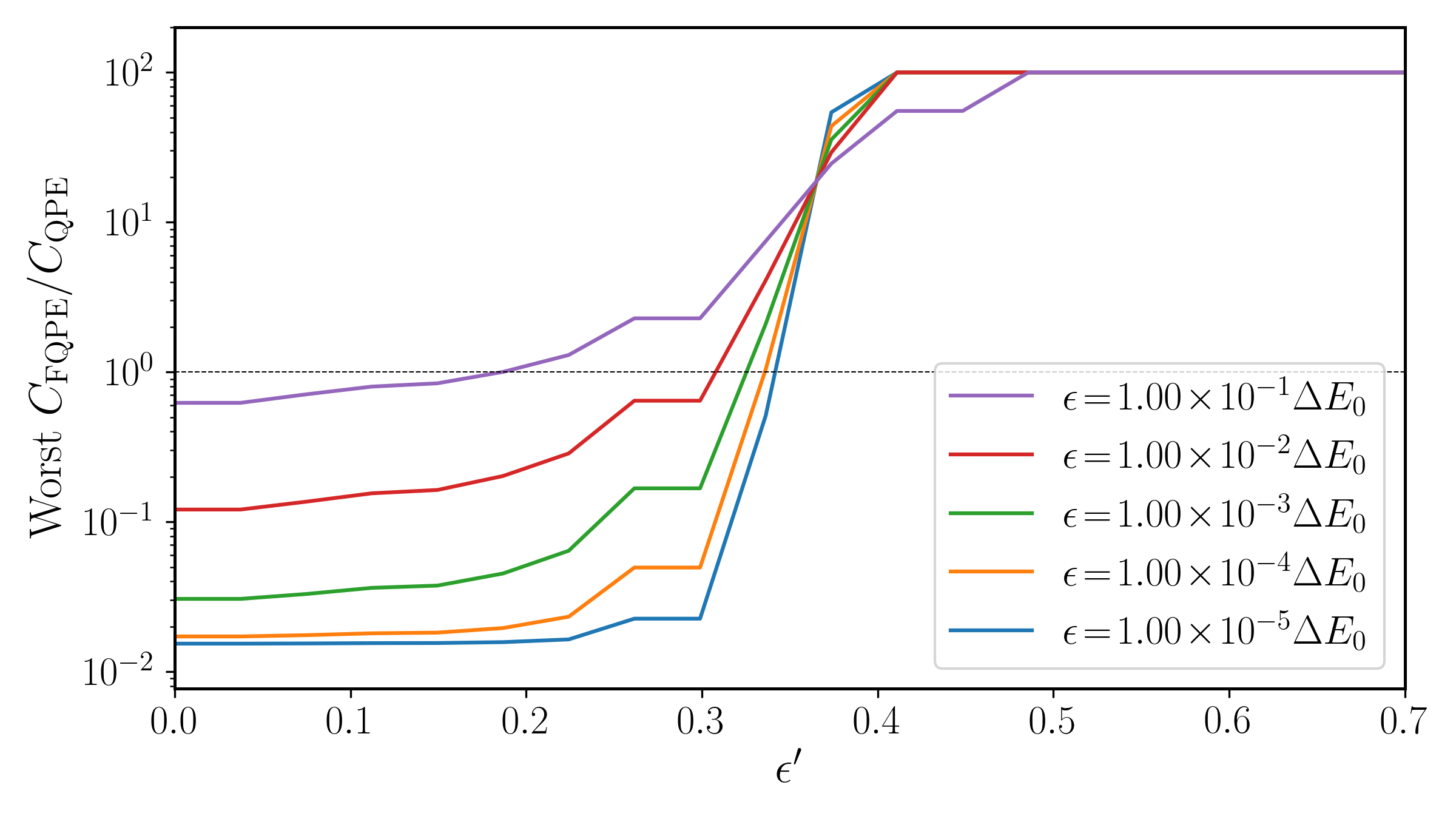
        }
    }
    \subfigure[~$N_{\mathrm{site}}=6$, Polynomial]{
        \includegraphics[width=0.47\textwidth]{
            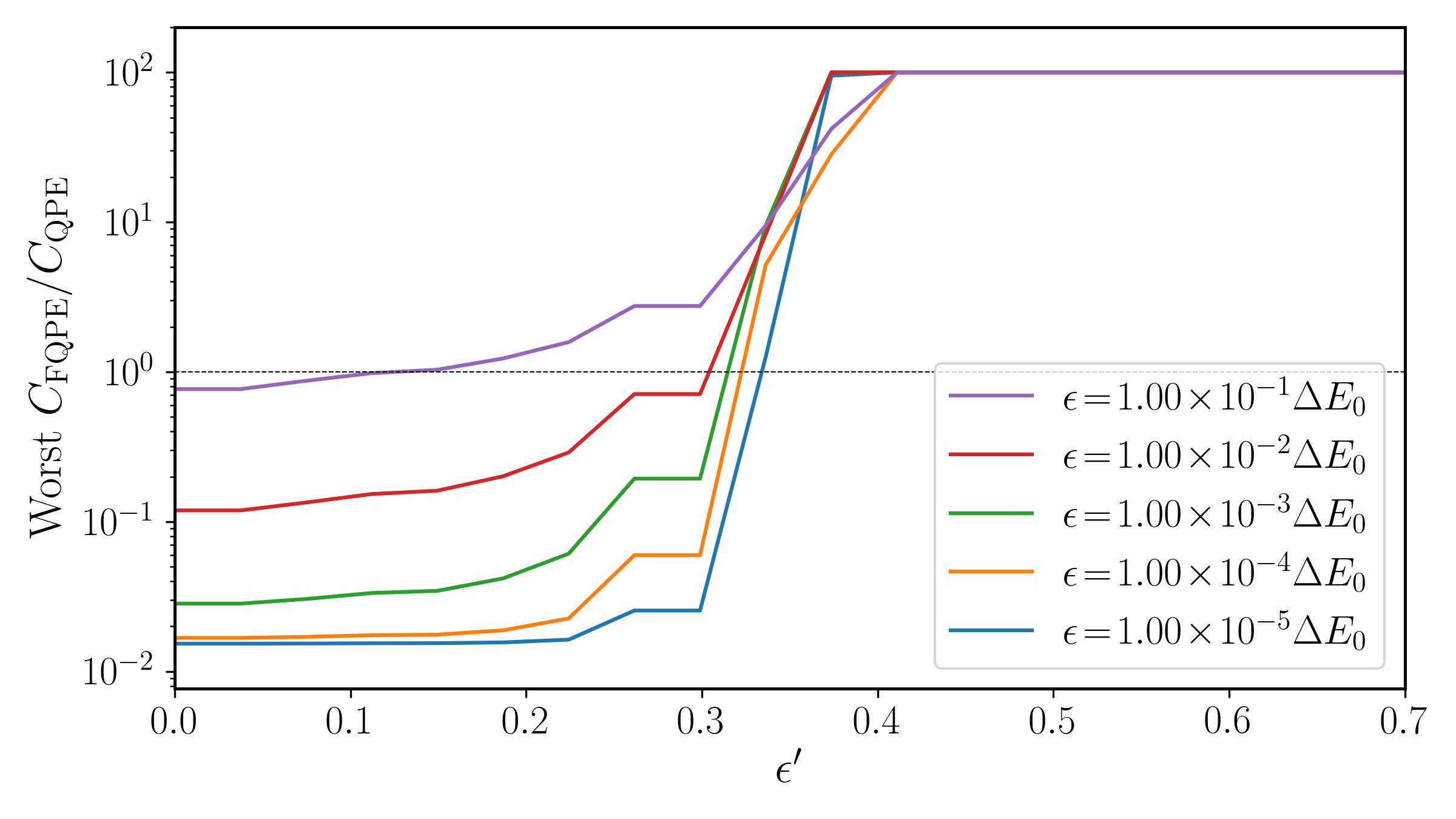
        }
    }\\
    \subfigure[~$N_{\mathrm{site}}=2\times3$, Trigonometric]{
        \includegraphics[width=0.47\textwidth]{
            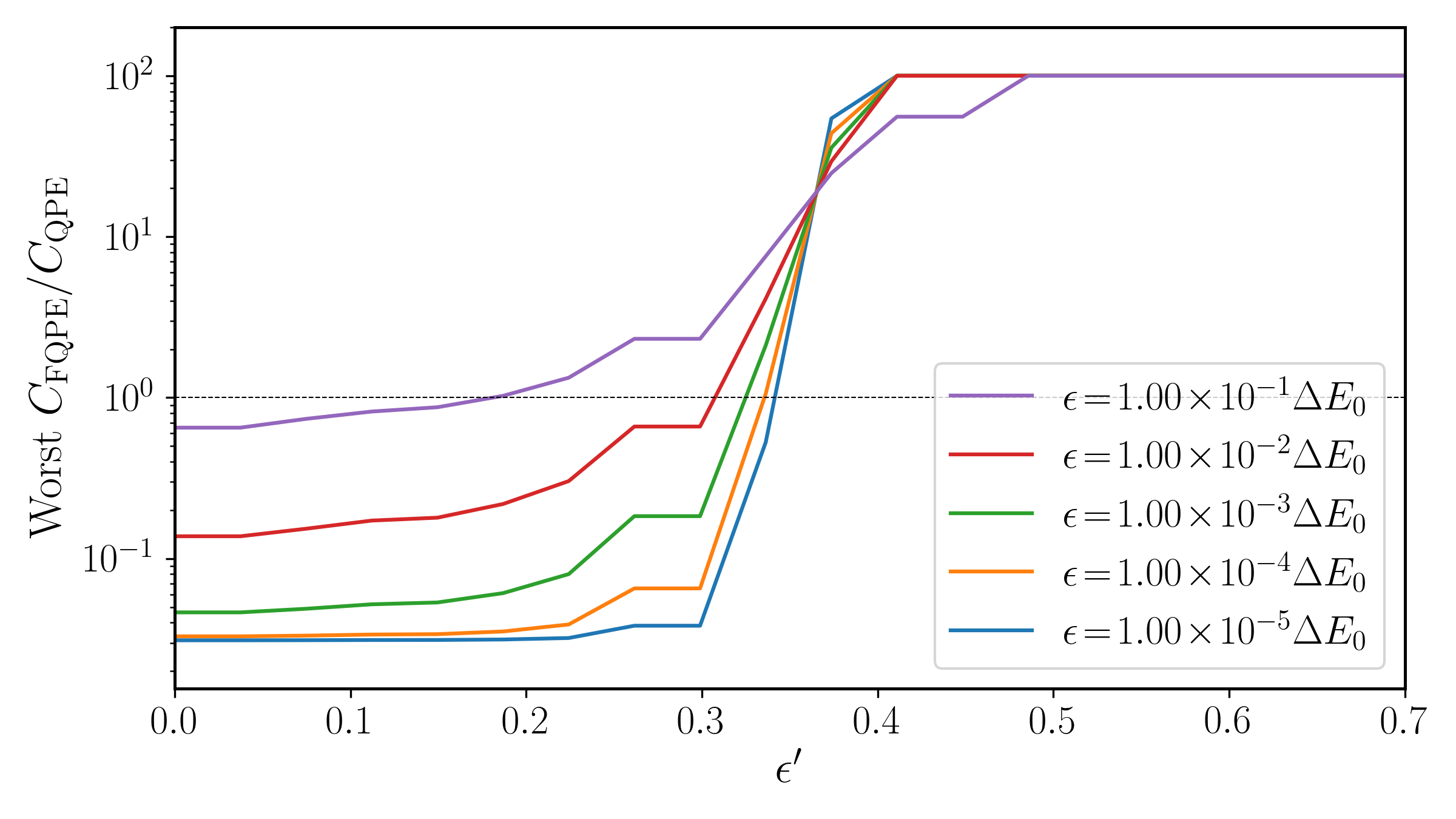
        }
    }
    \subfigure[~$N_{\mathrm{site}}=2\times3$, Polynomial]{
        \includegraphics[width=0.47\textwidth]{
            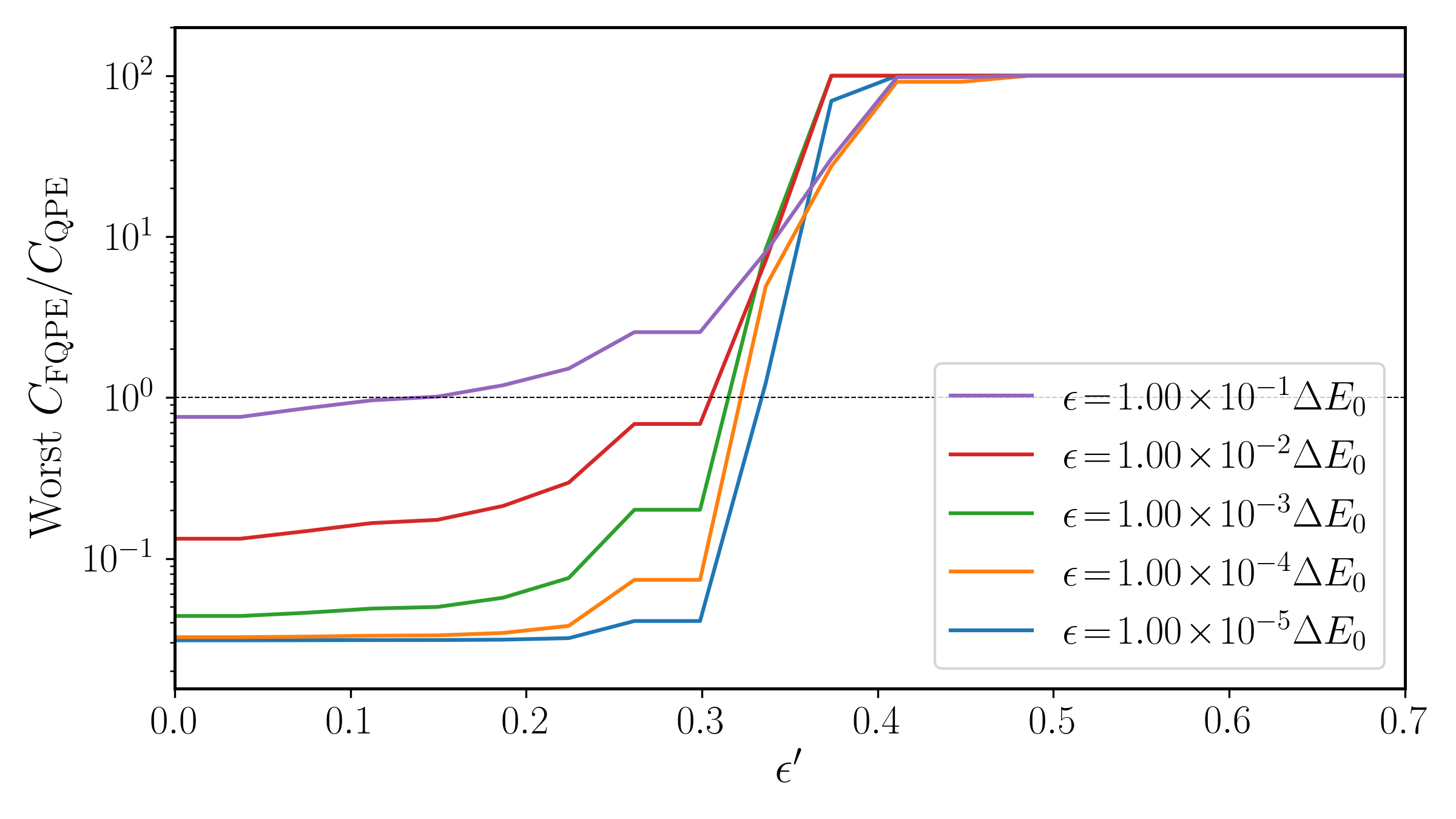
        }
    }
    \caption{
        Worst-case Gaussian FQPE cost as a function of prior estimate accuracy for the Hubbard model with $N_{\mathrm{site}}=6$ (1D) and $2\times3$ (2D), using trigonometric and polynomial Gaussian filters.
        Further description is analogous to the caption of Fig.~\ref{fig:fqpe_cost_worst}.
    }
\end{figure*}

\begin{figure*}
    \centering
    \subfigure[~$N_{\mathrm{site}}=6$, Trigonometric Krylov Filter]{
        \includegraphics[width=0.9\textwidth]{
            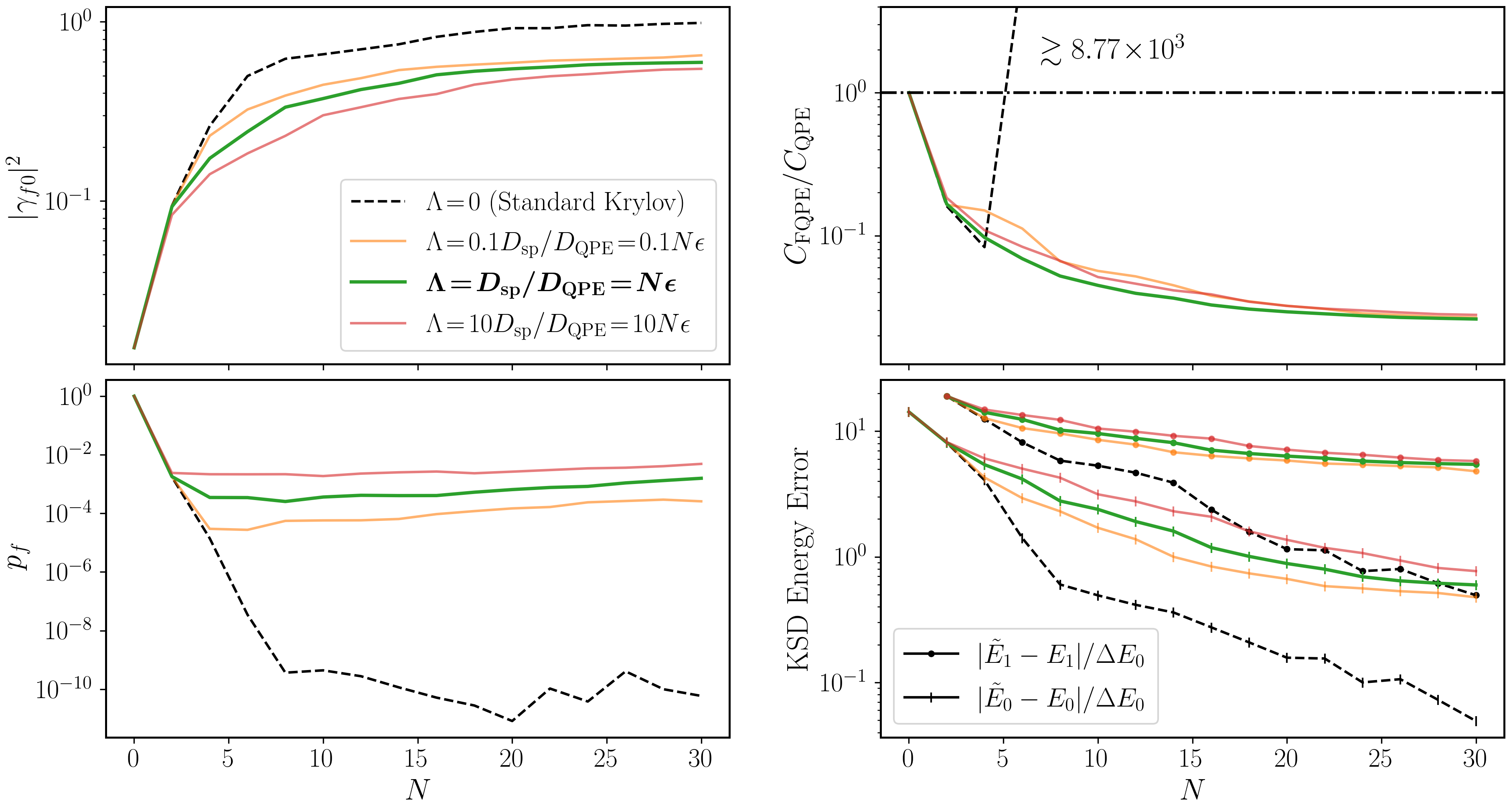
        }
    }\\
    \subfigure[~$N_{\mathrm{site}}=6$, Polynomial Krylov Filter]{
        \includegraphics[width=0.9\textwidth]{
            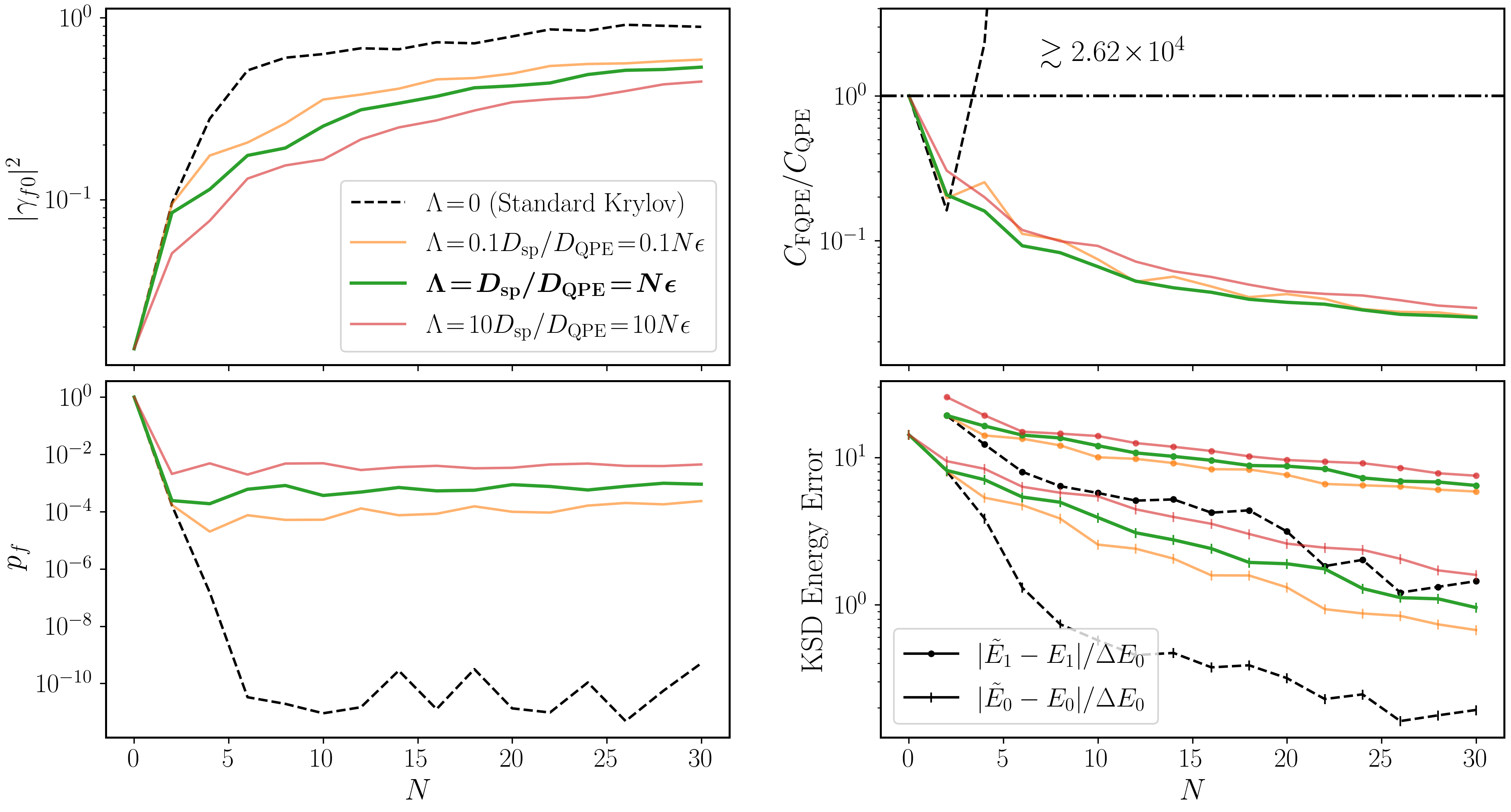
        }
    }
    \caption{
        Properties of trigonometric and polynomial (modified) Krylov filters applied to the Hubbard model with $N_{\mathrm{site}}=6$.
        Further description is analogous to the caption of Fig.~\ref{fig:krylov_filter}.
    }
\end{figure*}

\begin{figure*}
    \centering
    \subfigure[~$N_{\mathrm{site}}=2\times3$, Trigonometric Krylov Filter]{
        \includegraphics[width=0.9\textwidth]{
            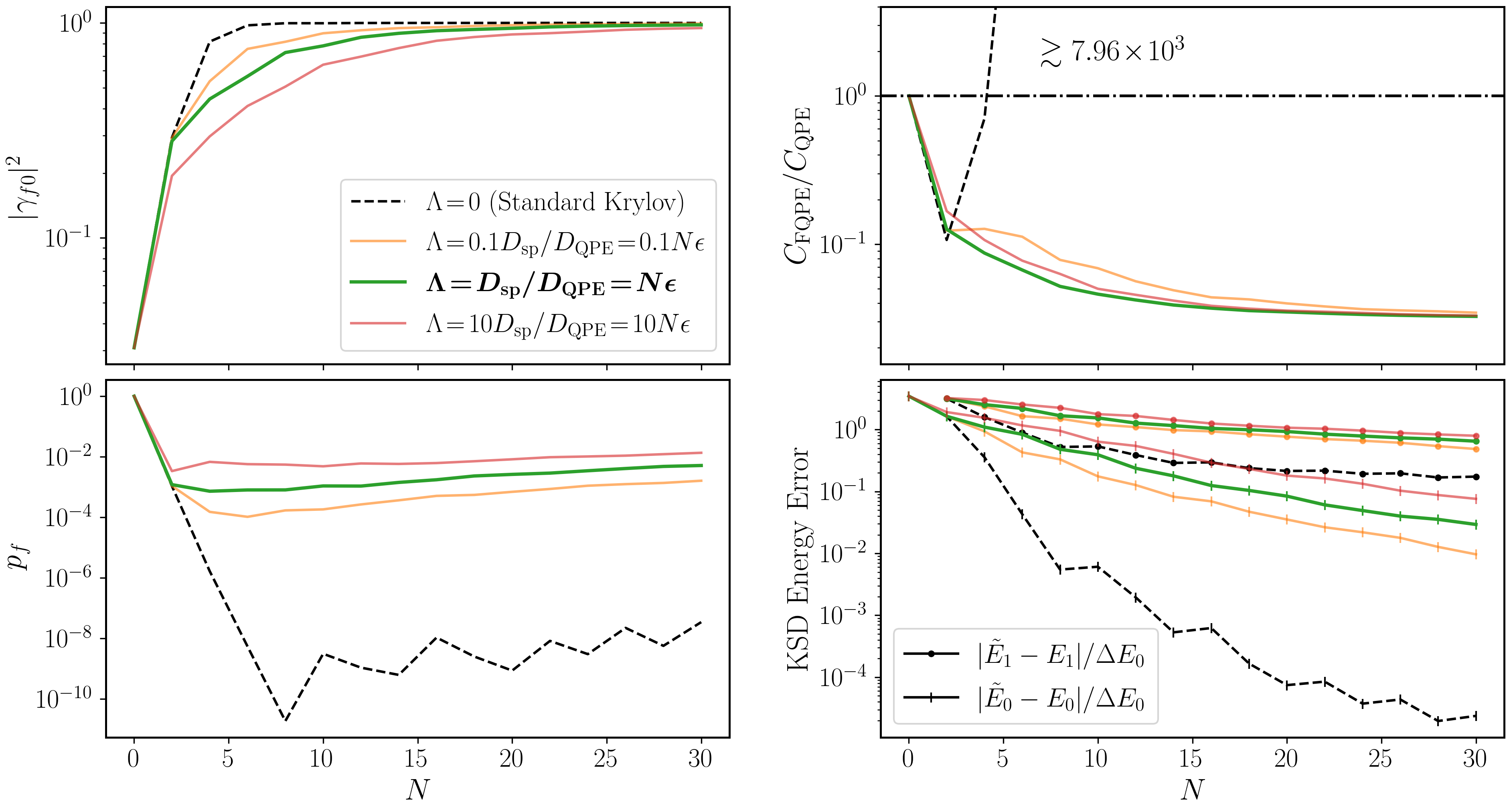
        }
    }\\
    \subfigure[~$N_{\mathrm{site}}=2\times3$, Polynomial Krylov Filter]{
        \includegraphics[width=0.9\textwidth]{
            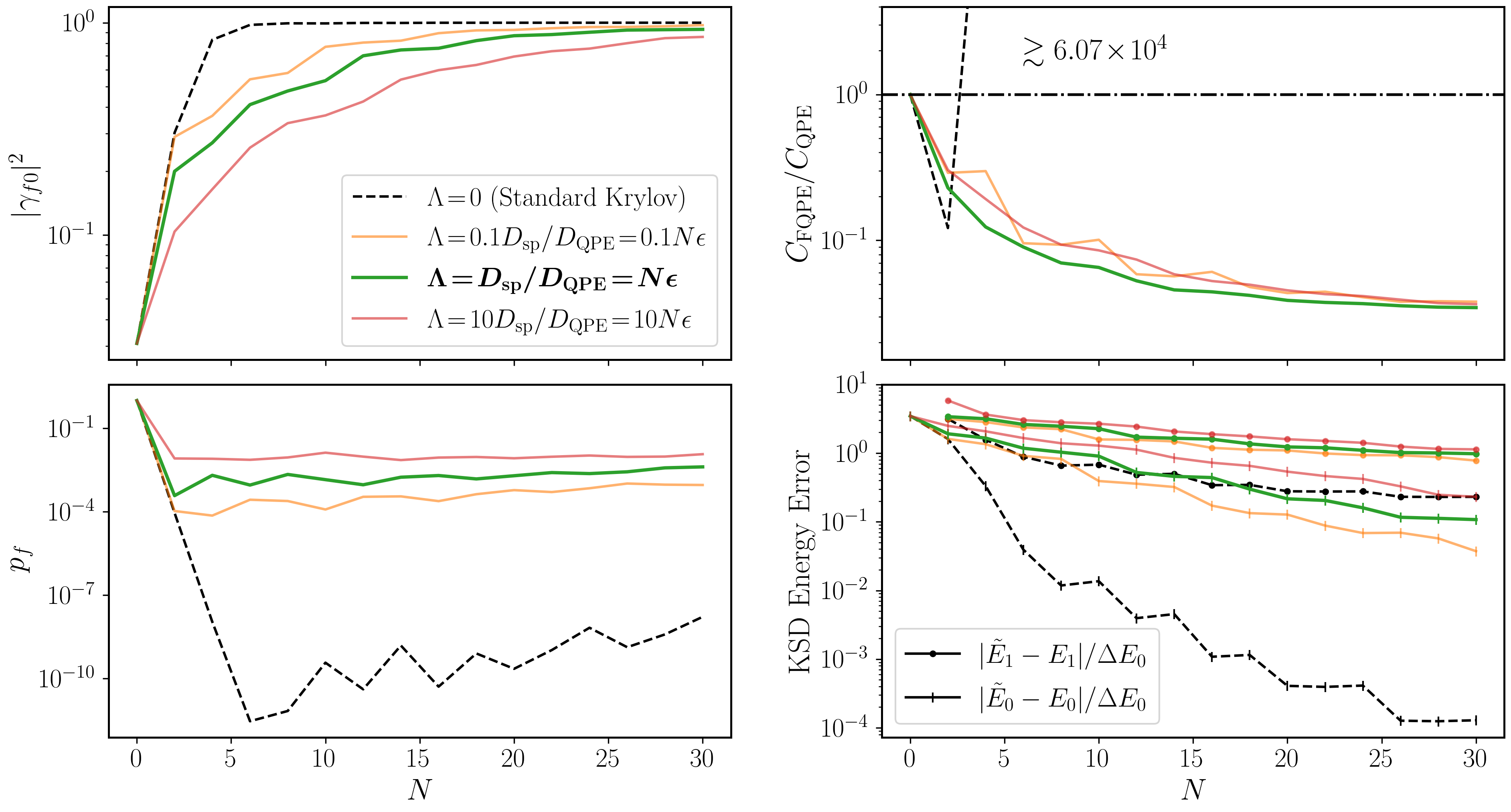
        }
    }
    \caption{
        Properties of trigonometric and polynomial (modified) Krylov filters applied to the Hubbard model with $N_{\mathrm{site}}=2\times3$.
        Further description is analogous to the caption of Fig.~\ref{fig:krylov_filter}.
    }
\end{figure*}

\newpage
\begin{figure*}
    \centering
    \subfigure[~$N_{\mathrm{site}}=7$, Polynomial Krylov filter and filtered state]{
    \includegraphics[width=0.96\textwidth]{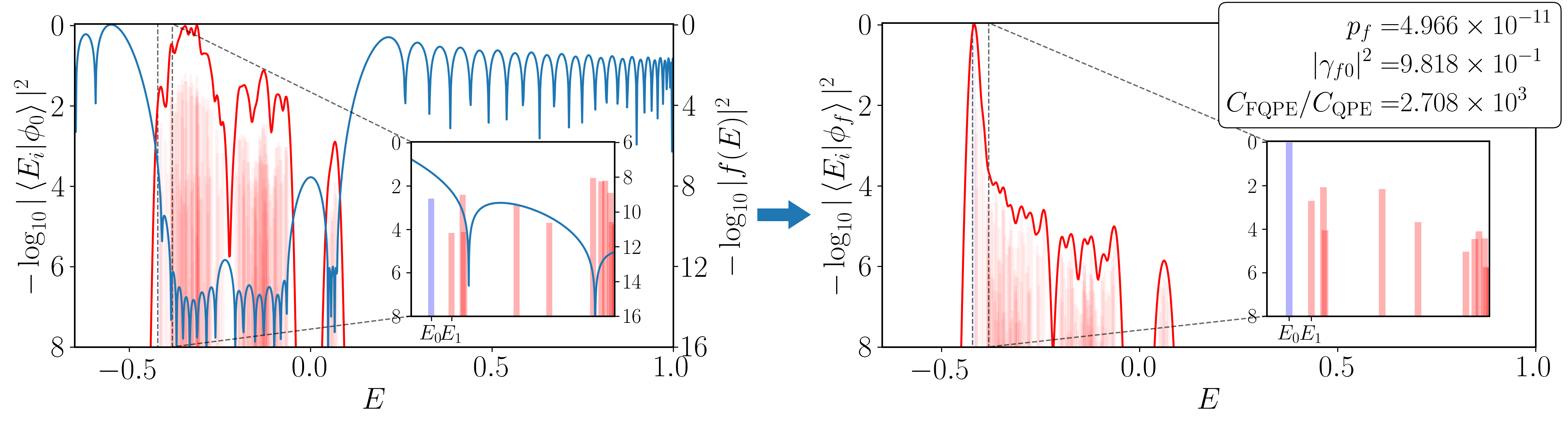}
    }\\
    \subfigure[~$N_{\mathrm{site}}=7$, Polynomial Modified Krylov filter and filtered state]{
    \includegraphics[width=0.96\textwidth]{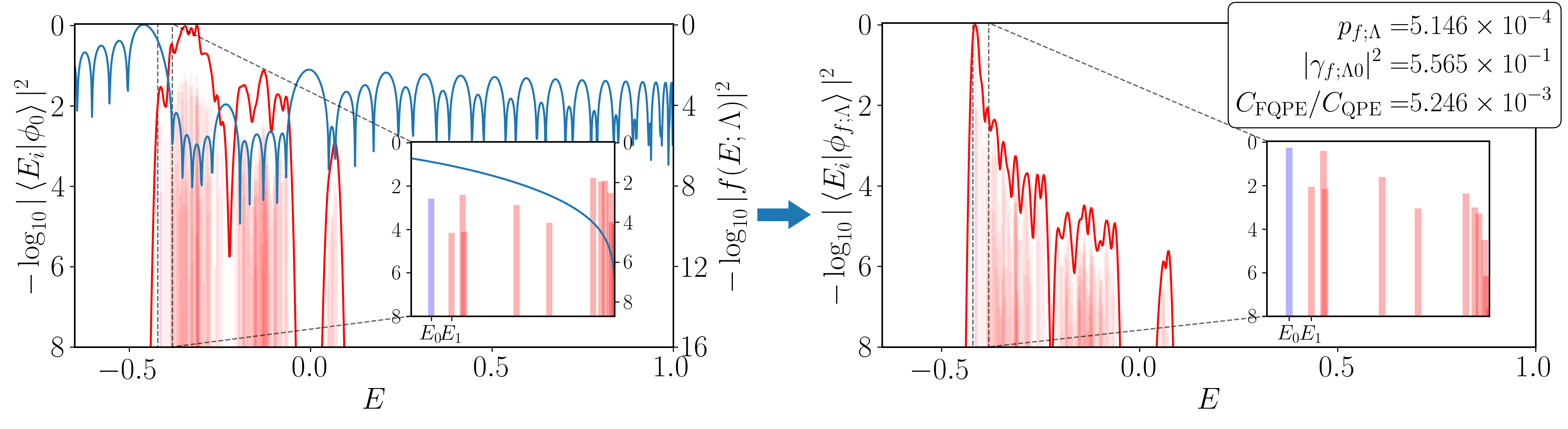}
    }\\
    \subfigure[~$N_{\mathrm{site}}=7$, Polynomial Gaussian filter and filtered state]{
    \includegraphics[width=0.96\textwidth]{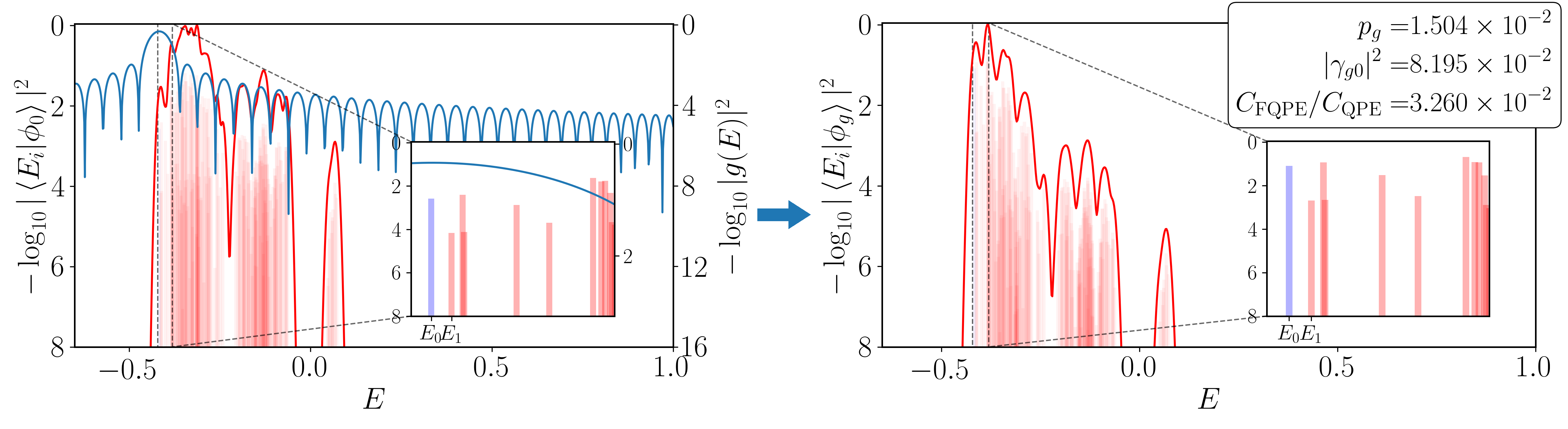}
    }
    \caption{
        Polynomial Krylov, modified Krylov, and Gaussian filter functions, together with the energy histograms of the reference state and the filtered states, including zoomed views near the ground-state energy.
        Further description is analogous to the caption of Fig.~\ref{fig:filter_comparison}.
    }
\end{figure*}

\clearpage
\clearpage
\bibliography{biblio}
\end{document}